# Unraveling multi-state molecular dynamics in single-molecule FRET experiments

# Part II: Quantitative analysis of multi-state kinetic networks

Oleg Opanasyuk[1,*], Anders Barth[1,a,*], Thomas-Otavio Peulen[1,b], Suren Felekyan[1], Stanislav Kalinin[1], Hugo Sanabria[2,‡], Claus A.M. Seidel[1,‡]

[1] Institut für Physikalische Chemie, Lehrstuhl für Molekulare Physikalische Chemie, Heinrich Heine Universität, Düsseldorf, Germany

[2] Department of Physics and Astronomy, Clemson University, Clemson, S.C., USA

[a] Present address: Department of Bionanoscience, Kavli Institute of Nanoscience, Delft University of Technology, Delft, The Netherlands

[b] Present address: Department of Bioengineering and Therapeutic Sciences, University of California, San Francisco, California, USA

*Contributed equally

[‡] Corresponding authors: cseidel@hhu.de, hsanabr@clemson.edu

**Abstract**

Single-molecule Förster Resonance Energy Transfer (smFRET) is ideally suited to resolve the dynamics of biomolecules. A significant challenge to date is capturing and quantifying the exchange between multiple conformational states, mainly when these dynamics occur on the sub-millisecond timescale. Many methods for the quantitative analysis are challenged if more than two states are involved, and the appropriate choice of the number of states in the kinetic network is difficult. An additional complication arises if dynamically active molecules coexist with solely static molecules in similar conformational states. To address these problems, we developed an integrative analysis framework that combines the information from FRET-lines, time-correlated single photon counting, and fluorescence correlation spectroscopy. While individually, these methodologies provide ambiguous results for quantitatively characterize the dynamics in complex kinetic networks, the global analysis enables accurate determination of the number of states, their kinetic connectivity, and the transition rate constants. To challenge the potential of smFRET to study multi-state kinetic networks, we benchmark the framework using synthetic data of three-state systems with different kinetic connectivity and interconversion rates.

**Table of Contents**

# 1 Analysis of multi-state kinetic networks

Biomolecular dynamics are often complex, involving multiple conformational states and sub-states that interconvert over a wide range of timescales from nanoseconds to minutes and hours. Single-molecule FRET (smFRET) experiments provide a wealth of information about the molecular system and are ideal for resolving these dynamics. In multiparameter fluorescence detection (MFD) experiments on freely diffusing molecules, this information is encoded in the time-ordered sequence of the detected photons recorded with picosecond resolution.

Various analysis methods have been developed over the years to obtain quantitative information on the structural dynamics of the biomolecular systems. The most widely used methods are the statistical analysis of FRET-efficiency histograms (Photon Distribution Analysis, PDA)[1-4], intensity-based fluorescence correlation spectroscopy (FCS)[5-8], and time-resolved fluorescence decay analysis (TCSPC)[9], but other approaches have been applied as well[10-12]. Each representation of the experimental data and the corresponding analysis method has its strengths and weaknesses to determine the fluorescence properties of the species, detect their kinetic connectivity, and quantify the rate constants (Table 1). Established methods such as FCS and TCSPC are computationally fast. They rely on established algorithms to find the optimal parameters of a physical or empirical model that describe the experimental data. While TCSPC is ideally suited to resolve the FRET efficiencies of the contributing states, conformational dynamics from nano- to milliseconds can be resolved by fluorescence correlation spectroscopy (FCS), fluorescence lifetime correlation spectroscopy (FLCS), and filtered-FCS, which utilize statistical weighting to recover species-specific correlation curves[6, 13, 14]. These correlation approaches work well for homogenous samples of dynamic molecules interconverting between two conformational states. However, they are challenged by the increased complexity of many biological systems that involve three or more states interconverting on different kinetic timescales or contain heterogeneous mixtures of static and dynamic molecules. Therefore, to unravel the complex dynamics of such systems, a holistic approach combining multiple methods is required.

Existing methods for the quantitative analysis of dynamics are applied to a reduced representation of the single-photon-counting data. At the same time, the full potential of the multidimensional dataset is not utilized. This multidimensional information is revealed in the pairwise histograms of averaged fluorescence observables - although the informational content is likewise reduced due to the averaging performed over each single molecule event[15]. One example of how the multidimensional information can be utilized is the pairwise plot of the intensity-based FRET efficiency $E$ and the intensity-weighted average donor fluorescence lifetime $\langle \tau_{D(A)} \rangle_F$. The correlation between these two FRET indicators enables the detection of conformational dynamics. Further, using FRET-lines to describe the exchange between different states, the connectivity within the kinetic network is revealed as described in Part I[3]. Ideally, a multi-state kinetic model would be directly fit this multidimensional data set. However, to our knowledge, a quantitative description of the complete multidimensional histogram is currently limited to computationally expensive stochastic simulations. The stochastic noise of Monte-Carlo simulations also makes this approach difficult to apply in optimization routines, which converge more rapidly if analytical expressions are employed. While such expressions are known for simple cases[16], they are currently unavailable for the multi-state networks discussed in this work.

Here, we take a step towards a holistic analysis framework by combining FRET-lines, TCSPC, and FCS in a global approach to quantify the exchange in multi-state kinetic networks. In a first step, the correct kinetic model is identified by a graphical analysis using FRET-lines, defining the number of states and their linkage. The exchange rates are then quantified using a global analysis of the donor fluorescence decay and the color correlation functions. The framework is applied to simulated data sets of multi-state systems with a binary exchange between two states in the presence of a background of static molecules. When only the TCSPC and FCS information is used, ambiguous solutions are obtained that differ in the kinetic connectivity of the species and the fraction of molecules participating in the dynamic exchange. To resolve this ambiguity, FRET-lines provide a graphical analysis of the kinetic connectivity of states and permits the estimation of the equilibrium constant from the peak of the dynamic population in binary systems. For systems involving a fast dynamic exchange between more than two states, additional information is required. Using simulations of three-state systems, we illustrate the potential of filtered-FCS to detect the direct exchange between different species in complex networks and deduce the kinetic linkage, even in this challenging case. Lastly, we derive relations between the correlation amplitudes and the single-molecule FRET indicators $E$ and $\langle \tau_{D(A)} \rangle_F$, highlighting the connections between the different representations of the data and the future possibility to extend this holistic approach to data analysis.

## 2 Nomenclature for multi-state systems in smFRET

Purely static or dynamic biomolecules are rarely found in nature. It is often observed that biomolecules can be activated through allosteric effects such as binding of cofactors or regulators, post-translational modifications, or conformational changes in associated domains, switching the molecule from a static into a dynamic state[17-19]. In such situations, molecules with the same FRET efficiency may either be static or participate in the conformational dynamics, introducing a degeneracy into the analysis in which the same observed FRET species may belong to different states that are either static or dynamic.

In FRET experiments on freely diffusing single molecules, the accessible timescale of dynamics is limited by the diffusion time to >0.1 ms$^{-1}$ causing additional complications because transitions between conformational states on slower timescales could appear as *pseudo-static* populations in the analysis.

To avoid confusion about the physical description of the biomolecular system as static or dynamic and to classify observed populations in the experiment, we propose a concise nomenclature for smFRET experiments in Figure 1. A *conformational state* $C^{(i)}$ is defined as a distinct structural state of the biomolecule that can be classified as static or be in dynamic exchange with other conformational states. The alternation between dynamic and static states of the biomolecule may be subject to allosteric regulation, biomolecular interactions, or covalent modifications. In the smFRET experiment, conformational states are observed indirectly through the fluorescence properties of the covalently linked dyes, such as the FRET efficiency, fluorescence lifetime, or fluorescence anisotropy of the donor and acceptor fluorophores. An *observed fluorescence species* $o^{(i)}$ is generally assigned to a single conformational state. However, due to quenching or sticking of the fluorophores, different fluorescence species may belong to the same conformational state. On the other hand, multiple conformational states may belong to the same fluorescence species if the fluorescence properties do not change significantly (see Supplementary Note 1 for an overview of potential ambiguities). In smFRET experiments, fluorescence species are observed as *populations* $p^{(i)}$ in the one- or two-dimensional histograms. Due to dynamic averaging during the diffusion time, a population may originate from a mixture of different fluorescence species. Dynamic and static populations may be distinguished in a plot of the FRET efficiency $E$ against the donor fluorescence lifetime $\langle \tau_{D(A)} \rangle_F$ (Figure 1, bottom). The static populations, originating from the fluorescence species $o^{(1)}$ and $o^{(2)}$, lie on the static FRET-line. In contrast, the dynamic population $p^{(1,2)}$ shows the characteristic dynamic shift (ds) from the static FRET-line, as introduced in Part I. The heterogeneity within the dynamic populations can be resolved by a sub-ensemble analysis of the fluorescence decays.

The assignment of static and dynamic populations is complicated when the sample contains a mixture of static and dynamic conformational states of identical FRET efficiencies (Figure 1, right). In the limit of fast dynamic exchange, a dynamic population is shifted from the static FRET-line and separated from static populations. As the conformational exchange becomes slower and approaches the diffusion time of the molecule, there is a probability that dynamic molecules do not undergo a conformational change during the observation time. While originating from dynamic conformational states, these single-molecule events will show as a *pseudo-static* population on the static FRET-line and are difficult to separate from actual static populations. As shown below, the fraction of dynamic molecules is a central parameter in the quantitative analysis of such heterogenous multi-state systems by correlation methods. In the next section, we will first show how a graphical analysis can be employed to estimate the equilibrium constant of the dynamic exchange in the background of static states.

**Table 1:** Comparison of methods to analyze conformational dynamics of multi-state kinetic networks in smFRET. The methods are compared to their ability to identify the different states (fluorescence properties and fractions), kinetic connectivity, dynamic exchange, and the accessible timescales. MLE: Gopich-Szabo photon trajectory analysis using maximum likelihood estimation[20], BVA: burst variance analysis[21], FRET-2CDE: FRET two-channel kernel density estimator[22]. * Timescale of dynamics ≥ 500 µs. ** The lower limit depends on the average inter-photon time. Faster timescales are accessible for higher signal count rates.

| Method | Identification of states | Kinetic connectivity | Quantification of dynamics | Accessible timescale |
|---|---|---|---|---|
| **TCSPC** | + | - | - | - |
| **FCS** | - | - | + | ns-ms |
| **MLE** | o | o | + | µs-ms** |
| **PDA/histogram analysis** | for slow dynamics* | - | + | 100 µs – 10 ms** |
| **BVA/FRET-2CDE** | for slow dynamics* | + | qualitatively | 100 µs – 10 ms** |
| **2D histogram: E vs.$\langle \tau_{D(A)} \rangle_F$** | for slow dynamics* | + | qualitatively | ns-ms |

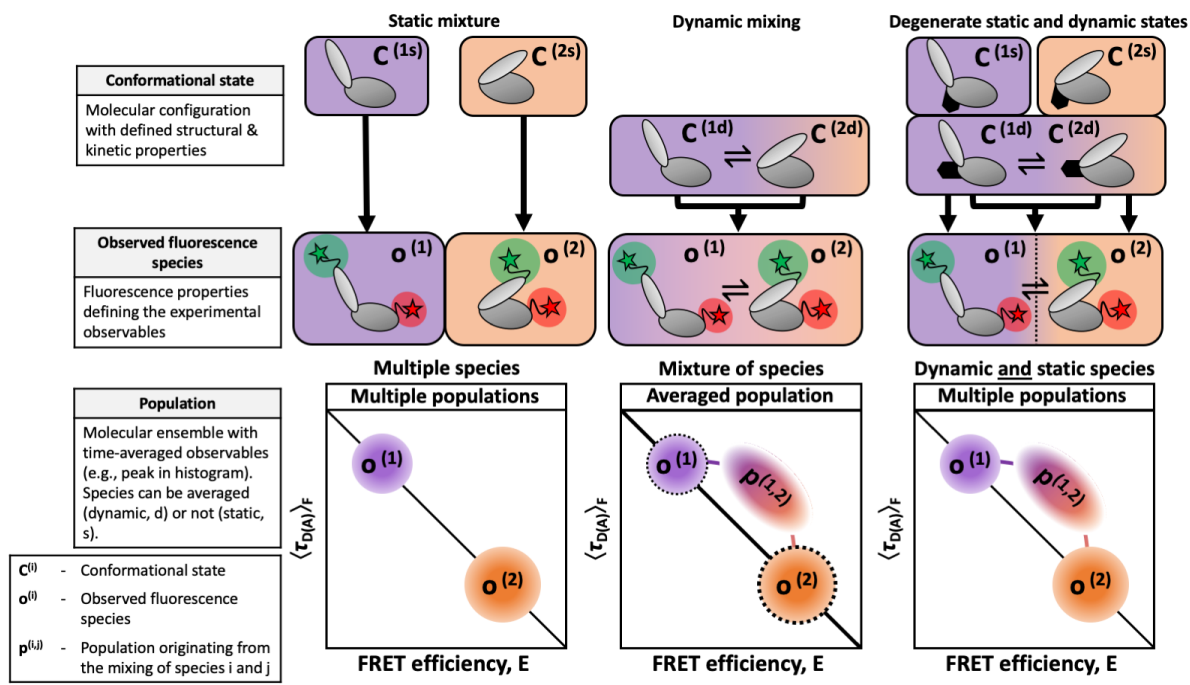

**Figure 1: Definitions in smFRET experiments of multi-state systems. Top row:** A *conformational state* is defined as a distinct structural state of a biomolecule, which may be found in dynamic exchange with other states or trapped in a static configuration. The dynamic behavior may be controlled by a conformational switch in an associated domain (black). **Middle row:** A *fluorescence species* is defined by the fluorescence properties such as the FRET efficiency, fluorescence lifetime, and fluorescence anisotropy of the donor and acceptor dyes. All members of a fluorescent species have identical fluorescence properties. Different conformational states may belong to the same fluorescent species if the structural change does not affect the properties of the fluorescent probe. Different fluorescent species are often found to belong to a single conformational state, e.g., due to interactions of the fluorophore with the biomolecular surface. Transient quenching or binding of the dyes to the surface would, for example, result in different fluorescent species that relate to the same conformational state (see Supplementary Note 1 for details). The assignment of fluorescent species to conformational states is an interpretative step in the analysis and usually requires prior structural knowledge. **Bottom row:** On the level of the experiment, one observes *populations* in the one- or two-dimensional histograms of the FRET efficiency $E$ against the donor fluorescence lifetime $\langle \tau_{D(A)} \rangle_F$. Populations are clusters of single-molecule events with identical observed fluorescence properties. A population may represent a single fluorescence species or, in the case of fast exchange between different fluorescent species, may originate from a heterogeneous mixture of different fluorescent species. Populations that fall on the static FRET-line (black diagonal line) represent a single fluorescence species, while shifted populations indicate dynamic exchange. Heterogeneity within the population may be revealed by a sub-ensemble analysis, e.g., of the fluorescence decay.

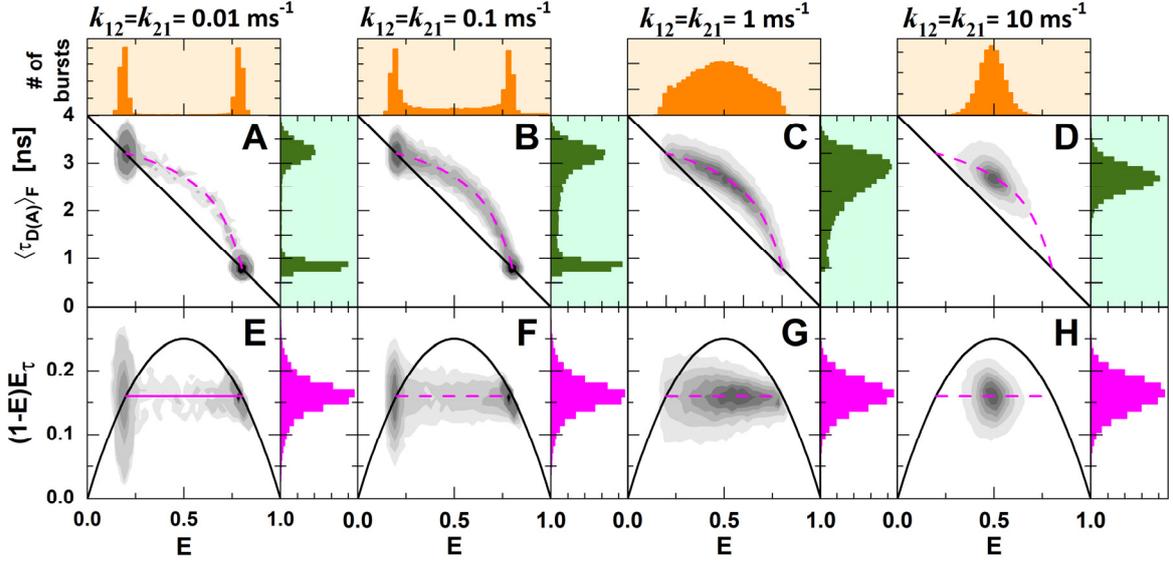

**Figure 2: Simulated smFRET experiments for a two-state system exchanging at increasing interconversion rates between states with FRET efficiencies of $E_1 = 0.2$ and $E_2 = 0.8$.** Shown are the two-dimensional histograms of $\langle \tau_{D(A)} \rangle_F$ vs. $E$. (A-D) and the difference between the first and second moments of the lifetime distribution, $(1-E)E_\tau$ vs. $E$ (E-H) for different timescales of conformational dynamics from slow ($k_{12} = k_{21} = 0.01$ ms$^{-1}$) to fast exchange ($k_{12} = k_{21} = 10$ ms$^{-1}$). The population's shape and location indicate the dynamic timescale, while the dynamic FRET-line is independent of the rate constants and describes all timescales. The diffusion time in all simulations was $t_{\text{diff}} = 1.5$ ms.

## 3 Graphical analysis of dynamic populations

Quantitative information on conformational dynamics is encoded in the shape of the FRET efficiency histogram, as in dynamic photon distribution analysis (PDA)[3, 23, 24]. These analyses, however, are challenged if the experiment contains a mixture of static and dynamic molecules due to the difficulty of distinguishing actual static and pseudo-static molecules. Pseudo-static molecules are dynamic molecules that, by chance, remained in one conformational state during the transit through the observation volume. This section will describe how the separation of static and dynamic molecules in the $E$-$\langle \tau_{D(A)} \rangle_F$ histogram can provide quantitative information on the equilibrium constant by a graphical analysis of the peak of the dynamic population.

In the description of FRET-lines, the timescales of the dynamics are not considered explicitly. For a dynamic system, the distribution of the state occupancies $x^{(i)}$ depends on the microscopic interconversion rates and the observation time[3, 16, 25]. For the calculation of dynamic FRET lines in Part I, we have instead considered all possible values for the state occupancy, $x^{(1)} \in \{0,1\}$. In other words, we have replaced the true distribution of the state occupancies by a uniform distribution with equal probability for all values of $x^{(1)}$ ($p(x^{(1)}) = $ const.). FRET lines may, however, still be used to address the timescale of dynamics qualitatively. In the absence of dynamics, the two-dimensional histogram will reveal distinct static populations as limiting states, which fall onto the static FRET line. In the case of fast exchange between distinct FRET states, the conformational dynamics are averaged for every single-molecule event, resulting in a single population representing the equilibrium. "Fast" interconversion relates to the timescale of diffusion (~1-5 ms), and generally classifies processes on a timescale of 100 μs and below. This single peak will be shifted from the static-FRET line in the two-dimensional histogram, as described before. The slow transition between limiting states, such as the dynamics on the timescale of diffusion or slower, leads to a broadening of the observed distributions; thus, the shape of the distribution depends on the timescale of the dynamics.

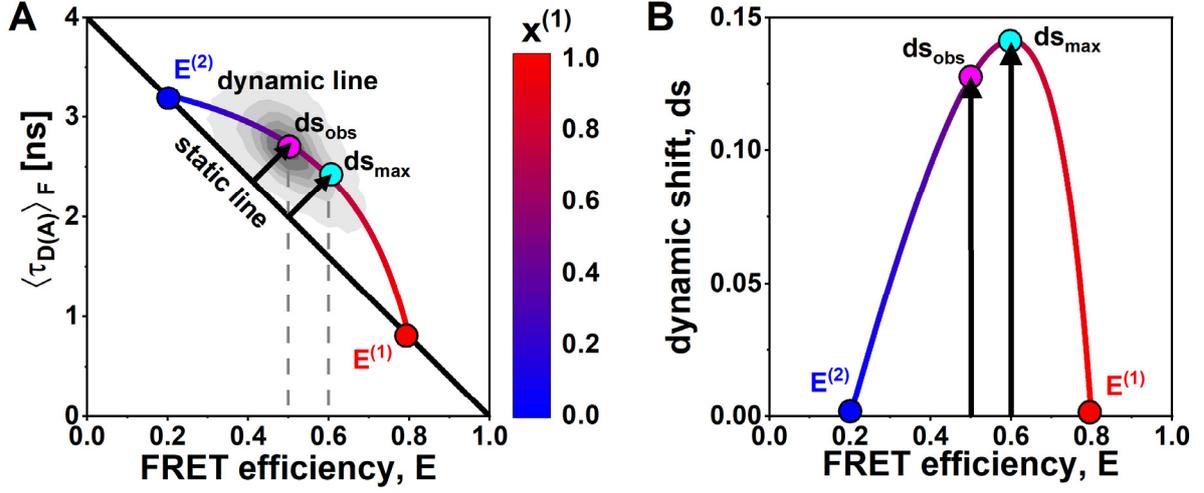

**Figure 3: The dependence of the observed dynamic shift on the average FRET efficiency. A)** The dynamic FRET-line for a two-state system with FRET efficiencies $E^{(1)} = 0.8$ (red) and $E^{(2)} = 0.2$ (blue) reaches a maximum dynamic shift, $ds_{max}$ (cyan), as indicated by the arrow, at a FRET efficiency of $E = 0.6$, corresponding to a species fraction of $x^{(1)} = 2/3$. A smaller dynamic shift, $ds_{obs}$ (magenta), is observed for the simulation with rate constants of $k_{12} = k_{21} = 10$ ms$^{-1}$ ($x^{(1)} = 0.5$, compare Figure 2D). The static FRET-line is given in black, and the dynamic FRET-line is color-coded according to the species fraction $x^{(1)}$. **B)** The dependence of the dynamic shift on the average FRET efficiency $E$ for the two-state system shown in panel A.

To illustrate this effect, we performed simulations of a two-state system with FRET efficiencies of 0.2 and 0.8 (Figure 2). We set the backward and forward rates equal ($k_{12} = k_{21}$) and varied them from 0.01 ms$^{-1}$ to 10 ms$^{-1}$, at a constant diffusion time $t_{diff} = 1.5$ ms. When the rate constants are significantly slower than the inverse diffusion time, $1/t_{diff} = 0.67$ ms$^{-1}$, the two subpopulations are separated because molecules do rarely interconvert during the observation time (Figure 2 A,E). With increasing rate constants, the molecules are more likely to change their state during the observation time, resulting in single-molecule events with intermediate FRET efficiencies (Figure 2 B-C, F-G), while complete averaging is observed at fast interconversion rates (Figure 2 D,H). However, the dynamic FRET line describes all possible mixing ratios between the involved states, regardless of the magnitude of the rate constants, and applies to all cases.

In the first part of the paper, we defined the observed deviation of a population perpendicular to the static FRET-line as the *dynamic shift* (Figure 3A). Interestingly, the dynamic shift of the population for the simulated system with $k_{12} = k_{21}$ does not reach its maximum possible value even for fast dynamics ($k_{12} = k_{21} = 10$ ms$^{-1}$, Figure 3A and Figure 2D). In the limiting case of fast dynamics, the observed dynamic shift of the dynamically-averaged population depends on the FRET efficiencies ($E^{(1)}$ and $E^{(2)}$) and the species fractions ($x^{(1)}$ and $x^{(2)}$) of the two states (see Supplementary Note 2):

$$ds(x^{(1)}) = \frac{1}{\sqrt{2}}\left(E^{(2)} - E^{(1)}\right)^2 \frac{x^{(1)}(1 - x^{(1)})}{x^{(1)}(1 - E^{(1)}) + (1 - x^{(1)})(1 - E^{(2)})}, \quad (1)$$

where $x^{(1)}$ is the species fractions of state 1. The dependence of the observed dynamic shift on the average FRET efficiency of the dynamic population, $E$, is illustrated in Figure 3B. The maximum dynamic shift, $ds_{max}$, is given by (see Part I):

$$ds_{max} = \frac{1}{\sqrt{2}}\left(\sqrt{1 - E^{(1)}} - \sqrt{1 - E^{(2)}}\right)^2, \quad (2)$$

which is obtained at a species fraction $x^{(1)}_{max}$ of:

$$x^{(1)}_{max} = \frac{\sqrt{1 - E^{(2)}}}{\sqrt{1 - E^{(1)}} + \sqrt{1 - E^{(2)}}}, \quad (3)$$

corresponding to a value of $x_{\max}^{(1)} = 2/3$ in the given example. The maximum dynamic shift defined by eq. (2) thus serves as a figure of merit to judge whether dynamics could be detected in the experiment in the ideal case, while the observed value of the dynamic shift also depends on the equilibrium constant according to eq. (1).

Beyond the qualitative assessment of dynamics by visual inspection of the two-dimensional histogram, it is possible to extract quantitative information about the dynamic equilibrium from the two-dimensional histograms. The equilibrium species fractions $x_d^{(1)}$ and $x_d^{(2)}$ (i.e., the fraction of molecules in state 1 or 2) can be determined from the average FRET efficiency for all single-molecule events, $\langle E \rangle_{\exp}$, by

$$x_d^{(1)} = \frac{k_{12}}{k_{21} + k_{12}} = \frac{\langle E \rangle_{\exp} - E^{(2)}}{E^{(1)} - E^{(2)}}; \qquad x_d^{(2)} = 1 - x_d^{(1)}, \tag{4}$$

where $E^{(1)}$ and $E^{(2)}$ are the FRET efficiencies of the limiting states. From the species fractions, the equilibrium constant can be calculated, which relates to the kinetic rates by:

$$K = \frac{x_d^{(2)}}{x_d^{(1)}} = \frac{k_{21}}{k_{12}}. \tag{5}$$

While for purely dynamic systems $\langle E \rangle_{\exp}$ is readily calculated from the experimental dataset, the presence of additional static states will result in incorrect values for the species fractions. In the case of a mixture of dynamic and static molecules, it would be advantageous if the equilibrium fraction could be obtained from the position of the dynamic population alone, which is most easily defined by its peak value or mode. For fast interconversion, the mode of the population in the two-dimensional histogram directly corresponds to the average FRET efficiency $\langle E \rangle_{\exp}$. If the timescale of kinetics becomes comparable to the diffusion time, the mode of the dynamic distribution deviates from the actual value of the species fraction $x^{(1)}$. To study this effect, we consider the distribution of FRET efficiencies for a dynamic system explicitly. The average FRET efficiency within a single-molecule event, $E$ depends on the fraction of time spent in the different states, equivalent to the species fractions $x^{(i)}$:

$$E = x^{(1)} E^{(1)} + (1 - x^{(1)}) E^{(2)} \tag{6}$$

Here, $x^{(1)}$ is the species fraction of state 1 in a single-molecule event, which is different from the equilibrium species fraction $x_d^{(1)}$ discussed before. If we know the state occupancy distribution of $x^{(1)}$, $P(x^{(1)})$, we can calculate the distribution of FRET efficiencies $P(E)$. In general, $P(x^{(1)})$ takes a complex mathematical form (a complete derivation is given in Supplementary Note 3), but can be simplified to the sum of three terms:

$$P(x^{(1)}) = \frac{k_{12}}{k_{12} + k_{21}} e^{-k_{21} T} \delta(1 - x^{(1)}) + \frac{k_{21}}{k_{12} + k_{21}} e^{-k_{12} T} \delta(x^{(1)}) + \xi_{12}(x^{(1)}) \tag{7}$$

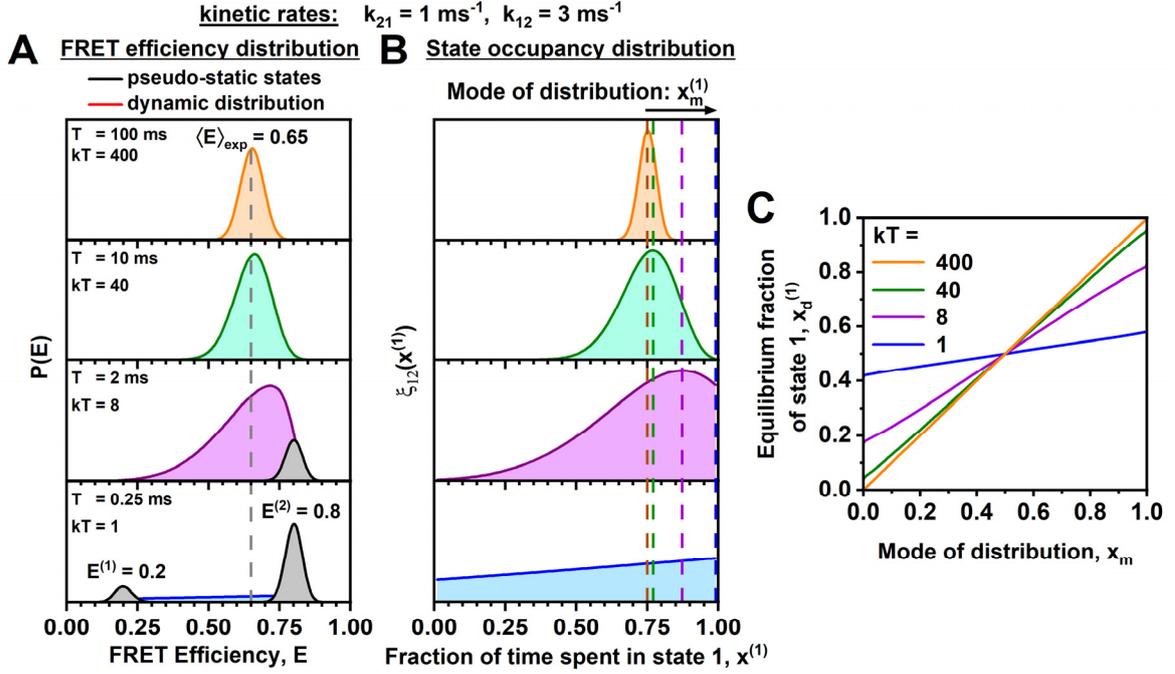

**Figure 4: Extracting equilibrium fractions from the dynamic population for a two-state system. A)** Simulated FRET efficiency distributions for a two-state system with interconversion rates of $k_{21} = 1$ ms$^{-1}$ and $k_{12} = 3$ ms$^{-1}$ and FRET states of $E^{(1)} = 0.2$ and $E^{(2)} = 0.8$ at varying observation times $T$. The distribution is split into the pseudo-static contribution of molecules that did not interconvert during the observation time (black) and the dynamic contribution of molecules that showed one or more transitions (red). The average FRET efficiency $\langle E \rangle_{exp} = 0.65$ is independent of the observation time and reflects the equilibrium between the two states. **B)** The dynamic part of the distribution of the occupancy of state 1, $\xi(x^{(1)})$, of the FRET efficiency histograms shown in A. At shorter observation times, the mode of the distribution $x_m$ deviates from the equilibrium fraction $x_d^{(1)} = \frac{k_{12}}{k_{12}+k_{21}} = 0.75$. **C)** The relationship between the mode of the exchange distribution and the equilibrium fraction is found to be approximately linear. This allows for a direct conversion of the measured modal value into the equilibrium fraction if the sum of the rates, $k = k_{12} + k_{21}$, is known.

Here, the first and the second terms describe the probability that the molecule was in state 1 or 2 at the beginning of the observation time $T$ and did not interconvert (pseudo-static states). Also, the term $\xi_{12}(x^{(1)})$ describes the dynamic part of the distribution of $x^{(1)}$, such as in the case that molecules switched between the states at least once during its observation time. This term is given by[3, 25-27]:

$$\xi_{12}(x^{(1)}) = \frac{T}{\frac{1}{k_{12}} + \frac{1}{k_{21}}} \times e^{-(k_{12}x^{(2)} + k_{21}x^{(1)})T}$$

$$\times \left[ 2I_0\left(2T\sqrt{k_{21}k_{12}x^{(1)}x^{(2)}}\right) + \frac{x^{(1)}k_{12} + x^{(2)}k_{21}}{\sqrt{k_{21}k_{12}x^{(1)}x^{(2)}}} I_1\left(2T\sqrt{k_{21}k_{12}x^{(1)}x^{(2)}}\right) \right] \quad (8)$$

where $x^{(2)} = 1 - x^{(1)}$ and $I_0$ and $I_1$ are the modified Bessel functions of the first kind of order zero and one, respectively. For fast interconversion (or long observation times), the pseudo-static terms vanish while only the dynamic term remains. On the other hand, for slow dynamics, the static terms tend to the equilibrium fractions of the two states, while the dynamic term vanishes.

Theoretical FRET efficiency distributions for a two-state dynamic system are given in Figure 4A for rate constants of $k_{21} = 1$ ms$^{-1}$ and $k_{12} = 3$ ms$^{-1}$ at different observation times $T$. For short observation times ($T = 0.25$ ms), only the pseudo-static peaks remain, while complete averaging is observed for long observation times ($T = 100$ ms). The average FRET efficiency $\langle E \rangle_{exp}$ of 0.65 relates to the equilibrium species fractions by eq. (4). Considering the case that additional static species contribute to the average FRET efficiency, we would like to infer the equilibrium species fraction only from the dynamic part of the distribution $\xi_{12}(x^{(1)})$ (Figure 4B). The property of $\xi_{12}(x^{(1)})$ that is most easily inferred from the two-dimensional histograms is its maximum or peak (compare Figure 2). For an observation time $T = 100$ ms, the modal value $x_m^{(1)}$ corresponds to the equilibrium fraction of state 1, $x^{(1)} = 0.75$ (Figure 4B, top). However, as the observation time decreases, $x_m^{(1)}$ deviates from the equilibrium fraction to the point where the modal value coincides with the pseudo-static population ($x_m^{(1)} = 1$, Figure 4B bottom). Fortunately, the relationship between the actual equilibrium fraction of state 1, $x_d^{(1)}$, and the modal value of the distribution, $x_m^{(1)}$, is approximately linear, enabling a simple conversion between the two quantities by (Figure 4C and Supplementary Note 3):

$$x_d^{(1)} \approx x_{d,\lim}^{(1)} + (1 - 2\, x_{d,\lim}^{(1)}) x_m^{(1)} \tag{9}$$

where $x_{d,\lim}^{(1)}$ is the limiting equilibrium fraction at $x_m = 0$ (i.e., the ordinate intercept in Figure 4C) that depends only on the average number of transitions during the observation time given by $(k_{21} + k_{12})T$:

$$x_{d,\lim}^{(1)}(kT) = \frac{3}{2}\left(1 + \frac{kT}{2}\left(1 + \frac{I_0\left(\frac{kT}{2}\right)}{I_1\left(\frac{kT}{2}\right)}\right)\right)^{-1} \tag{10}$$

where $k = k_{21} + k_{12}$.

The model value of the species fraction, $x_m^{(1)}$, can be obtained from the modal value of the FRET efficiency distribution, $E_m$, obtained by graphical analysis if the FRET efficiencies of the limiting states are known:

$$x_m^{(1)} = \frac{E_m - E^{(2)}}{E^{(1)} - E^{(2)}} \tag{11}$$

If the sum of rates $k$ is known from FCS analysis or other methods, the equilibrium fraction and the equilibrium constant can be determined from equations (9) and (10), which allows determining the microscopic rate constants quantitatively using the equilibrium information obtained from the graphical analysis of the two-dimensional histograms. While the approach in principle requires observation time windows of identical length $T$, it may, as an approximation, be set to the diffusion time for datasets of single-molecule events of freely diffusing molecules.

## 4 Global analysis of multi-state dynamics

### 4.1 Analytical description of FCS curves

Fluorescence Correlations Spectroscopy (FCS) relies on the fluctuations of recorded signals to characterize molecular interactions such as binding and unbinding, chemical kinetics, and diffusion of fluorescent molecules[5, 8, 28]. Also, when combined with FRET, FCS, the conformational dynamics can be quantitatively determined[29-33]. Typically, the fluorescence signals are collected over specific spectral detection windows. Here, we refer to the correlation analysis of the fluorescence intensities of a donor and an acceptor fluorophore (monitored in two detection channels commonly named "green" and "red") as *color-FCS*. We avoid the conventional abbreviation FRET-FCS to differentiate it from the related method of filtered-FCS, which relies on FRET to distinguish different species but does not explicitly use color channels. Analytical models for color-FCS are usually limited to kinetic networks involving two states due to the increased number of parameters of multi-state systems and the limited experimental information available[32]. Advanced correlation methods take advantage of the lifetime information available with pulsed laser excitation, allowing one to interrogate biomolecular dynamics by two-dimensional maps of fluorescence decays[34, 35] or filtered correlation algorithms (fFCS)[6, 13, 36, 37]. By using an additional dimension of the collected data, these methods offer the potential to interrogate more complex kinetic networks. Quantitative analysis of FCS experiments requires a set of model-specific analytic functions that describe the time-evolution of the correlations. Often, smFRET and FCS experiments are used to study unimolecular reactions, wherein a biomolecule switches between different conformational states during the observation time. These dynamic molecules may be found together with molecules that are stable on the timescale of seconds to minutes and are thus considered as static in the single-molecule experiment.

In our model, we thus consider the coexistence of static and dynamic molecules with identical properties of the respective conformational states (Figure 1). Dynamic molecules may change between different conformational states during the observation time, resulting in variations of the fluorescence properties such as the FRET efficiency and donor fluorescence lifetime, while these properties remain constant for static molecules. For unimolecular reactions, the time evolution of the different states is described by a system of linear differential equations, which for a system with three dynamic states is given by:

$$\frac{d}{dt}\begin{pmatrix} x^{(1)} \\ x^{(2)} \\ x^{(3)} \end{pmatrix} = \begin{pmatrix} -(k_{21} + k_{31}) & k_{12} & k_{13} \\ k_{21} & -(k_{12} + k_{32}) & k_{23} \\ k_{31} & k_{32} & -(k_{13} + k_{23}) \end{pmatrix} \begin{pmatrix} x^{(1)} \\ x^{(2)} \\ x^{(3)} \end{pmatrix}$$

The rate constants $k_{ij}$ describe the rates of transition from state $j$ to state $i$. In general, this is expressed in matrix notation as:

$$\frac{dx}{dt} = \mathbf{K}\, x, \tag{12}$$

where $\mathbf{K}$ is the transition rate matrix and $x$ is the vector of the total fractions of the species. In the following, we denote the fraction of static molecules by $x_s^{(i)}$, and the fraction of dynamic molecules by $x_d^{(i)}$. Both fractions are normalized to one, i.e. $\sum_i x_s^{(i)} = 1$ and $\sum_i x_d^{(i)} = 1$. Hence, the total fraction of a species $x^{(i)}$ is given by the sum of the fractions $x_d^{(i)}$ and $x_s^{(i)}$, weighted by the total fraction of dynamic molecules $p_d$ as:

$$x = p_d x_d + (1 - p_d) x_s. \tag{13}$$

where $x_d$ and $x_s$, are the vectors of the fractions of the dynamic and the static states, and $p_d$ describes the fraction of molecules that participate in dynamic exchange.

The correlation function, $G_{ab}$, is modeled based on the set of reaction rate constants, fluorescence properties, and the population of the static states. The general definition of the correlation function between two time-dependent signals $S_a(t)$ and $S_b(t)$ is given by:

$$G_{ab}(t_c) = \frac{\langle S_a(t) S_b(t + t_c) \rangle}{\langle S_a(t) \rangle \langle S_b(t + t_c) \rangle}, \tag{14}$$

where $\langle \ldots \rangle$ denotes the time average over a long measurement. In the following, we assume that the signal fluctuations due to the diffusion of molecules and the conformational dynamics arise from independent sources (i.e., that they are statistically independent), which is generally fulfilled for experimental systems. The contributions of diffusion and dynamics can thus be treated separately, and the correlation function $G_{ab}(t_c)$ is given by the product:

$$G_{ab}(t_c) = \frac{1}{N} G_{\text{diff}}(t_c)\, G_{k,ab}(t_c) + 1, \tag{15}$$

where $t_c$ is the correlation time, $N$ is the average number of molecules in the observation volume, the factor $G_{k,ab}(t_c)$ describes the kinetic exchange and $G_{\text{diff}}(t_c)$ the diffusion of the molecules. For a 3D Gaussian detection profile, the factor $G_{\text{diff}}(t_c)$ is given by:

$$G_{\text{diff}}(t_c) = \left(1 + \frac{t_c}{t_{\text{diff}}}\right)^{-1} \left(1 + \left(\frac{w_0}{z_0}\right)^2 \frac{t_c}{t_{\text{diff}}}\right)^{-1/2}, \tag{16}$$

where $t_{\text{diff}}$ is the diffusion time. The parameters $w_0$ and $z_0$ are the width of the focal and the axial plane of the detection volume, respectively, where the intensity decays to $1/e^2$ of the maximum value.

The kinetic factor $G_{k,ab}(t_c)$ depends on the correlation matrix $\mathbf{G}(t_c)$ that describes the time evolution of the fluctuations of the species populations, and the signals $\mathbf{S}_a$ and $\mathbf{S}_b$ of the observed signal of the different species (given as column vectors). Then, the kinetic part of correlation can be expressed in matrix notation as:

$$G_{k,ab}(t_c) = \frac{\mathbf{S}_a^{\text{T}} \mathbf{G}(t_c) \mathbf{S}_b}{\overline{S}_a \overline{S}_b} \tag{17}$$

where $\overline{S}_i = (\mathbf{S} \cdot \mathbf{x}) = \sum_{i=1}^{n} S^{(i)} x^{(i)}$ is the average of signal $\mathbf{S}$ over the species fractions $\mathbf{x}$, which corresponds to the time average of the signals under the assumption that the system is ergodic. In color-FCS, the vectors $\mathbf{S}_a$ and $\mathbf{S}_b$ correspond to the green and red signal intensities $\mathbf{S}_G$ and $\mathbf{S}_R$. In filtered-FCS, $\mathbf{S}_a$ and $\mathbf{S}_b$ are the fractional fluorescence intensities of the species, obtained by weighting the signal based on the fluorescence decay using the filter functions for each species $a$ and $b$, respectively. The general solution for the correlation matrix $\mathbf{G}(t_c)$ of the kinetic network in the presence of static states is given by:

$$\mathbf{G}(t_c) = p_d \, e^{\mathbf{K} t_c} \, \mathbf{X}_d + (1 - p_d) \mathbf{X}_s, \tag{18}$$

where $e^{\mathbf{K} t_c}$ describes the time evolution of the system, $\mathbf{X}_d$ and $\mathbf{X}_s$ are the diagonal matrices of the dynamic fractions $\mathbf{x}_d$ and static fractions $\mathbf{x}_s$ (for details, see Supplementary Note 4). The matrix exponential $e^{\mathbf{K} t_c}$ can be solved using the eigenvalue decomposition (EVD) of the transition rate matrix:

$$\mathbf{K} = \sum_{l=0}^{n-1} \mathbf{\Gamma}^{(l)} \lambda^{(l)} \quad \Rightarrow \quad e^{\mathbf{K} t_c} = \sum_{l=0}^{n-1} \mathbf{\Gamma}^{(l)} e^{\lambda^{(l)} t_c}, \tag{19}$$

where $\mathbf{\Gamma}^{(l)}$ are the eigen-matrices and $\lambda^{(l)}$ the eigenvalues of $\mathbf{K}$ which relate to the measured FCS relaxation times $t_R^{(l)}$ by:

$$\lambda^{(l)} = -1/t_R^{(l)} \tag{20}$$

Equation (18) can then be expanded into a sum of exponential terms and substituted into eq. (17) to obtain the following general expression for the kinetic correlation function in the presence of static states:

$$\begin{aligned}
G_{k,ab}(t_c) &= 1 + A_{ab}^{(0)} + \sum_{l=1}^{n-1} A_{ab}^{(l)} e^{\lambda^{(l)} t_c}; \\
A_{ab}^{(0)} &= \sum_{i<j} \partial_{ab}^{(ij)} \left( x^{(i)} x^{(j)} - p_d x_d^{(i)} x_d^{(j)} \right); \\
A_{ab}^{(l)} &= \sum_{i<j} \partial_{ab}^{(ij)} G_{ij}^{(l)}, l = \{1 \ldots n-1\};
\end{aligned} \tag{21}$$

where the matrix elements $G_{ij}^{(l)}$ are given by:

$$G_{ij}^{(l)} = p_d [\mathbf{\Gamma}^{(l)} \mathbf{X}_d]_{ij} \tag{22}$$

and the factors $\partial_{ab}^{(ij)}$ describe the contrast between the species $i$ and $j$ and are given by:

$$\partial_{ab}^{(ij)} = \frac{\left(S_a^{(i)} - S_a^{(j)}\right)\left(S_b^{(i)} - S_b^{(j)}\right)}{\overline{S}_a \, \overline{S}_b} \tag{23}$$

For details on the derivations, see Supplementary Note 4.

To obtain the correlation functions, we need to define the correlated signal vectors $S_a$ and $S_b$. In color-FCS, these signals are the detected "green" (donor) and "red" (acceptor) signal intensities $S_G$ and $S_R$:

$$S_a, S_b = \begin{cases} q_G = Q_0(1 - E) \\ q_R = Q_0(\gamma E + \alpha(1 - E)) \end{cases} \quad (24)$$

where $E$ is a vector whose elements correspond to the FRET efficiencies of the fluorescence species, $Q_0$ is the molecular brightness of the donor in the absence of FRET, $\alpha$ is the crosstalk from the donor fluorophore into the red detection channel. $\gamma$ is a combined correction parameter relating the donor and acceptor fluorescence quantum yield and the detection efficiencies of the green and red channels[38-40]. For simplicity, we assume that the crosstalk $\alpha$ of the donor fluorescence into the red detection channel of the acceptor is zero, that the $\gamma$-factor is one, and consider no background noise. In the expressions of the normalized correlation function, the scaling factor $Q_0$ cancels out. For the simplest case of a two-state dynamic system in the presence of static states, we then obtain the general expression for the kinetic correlation function:

$$G_{k,ab}(t_c) = 1 + \partial_{ab}^{(12)} \left( x^{(1)}(1 - x^{(1)}) + p_d\, x_d^{(1)}(1 - x_d^{(1)}) \left(e^{-(k_{12}+k_{21})\,t_c} - 1\right) \right) \quad (25)$$

where the pre-factor $\partial_{ab}^{(12)}$ depends on the channels that are correlated:

$$\partial_{ab}^{(12)} = \begin{cases} (E^{(1)} - E^{(2)})^2/(1 - \bar{E})^2 &, ab = GG \\ (E^{(1)} - E^{(2)})^2/\bar{E}^2 &, ab = RR \\ -(E^{(1)} - E^{(2)})^2/(\bar{E}(1 - \bar{E})), & ab = RG, GR \end{cases} \quad (26)$$

where the average FRET efficiency $\bar{E}$ is given by the average over the total species fractions $x^{(i)}$. The complete derivation of the analytical form of the correlation function for two- and three-state systems is outlined in Supplementary Note 4. Corresponding expressions had previously been obtained for two-state dynamic systems in the presence of a third static state[31].

### 4.1.1 Ambiguities in color-FCS

Before applying the formalism derived in the previous section for the quantitative analysis of the simulated datasets, we emphasize why the combination of FCS and TCSPC is needed. Color-FCS (or FRET-FCS) is generally underdetermined as there are more model parameters than experimentally accessible parameters. Thus, it is required that the FRET efficiency of at least one of the two states is known, but better results are obtained if both FRET efficiencies are restraint[6]. The origin of this ambiguity is outlined in the following.

For a purely dynamic two-state system, i.e., in the absence of static molecules, the expression for the correlation function given in equation (25) simplifies to:

$$G_{k,ab}(t_c) = 1 + \partial_{ab}^{(12)} x_d^{(1)}(1 - x_d^{(1)})\, e^{-(k_{12}+k_{21})\,t_c} \quad (27)$$

In an experiment, we measure two autocorrelation functions (GG, RR) and two cross-correlation functions (GR, RG). The cross-correlation functions contain identical information, and the time constant of the exponential term is shared between all correlation functions. Thus, we determine three correlation amplitudes, $\partial_{ab}^{(12)} x_d^{(1)}(1 - x_d^{(1)})$, and the decay rate of the exponential term, $k_{12} + k_{21}$. As is evident from eq. (26), however, the cross-correlation amplitude relates to the auto-correlation amplitudes by:

$$\partial_{GR}^{(12)} = -\sqrt{\partial_{GG}^{(12)} \partial_{RR}^{(12)}}, \quad (28)$$

and contains no independent information. The system is thus underdetermined, as we have access to only three experimental observables compared to the four parameters of the model ($E^{(1)}$, $E^{(2)}$, $k_{12}$ and $k_{21}$). The ambiguity between the model parameters takes a complex form and is illustrated in Supplementary Note 5. This ambiguity is resolved if the FRET efficiencies of the states are known from single-molecule FRET efficiency histograms or fluorescence decay analysis. In the following, we explore the combination of FCS with TCSPC to restrain the FRET efficiencies of the states using the information provided by the fluorescence decays, which enables quantitative analysis of the kinetics by FCS.

### 4.1.2 Joint analysis of fluorescence decays and FCS

To unambiguously resolve all contributing states and the interconversion, we combine the information provided by FCS and TCSPC and optimize all model parameters globally. While FCS is sensitive to the relaxation rate constants, TCSPC informs the FRET efficiencies and the total species fractions. Thus, the two methods provide orthogonal information that defines the FRET efficiencies of the states and the transition rate matrix. The global analysis is also expected to stabilize the optimization algorithm and reduce the uncertainty of the model parameters.

The donor fluorescence decay of the FRET sample $f_{D|D}^{(DA)}(t)$ depends on the FRET efficiencies of the species and the associated total species fractions $x^{(i)}$. We assume that the time scale of fluorescence and the time scale of dynamics are decoupled. In other words, the fluorescence lifetime is much shorter than the relaxation time of the kinetic processes. Therefore, the fluorescence decay of the ensemble of molecules can be described by the total fractions of the FRET species $x^{(i)}$ and their FRET efficiencies $E^{(i)}$:

$$f_{D|D}^{(DA)}(t) = \sum x^{(i)} \exp\left(-\frac{t}{\tau_{D(0)}(1-E^{(i)})}\right), \qquad (29)$$
$$\text{where } x^{(i)} = (1-p_d)x_s^{(i)} + p_d x_d^{(i)}$$

To optimize the model parameters, we define a global goodness-of-fit function, $\chi^2_{\text{global}}$, as the sum of the squared weighted deviations for TCSPC, $\chi^2_{TCSPC}$, and FCS, $\chi^2_{FCS}$:

$$\chi^2_{\text{global}} = \chi^2_{TCSPC} + \chi^2_{FCS}$$
$$= \sum_t \left(\frac{f_{D|D}^{(DA),\exp.}(t) - f_{D|D}^{(DA)}(t)}{\sigma_{TCSPC}(t)}\right)^2 + \sum_{a,b}\sum_{t_c}\left(\frac{G_{ab}^{\exp.}(t_c) - G_{ab}(t_c)}{\sigma_{FCS}(t_c)}\right)^2 \qquad (30)$$

Here, the weighting factors, $\sigma_{TCSPC}$ and $\sigma_{FCS}$, account for the non-uniform noise in the data. For TCSPC, the weighting factor is estimated based on the experimental counts under the assumption of Poissonian counting statistics as $\sigma_{TCSPC}(t) = \sqrt{f_{D|D}^{(DA),\exp.}(t)}$. For FCS; the weights are estimated based on the recorded data as described in Kask et al.[41] Overall, one experimental fluorescence decay $f_{D|D}^{(DA)}(t)$ and four correlation curves ($G_{GG}(t_c)$, $G_{GR}(t_c)$, $G_{RG}(t_c)$ and $G_{RR}(t_c)$) contribute to $\chi^2_{\text{global}}$.

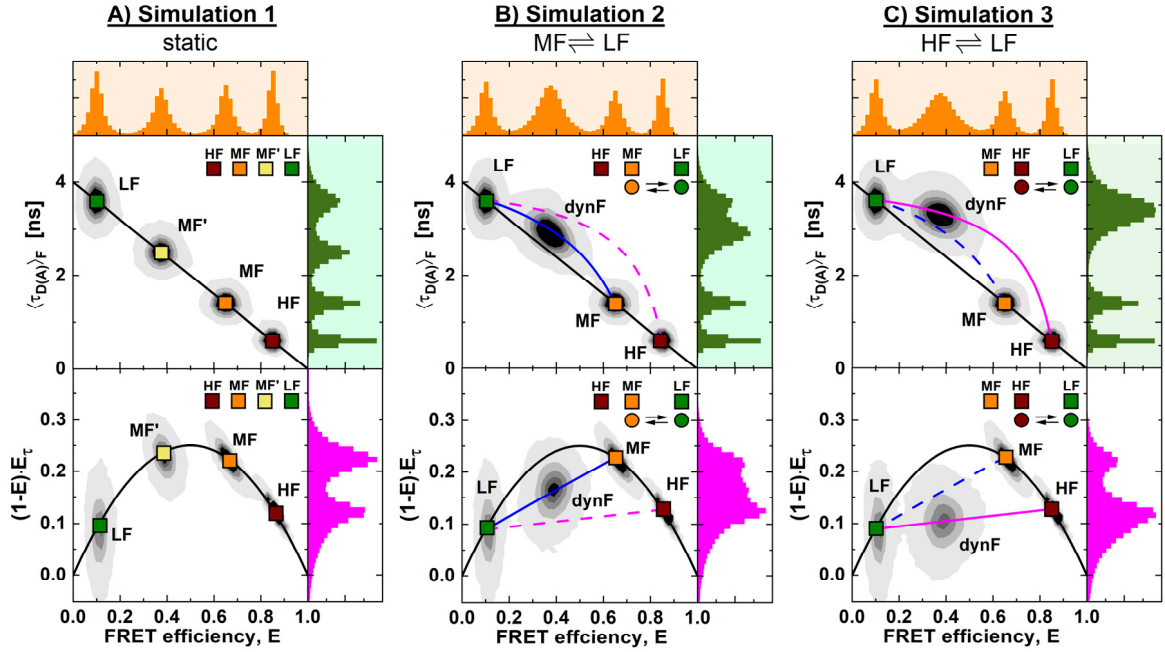

**Figure 5: Simulations of heterogeneous mixtures of static and dynamic molecules.** Three different simulations were performed and are displayed in the $(E, \langle \tau_{D(A)} \rangle_F)$ (top) and moment representations (bottom). (**A**) Simulation 1: A mixture of four static FRET species with high (HF), medium (MF; MF') and low (LF) FRET efficiency, resulting in a 1D FRET efficiency histogram with four distinct peaks. (**B**) Simulation 2: A mixture of three static species and one dynamic FRET species fluctuating between MF and LF states with exchange rate constants $k_{MF \to LF} = k_{LF \to MF} = 5$ ms$^{-1}$. This exchange results in a heterogeneous dynamic population (dynF) with an average FRET efficiency equal to the MF' species of simulation 1. (**C**) Simulation 3: A mixture of three static species and one dynamic FRET species fluctuating between an HF and an LF state. The exchange rate constants are $k_{HF \to LF} = 6.3$ ms$^{-1}$ and $k_{LF \to HF} = 3.7$ ms$^{-1}$. The black line corresponds to a static FRET line. The magenta and blue lines correspond to dynamic FRET lines describing HF/LF and MF/LF mixtures. Solid lines indicate the simulated exchange, while dashed lines correspond to the kinetic transition that was not considered in the simulation. The set of simulation parameters is summarized in Tables S1 and S2.

## 4.2 Analysis of simulations

### 4.2.1 Analysis of three-state kinetic networks

To benchmark the global analysis framework for the analysis of multi-state systems, we simulated a series of experiments. We consider a heterogeneous mixture of various static and dynamic FRET species (Figure 5, Tables S1-S2). No linker dynamics were included in the simulations. First, we consider four distinct static species with low, medium, and high FRET efficiency (LF, MF, MF' and HF species, respectively, Figure 5A and S9). As expected, the four populations in the two-dimensional histograms lie on the static FRET-line, and no indication for conformational dynamics is seen. We then simulated heterogeneous mixtures of three FRET states, two of which are in dynamic exchange (Figure 5 B-C). The kinetic rates were chosen such that the resulting dynamic population has an average FRET efficiency that is identical to the MF' population in the static mixture, resulting in almost indistinguishable one-dimensional FRET efficiency histograms for the three simulations. By overlaying the dynamic FRET-lines connecting the static species, the interconverting states of the dynamic population can be assigned (solid lines). Dynamic FRET-lines of species that are not in dynamic exchange do not intersect with the dynamic population (dashed lines). Moreover, as the dynamic populations are positioned directly on the limiting binary dynamic FRET-lines, we can exclude the possibility of ternary exchange between all three states, which would instead result in a population positioned within the area defined by the three limiting lines (compare section 3.5 of Part I). The graphical analysis by FRET lines thus provides a simple approach to determine the kinetic connectivity of the network. For the moment representation (Figure 5, bottom), dynamic FRET-lines can be drawn as simple lines. While the equilibrium constant may be extracted from the plots given the fast dynamics in these examples (as described in section 3), additional information is required to quantify the interconversion rates of the kinetic network.

In the following, we apply the global analysis of FCS and TCSPC to the simulated datasets shown in Figure 5. For the simulation of four static species (Figure 5A), the absence of conformational dynamics is confirmed by the absence of a kinetic contribution in the FCS curves (SI Figure G2). For the simulations with dynamics between two species (Figure 5 B-C), static and dynamic species are indistinguishable in analyzing the fluorescence decays, which are well described by a three-state model (Figure 6 A,C).

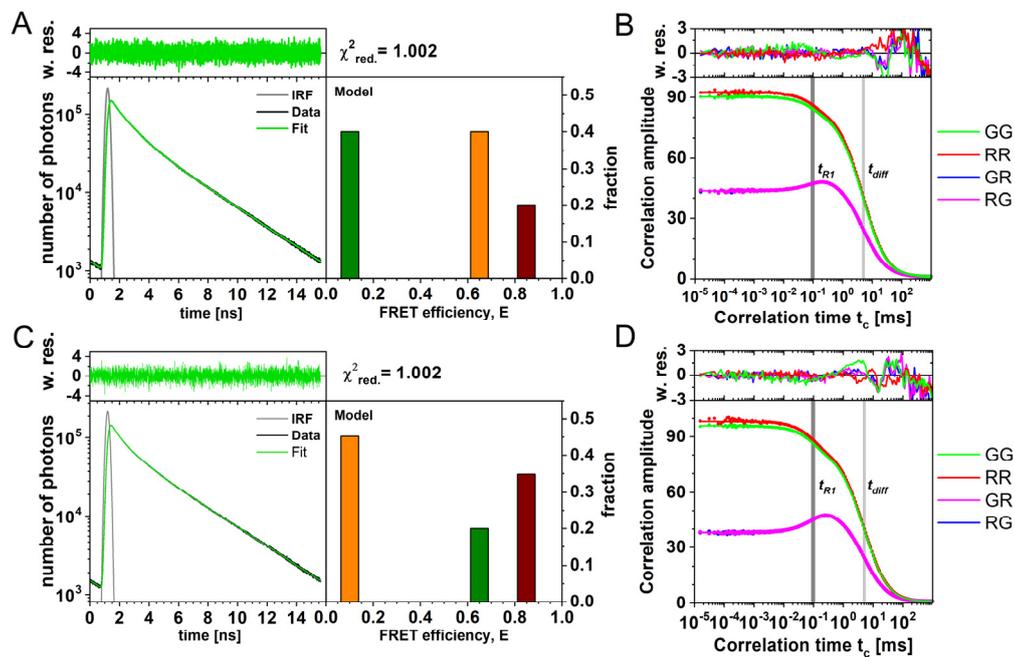

**Figure 6: Global analysis of simulation 2 (A-B) and 3 (C-D). A, C)** Ensemble fluorescence decays of the donor fluorophore (left) and recovered FRET efficiency components (right). **B, D)** Color-FCS autocorrelation and cross-correlation functions reveal a single relaxation time $t_{R1} = 100$ μs for both simulations. The kinetics are superimposed on the diffusion term of the correlation function with a diffusion time $t_{diff} = 5$ ms. Weighted residuals (w. res.) of the fits are given above. The simulation parameters are given in Table S1 and S2.

The green/red cross-correlation curves show a pronounced anti-correlation, which reveals conformational dynamics (Figure 6 B,D). In the analysis of the correlation curves, a single relaxation time of ~100 μs is sufficient to describe the data in both cases. This implies that only two of the three species are in dynamic exchange, consistent with the graphical analysis performed by FRET-lines. We thus consider three possible kinetic schemes that differ in the assignment of the purely static state:

$$\text{MF, HF} \rightleftharpoons \text{LF,}$$
$$\text{HF, MF} \rightleftharpoons \text{LF,} \qquad (31)$$
$$\text{LF, HF} \rightleftharpoons \text{MF.}$$

In this notation, MF, HF ⇌ LF refers to the scheme where the HF and the LF species are exchanging, and the MF species is purely static.

To distinguish amongst the possible kinetic schemes in eq. (31) and to quantify the microscopic parameters, we use the global analysis of FCS and TCSPC. In total, there are nine model parameters: three FRET efficiencies $E^{(i)}$ for the FRET species HF, MF, and LF; two exchange rate constants $k_{ij}$ to describe the exchange among the two dynamic states, two independent fractions $x_s^{(i)}$ of the static states (the third is determined by the other two), and the probability that a molecule is in the dynamic state $p_d$. These *microscopic parameters* define the experimental observables such as the slope of the fluorescence decays or the amplitudes and relaxation timescales of the FCS curves. The fluorescence decays are fully described by the FRET efficiencies $E^{(i)}$ and total species fractions $x^{(i)}$ (see eq. (29)), while the relation between the microscopic parameters and the amplitudes of the FCS curves is more complex. In the description of the FCS model function, we had split the contributions of the FRET efficiencies of the different states from the quantities that depend only on the parameters of the kinetic network (see eq. (25)). The relaxation time of the kinetic amplitude is given by the inverse of the sum of the interconversion rates, $t_R = (k_{12} + k_{21})^{-1}$. The amplitudes of the auto- and cross-correlation curves depend mainly on the total species fractions and FRET efficiencies, which are determined from the information provided by TCSPC. As described in detail in Supplementary Note 6, the only new information obtained from the amplitudes of the FCS curves is the relative amplitude of the kinetic term (the pre-exponential factor in eq. (94)), given by $p_d x_d^{(1)} x_d^{(2)}$. Thus, only seven parameters are available from the experiment: the FRET efficiencies and static fractions obtained from TCSPC, and the sum of the rates and the product $p_d x_d^{(1)} x_d^{(2)}$ from the FCS curves. The system is hence inherently underdetermined, and ambiguity is expected between the fraction of dynamic molecules $p_d$ and the interconversion rates $k_{ij}$ that define the dynamic fractions $x_d$:

$$p_d x_d^{(1)} x_d^{(2)} \propto p_d k_{12} k_{21} = p_d k_{12} (t_R^{-1} - k_{12}) \qquad (32)$$

### 4.2.2 Resolving complex kinetic networks using the global analysis framework

To test this prediction, we sampled the probability distribution of the parameters for the possible kinetic networks given in eq. (31) using a Markov chain Monte Carlo (MCMC) approach. The resulting distributions are shown as two-dimensional contour plots in Figure 7A. Indeed, the experimentally accessible parameters show defined, narrow distributions due to the high signal-to-noise ratio of the simulated data. However, the distributions of the microscopic parameters exhibit the expected ambiguity (Figure 7B) which arises because only the product $p_d x_d^{(1)} x_d^{(2)}$ can be quantified. Additionally, a second ambiguity arises between the different realizations of the kinetic network in eq. (31). By considering the permutations of the kinetic scheme, different assignments of the FRET efficiencies to the static and dynamic states can result in identical FCS amplitudes, splitting the solution space into two branches (see Supplementary Note 6). For the given example of simulation 2, two competing solutions exist between the schemes LF⇌MF, HF, and LF⇌HF, MF (colored blue and red in Figure 7B). In contrast, the third permutation results in non-physical solutions for the microscopic parameters. These two solutions are indistinguishable in the analysis framework, as is evident from the identical reduced $\chi^2_{\text{global}}$ (Figure 7 B). The observed ambiguities can be described analytically based on the analytical model functions if the actual parameters are known. The resulting relation between the parameters $p_d$ and $k_{12}$ is shown in Figure 7C and described in detail in Supplementary Note 6.

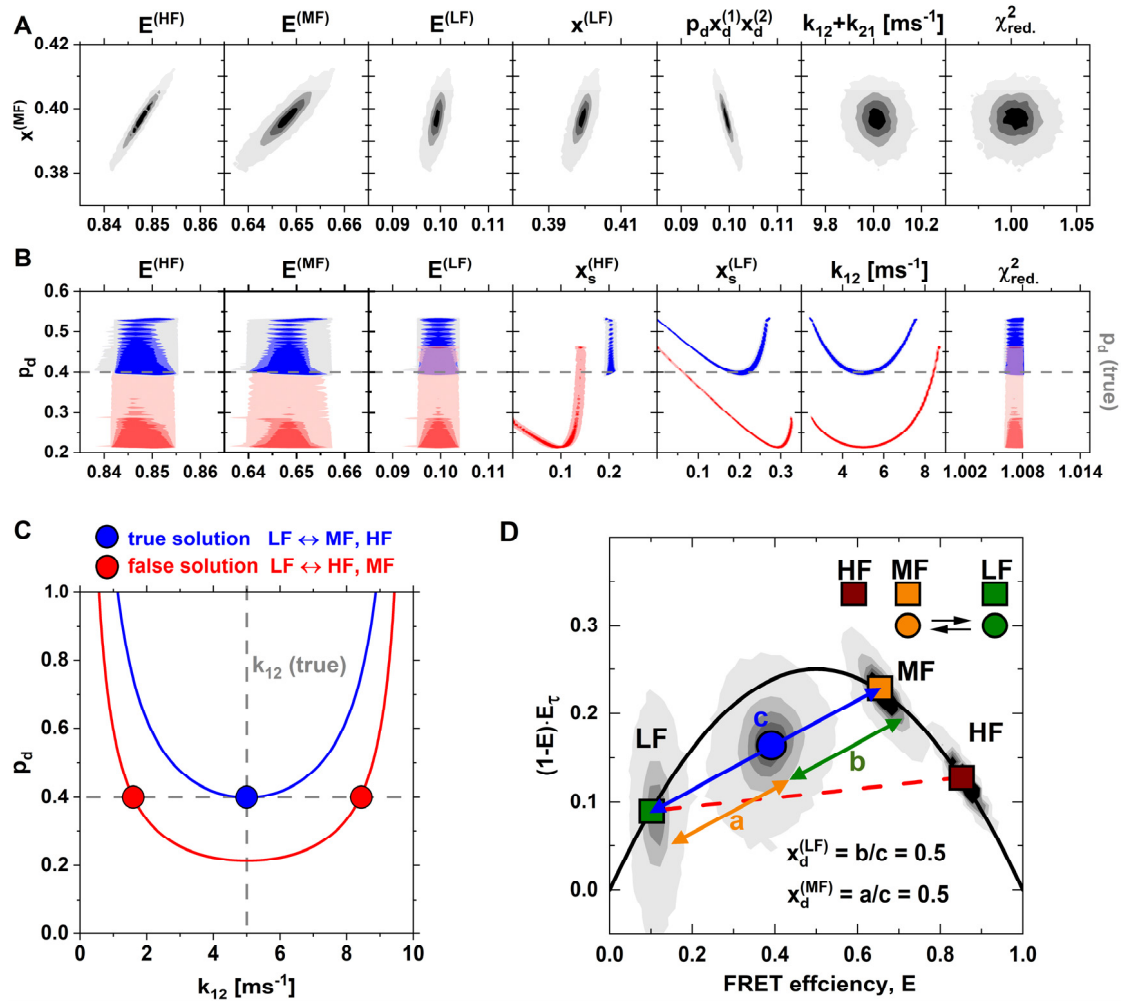

**Figure 7: Global analysis of simulation 2. A)** Pairwise distributions of the independent experimental parameters for the total fraction of molecules in the medium FRET state $x^{(MF)}$ determined by Markov chain Monte Carlo sampling. The parameters show narrow distributions centered at the ideal values due to the ideal signal-to-noise ratio in the simulations. **B)** Pairwise distributions of the microscopic model parameters and the $\chi_r^2$ for the total fraction of dynamic molecules $p_d$. Due to the ambiguity of the experimental parameters, two different kinetic schemes (blue: LF⇌MF, red: LF⇌HF) are consistent with the data as indicated by the identical $\chi_r^2$ (right). Each of these two kinetic schemes is additionally consistent with a range of model parameters. If the correct kinetic scheme (blue: LF⇌MF) and the total fraction of dynamic molecules are known, the unique solution can be identified (gray dashed line). **C)** Ambiguities arising in the global analysis. The first ambiguity arises from the fact that the system is inherently underdetermined, preventing both $p_d$ and $k_{12}$ to be resolved at the same time (dashed lines). Additionally, due to the ambiguity of the FCS amplitudes, different kinetic schemes can be compatible with the data, resulting in two branches of the solution space in the given case (red: false solution, blue: true solution). **D)** A graphical analysis by FRET-lines resolves the connectivity within the kinetic network. The position of the dynamic population on the dynamic FRET-line between the states LF and MF shows that the kinetics occur between the LF and MF states, while the HF state is static (blue line in C). The dynamic FRET-line for the competing solution (red line C) is shown as a red dashed line. To resolve the ambiguity between the parameters $p_d$ and $k_{12}$ in the general case of $k_{12} \neq k_{21}$, it is additionally required to determine the dynamic species fractions $x_d^{(i)}$. In the case of fast dynamics as given here, this parameter can be estimated from the position of the dynamic population along the dynamic FRET-lines. The quantities a, b and c represent the length of the vectors connecting the LF state and the dynamic population (orange), the MF state and the dynamic population (green), and the LF and MF states (blue), respectively. For slower dynamics, the mode of the population deviates from the true mean and the procedure described in section 3 should be applied.

The question remains how these ambiguities can be resolved. To decide between the two branches corresponding to the different state assignments, FRET-lines provide the required information by identifying the kinetic connectivity of the network. From the dynamic FRET-lines, we had identified LF⇌ MF, HF as the true solution (Figure 5B and Figure 7D), allowing us to eliminate the competing solution LF⇌ HF, MF. To resolve the ambiguity arising from the presence of purely static states (eq. (32)), it is required to determine either the total fraction of dynamic molecules $p_d$ or the rate constant $k_{12}$ (see dashed lines in Figure 7C). For fast dynamics, $p_d$ is directly accessible from the two-dimensional histograms as the fraction of molecules in the dynamic population that deviates from the static FRET-line. However, this approach does not apply to slower dynamics due to pseudo-static states on the static FRET-line. The fraction of dynamic molecules $p_d$ may also be obtained from a detailed analysis of the FRET efficiency histogram by PDA using a combination of static and dynamic populations[3]. In the given example, the knowledge of $p_d$ resolves the ambiguity because the interconversion rates were chosen equal ($k_{12} = k_{21} = 5$ ms$^{-1}$). However, an ambiguity remains for the general case of $k_{12} \neq k_{21}$ regarding the assignment of the interconversion rates to the dynamic states (see Supplementary Note 6 for details). Knowledge of the rate $k_{12}$ (or $k_{21}$) resolves the ambiguity in the analysis in all cases. To define the rate $k_{12}$, it is sufficient to know the FCS relaxation time $t_R$ and the dynamic fraction $x_d^{(1)}$. As we have shown in section 3, this information may be obtained from the mode of the dynamic distribution for intermediate to fast dynamics. For the fast dynamics of the system discussed here, one may also estimate the dynamic fraction $x_d^{(1)}$ directly from a graphical analysis in the moment representation from the position of the dynamic population along the dynamic FRET-line connection the LF and MF states (Figure 7D). The dynamic population is positioned at the center of the line, and the resulting dynamic fractions are $x_d^{(LF)} = x_d^{(MF)} = 0.5$.

In summary, we demonstrated that, even for simple two-state kinetic networks in the presence of a background of static molecules, a global analysis of TCSPC and FCS provides ambiguous solutions. These ambiguities can partially be resolved using FRET-lines to eliminate models that are incompatible with the data. Also, it is required to know either the total fraction of dynamic molecules $p_d$ or the equilibrium constant of the dynamic process to fully determine the microscopic parameters.

### 4.2.3 Resolving complex kinetic networks by filtered FCS

In the previous section, we showed how binary kinetic exchange in the presence of a background of static molecules could be resolved by integrative analysis of fluorescence decays, FCS curves, and FRET-lines. To provide more challenging test cases, we performed a series of simulations with an exchange between three species in a linear reaction scheme. The possible kinetic schemes are MF⇌HF⇌LF, HF⇌MF⇌LF, and MF⇌LF⇌HF (Figure 8 B-D). Due to the fast kinetic exchange, complete averaging is observed for the dynamic population (dynF). The interconversion rates were chosen such that the FRET efficiency $E$ and fluorescence-weighted average lifetime $\langle \tau_{D(A)} \rangle_F$ are identical for the dynamic population. Consequentially, the four scenarios are in principle indistinguishable for the two-dimensional histograms of $\langle \tau_{D(A)} \rangle_F$ vs. $E$ (top row) or the moment representation (bottom row). While the corresponding dynamic FRET-lines indicate the correct kinetic pathways in Figure 8 B-D, the fast exchange renders it impossible to resolve the kinetic network in this case. However, the position of the dynamic population between the limiting FRET-lines of the binary exchanges is a clear indication for a three-state exchange. As described in section 3 the equilibrium fractions of the contributing species can be determined by a graphical analysis in this case.

To analyze the kinetics in the complex scenario of fast three-state dynamics, we apply filtered fluorescence correlation spectroscopy (fFCS)[6, 13]. Filtered FCS exploits the information contained in the fluorescence decays to increase the selectivity and contrast of the correlation functions. By characterizing the different species in the mixture by their fluorescence decay patterns, filters are constructed that allow one to separate the contributions of the different species to the correlation function using statistical weights. Due to the orthogonality of the filters, the resulting correlation functions are species-specific. Thus, it is possible to resolve the binary exchange between different species even in complex mixtures and obtain information on the respective relaxation times $t_R$. A distinct advantage of fFCS is that the number of correlation curves increases to the square of the number of contributing species. In contrast, for cFCS the number of correlation functions is limited by the number of color detection channels (four for two-color detection). However, fFCS requires prior knowledge of the number of species and depends crucially on the quality of the filters, which require precise knowledge of the fluorescent properties of each species. Given these prerequisites, fFCS can reveal the kinetic connectivity and quantify the exchange rate constant of the kinetic network.

To show the potential of fFCS, we return to the previous test cases of binary exchange between two species in the kinetic scheme MF, LF ⇌ HF (Figure 8A). For this simulation, the two-dimensional histogram reveals four peaks. Three of the four peaks (HF, MF, and LF) are located on the static FRET line, corresponding to molecules with constant fluorescence properties during the observation time. The dynamic population (dynF) is positioned on the dynamic FRET-line describing the exchange between the LF and HF populations and reveals the dynamic exchange between these species. Only the cross-correlation function between the LF and HF species shows a positive signal (Figure 8E). Also, the corresponding species autocorrelation functions reveal a positive correlation term that matches the timescale of the rise of the cross-correlation function (SI Figure G3). From a global analysis of the species auto- and cross-correlation functions, we obtain a single relaxation time, $t_R$, which relates to the interconversion rates by $t_R = 1/(k_{LF \to HF} + k_{HF \to LF})$. The cross-correlation functions that interrogate the MF⇌HF or LF⇌MF transitions, on the other hand, show no amplitude, proving that there is no exchange between these species.

For fast exchanging processes, it is not possible to resolve the kinetic network from visual inspection of the two-dimensional histograms (Figure 8 B-C). To address this problem, we computed all possible cross-correlation functions using specific filters for the three species HF, MF, and LF (Figure 8 F-G). There is no direct connection between the LF and MF species for the linear kinetic scheme MF⇌HF⇌LF (Figure 8 F). Correspondingly, the exchange between these species is delayed compared to the direct transitions between the MF/HF and HF/LF species, as is evident from the delayed rise of the LF-MF species cross-correlation function (sCCF) compared to the HF-LF and the HF-MF sCCF. Identical observations are made for the kinetic networks LF⇌MF⇌HF and MF⇌LF⇌HF (Figure 8 G-H), showing a delayed rise of the cross-correlation for the indirect pathway. We can thus obtain qualitative information about the connectivity in the network through the relaxation times of the species cross-correlation functions and identify dominant pathways in the network. However, it is impossible to exclude the possibility of exchange between species from the delayed interconversion alone. Similar results would, for example, be obtained for fully connected kinetic networks with a slow exchange between two species. The number of relaxation times required to describe the data provides information about the minimum number of states of the network. Since the relaxation times correspond to the inverse of the non-zero eigenvalues of the transition rate matrix, a kinetic network involving $N$ states shows $N-1$ relaxation times in the correlation functions.

In summary, fFCS enables the direct interrogation of transitions between distinct species and allows to recover the relaxation times and the connectivity within the kinetic network. Indirect transitions show a delayed rise of the SCCF compared to direct transitions. Non-unique solutions of the joint analysis presented in section 4.1.2 are thus resolved, enabling the analysis of complex kinetic networks by smFRET. Practically, the analysis can be limited by the noise of the experimental data, which affects the quality of the resulting filters. Thus, the separation of the species and the signal-to-noise ratio of the resulting correlation functions. Another practical challenge is the identification of the contributing species for the design of the filters. Here, FRET-lines are essential to identify and assign static and dynamic species from the two-dimensional histograms. To verify that the proper solution is attained, control simulations of ambiguous solutions could be performed, enabling a direct comparison of the experimental two-dimensional histograms to that of the obtained solutions[42].

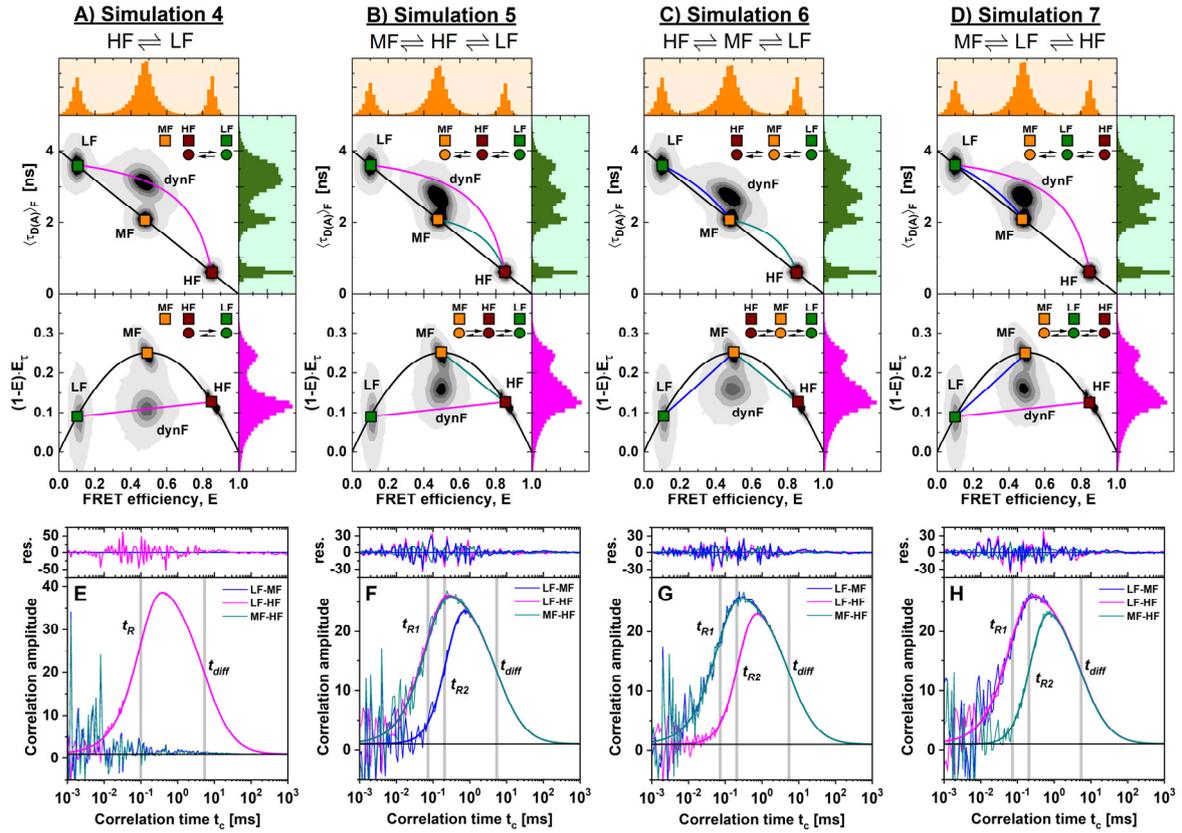

**Figure 8: Simulations of heterogeneous mixtures of static and dynamic molecules involving three-state kinetic networks.** Four different simulations were performed and are displayed in the $(E, \langle \tau_{D(A)} \rangle_F)$ (top) and moment (bottom) representations. (**A**) Simulation 4: A mixture of three static FRET species with high (HF), medium (MF), and low (LF) FRET efficiency and one dynamic species (dynF) with binary dynamics between the LF and HF states with exchange rate constants $k_{HF \to LF} = k_{LF \to HF} = 5$ ms$^{-1}$. (**B-D**) Simulations 5-7: Mixture of three static species (LF, MF, HF) and one dynamic FRET species fluctuating between all three states with exchange rate constants $k_{i \to j} = 5$ ms$^{-1}$ in linear kinetic schemes with different connectivity. The fast exchange results in a defined heterogeneous dynamic population (dynF) with equal average FRET efficiency for the three different kinetic networks. The black line corresponds to a static FRET line. The magenta and blue lines correspond to dynamic FRET lines describing HF/LF and MF/LF mixtures. Solid lines indicate the simulated exchange, while dashed lines correspond to the kinetic transition that was not considered in the simulation. The set of simulation parameters is summarized in Table S1-2. **E-H)** Corresponding species cross-correlation functions between the three species LF, MF, and HF for the different simulations. For the binary exchange (A,E), a positive cross-correlation signal is obtained only for the exchanging species (LF-HF). For the linear three-state kinetic networks, the delayed exchange is detected between states that show no direct connectivity, while connected states exchange faster. The correlation curves are fitted to a kinetic model involving one (E) or two (F-H) relaxation times $t_R$ and a diffusion model with a global diffusion time $t_{\text{diff}}$. Weighted residuals of the fits are given above. The corresponding autocorrelation curves are given in Figure S10.

## 4.3 Connecting FCS amplitudes and FRET indicators

We have shown how the combined information from FCS, TCSPC, and FRET-lines can resolve ambiguities in the analysis. The single-molecule information encoded in the two-dimensional histograms has been used to estimate the kinetic connectivity graphically, but its full potential has not been exploited. We derive the relationship between the two-dimensional histograms and the correlation amplitudes, which provides an additional restraint to the analysis.

The kinetic correlation functions represent the time-dependent (co)variance of the signals. The initial amplitude at zero lag time $t_C$ may thus be expressed as:

$$G_{ab}(t_c = 0) = \frac{\langle S_a(t)S_b(t)\rangle}{\langle S_a(t)\rangle\langle S_b(t)\rangle} = \frac{\text{Cov}(S_a, S_b)}{\bar{S}_a \bar{S}_b} \tag{33}$$

where $\text{Cov}(S_a, S_b)$ is the covariance between the signals $S_a$ and $S_b$. In the ideal case, the signals in the donor and acceptor channel are defined by the FRET efficiency:

$$\begin{aligned} S_G &\propto 1 - E \\ S_R &\propto E \end{aligned}, \tag{34}$$

resulting in the following expressions for the amplitudes of the correlation functions:

$$\begin{aligned} G_{k,GG}(0) &= \frac{\text{Var}(E)}{(1-E)^2} \\ G_{k,RR}(0) &= \frac{\text{Var}(E)}{E^2} \\ G_{k,GR}(0) &= -\frac{\text{Var}(E)}{(1-E)E} \end{aligned} \tag{35}$$

where we have used the relationship $\text{Var}(E) = \text{Var}(1-E) = -\text{Cov}(E, 1-E)$. Thus, the correlation amplitudes represent the normalized variance of the FRET efficiency.

In the first part of the paper, we have shown that the variance of the lifetime or FRET efficiency distribution can be determined from the FRET observables $E$ and $\langle \tau_{D(A)} \rangle_F$ by:

$$\text{Var}(E) = (1-E)(E - E_\tau) \tag{36}$$

where $E_\tau = 1 - \frac{\langle \tau_{D(A)} \rangle_F}{\tau_{D(0)}}$ is the FRET efficiency calculated from the intensity-weighted average lifetime.

This implies that we can calculate the amplitudes of the correlation function directly from the single-molecule FRET indicators, under the condition that the dynamics are fast compared to the diffusion time. For slower dynamics, the variance is underestimated due to the limited sampling within a single-molecule event (see section 3).

The connection between single-molecule FRET indicators and the correlation amplitudes is illustrated in Figure 9 for a three-state system with fast dynamics in the absence of static states, showing a single population that falls between the binary FRET-lines in the $(E, \langle \tau_{D(A)} \rangle_F)$, moment or variance representations (Figure 9 A-C). Using equations (35), a molecule-wise estimate of the correlation amplitudes is obtained that can be compared to the actual correlation amplitudes obtained from FCS analysis (Figure 9 D-G). The static and dynamic FRET-lines can likewise be converted into the equivalent of FCS amplitudes (Figure 9 D-F). The information encoded in the single-molecule FRET indicators could be used as an additional restraint in the analysis.

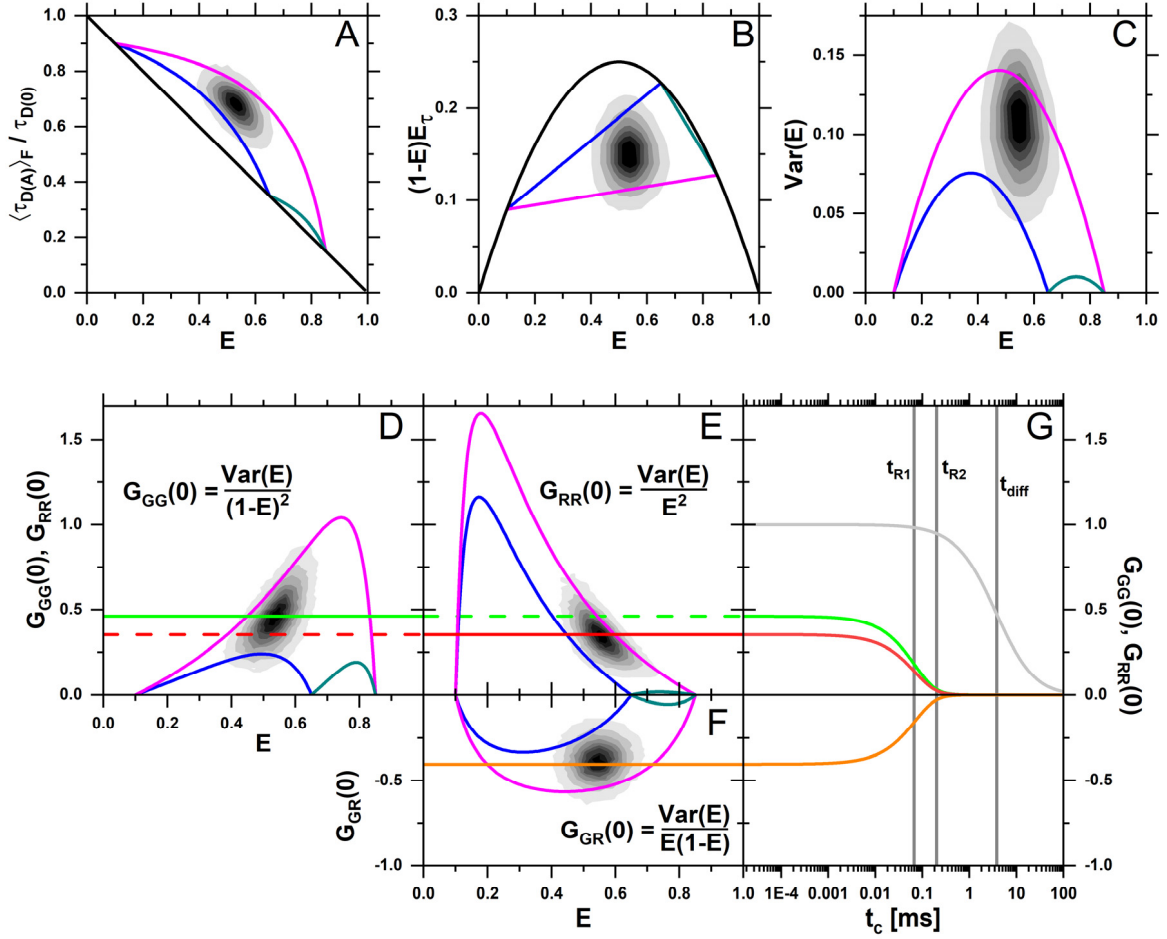

**Figure 9: The relationship between correlation amplitudes and the FRET indicators $E$ and $\langle \tau_{D(A)} \rangle_F$.** A linear three-state system MF⇌LF⇌HF with interconversion rates $k_{i \to j} = 5$ ms$^{-1}$ (Simulation 7) and a diffusion time $t_{\text{diff}} = 3.8$ ms shows a single population that falls between the limiting dynamic FRET-lines in the $(E, \langle \tau_{D(A)} \rangle_F)$ parameter space (A), the moment representation (B) and the variance representation (C). By normalizing the variance by the product of the average signals, the correlation amplitudes $G_{GG}(0)$, $G_{RR}(0)$ and $G_{GR}(0)$ can be estimated for each single-molecule event (D-F). The estimate of the amplitudes obtained from the single-molecule data agrees with the correlation amplitudes of the FCS curves (G. Static FRET-lines are given in black, and binary exchange lines in the various representation are colored blue for the exchange LF⇌MF, magenta for LF⇌HF, and turquoise for MF⇌HF. The donor and acceptor auto- and cross-correlation functions are given in green ($G_{GG}$), red ($G_{RR}$), and orange ($G_{GR}$). The relaxation times of the FCS curves are $t_{R1} = 67$ μs and $t_{R2} = 200$ μs.

## 5 Conclusions

Using synthetic datasets, we challenged the capabilities of conventional analysis methods used in smFRET experiments that rely on one-dimensional data representation. In our integrative approach, FRET-lines serve as visual guidelines for interpreting experiments and the classification of populations of single molecules as static or dynamic. For slow exchange kinetics, FRET-lines directly resolve the connectivity of states. For kinetics that is significantly faster than the integration time, molecule-wise histograms, combined with FRET-lines, help distinguish dynamic averages from static populations. We developed a global analysis framework of FCS and TCSPC, which was not sufficient to identify unique solutions for two-state kinetic networks in the presence of static states. While it was possible to detect the presence of dynamics and quantify their timescale, the network connectivity and the corresponding static and dynamic fractions were not unambiguously recovered. Here, FRET-lines provided the required information to identify the limiting states of the dynamic exchange and their connectivity within the kinetic network. As a next step, the global analysis framework could be extended to utilize species-correlation function from filtered-FCS and include photon distribution analysis. We also showed that the equilibrium constant of dynamic processes could be estimated from a graphical analysis of the $E$-$\langle \tau_{D(A)} \rangle_F$ plot even in the presence of a background of static molecules. Together with the relaxation time obtained by FCS, the microscopic rate constants can thus be quantified without requiring a precise determination of the FCS amplitudes. More complex kinetic networks consisting of three fast-exchanging states could be resolved by filtered-FCS, providing species-specific auto- and cross-correlation curves that reveal the connectivity from the patterns of the relaxation times and amplitudes. Unlike color-FCS, the number of species correlation functions in fFCS increases with the number of states, allowing robust analyses of multi-state networks.

Our global analysis framework is a step towards a self-consistent, holistic model of the two-dimensional histogram of the observables $E$ and $\langle \tau_{D(A)} \rangle_F$ in smFRET experiments. The prediction of the molecule-wise distribution of these parameters currently relies on Monte Carlo simulations, which introduce stochastic noise into the analysis that poses a problem for most optimization algorithms. Although specialized algorithms from the field of machine learning, such as evolutionary algorithms or simulated annealing, may be used to overcome the convergence problem, these algorithms require many iterations for convergence. Deterministic and efficient algorithms are therefore needed. We envision that the future of smFRET will rely on a holistic analysis of the complete experimental information, wherein the kinetic information encoded in the multidimensional histograms of molecule-wise parameters will be an essential first step for proposing candidate models that are subjected to further analysis. Using this approach, it will be possible to quantify the kinetics in complex networks, paving the way towards understanding the intrinsic dynamics of biomolecules and addressing fundamental questions relating to their function.

**Table 3.** Used symbols and definitions

| Definitions in smFRET experiments of multi-state systems | |
|---|---|
| $C^{(1s)}, C^{(2s)}$ | static conformational (structural) state of the biomolecule |
| $C^{(1d)}, C^{(2d)}$ | dynamic conformational (structural) state of the biomolecule |
| $o^{(1)}, o^{(2)}$ | observed fluorescence species defined by a unique set of fluorescence properties |
| $p^{(1,2)}$ | population in the experiment originating from dynamic mixing of species 1 and 2 |
| **Graphical analysis of kinetics** | |
| ds | dynamic shift of a population orthogonal to the static FRET-line |
| $k_{12}, k_{21}$ | interconversion rates between states 1 and 2 |
| $E^{(i)}$ | FRET efficiency of species $i$ |
| $x^{(i)}$ | state occupancy (fraction) of species $i$ |
| $x_d^{(i)}$ | equilibrium species fraction of species $i$ in a dynamic system |
| $\langle E \rangle_{\text{exp}}$ | average FRET efficiency over all single molecule events of the experiment |
| $K$ | equilibrium constant of dynamics |
| $x_m$ | modal value of the state occupancy distribution |
| $E_m$ | modal value of the FRET efficiency distribution |
| $P(E), P(x^{(1)})$ | probability to observe a given average FRET efficiency or state occupancy of state 1 within a single-molecule event |
| $\xi_{12}(x^{(1)})$ | dynamic part of the state occupancy distribution |
| $I_0(x), I_1(x)$ | modified Bessel functions of the first kind of order zero and one |
| $T$ | observation time / integration time |
| $x_{d,\text{lim}}^{(i)}$ | limiting equilibrium fraction at $x_m = 0$ |
| $k$ | sum of the microscopic rates $k = k_{12} + k_{21}$ |
| **Description of correlation functions** | |
| $\mathbf{x}$ | vector of the total species fractions $x^{(i)}$ |
| $\mathbf{K}$ | transition rate matrix |
| $\mathbf{x}_d, \mathbf{x}_s$ | vector of the dynamic and static species fractions |
| $p_d$ | global fraction of dynamic molecules |
| $t_c$ | correlation lag time |
| $G_{ab}(t_c)$ | correlation function of channels $a$ and $b$ |
| $S_a(t)$ | signal in channel $a$ at time $t$ |
| $\langle ... \rangle$ | time-average over a long measurement |
| $G_{\text{diff}}(t_c)$ | diffusion correlation function |
| $G_{k,ab}(t_c)$ | kinetic correlation function between the channels $a$ and $b$ |
| $N$ | average number of particles in the confocal volume |
| $t_{\text{diff}}$ | diffusion time |
| $w_0, z_0$ | lateral and axial width of the confocal volume |
| $\mathbf{S}_a$ | vector of the signals of the different species in channel $a$ |
| $\overline{S_a}$ | species-averaged signal in channel $a$ |
| $\mathbf{X}_d, \mathbf{X}_s$ | diagonal matrices of the dynamic and static species fractions |
| $\lambda^{(l)}$ | $l$-th eigen-value of the transition rate matrix |
| $\mathbf{\Gamma}^{(l)}$ | $l$-th eigen-matrix of the transition rate matrix |
| $t_R^{(l)}$ | $l$-th relaxation time (inverse of the negated $l$-th eigenvalue of the transition rate matrix) |
| $A_{ab}^{(l)}$ | $l$-th pre-exponential factor of the kinetic correlation function between signals $a$ and $b$ |
| $\partial_{ab}^{(ij)}$ | normalized contrast factor between species $i$ and $j$ of the correlation function between signals $a$ and $b$ |
| $\bar{E}$ | species-averaged FRET efficiency |
| $\mathbf{q}_G, \mathbf{q}_R$ | apparent brightness in the green (G) and red (R) detection channels |
| $Q_0$ | molecular brightness of the donor in the absence of FRET |
| $\chi^2_{\text{global}}, \chi^2_{TCSPC}, \chi^2_{FCS}$ | reduced chi-squared values of the global analysis |
| $w_{TCSPC}, w_{FCS}$ | weights used in the analysis of TCPSC decays and FCS curves |
| LF, MF, HF | low-FRET, medium-FRET and high-FRET species |
| $\text{Var}(x)$ | variance of the quantity $x$ |

| Cov(x,y) | covariance of the quantities $x$ and $y$ |

## Acknowledgements

We are thankful to Don C. Lamb and Mark Bowen for their comments. HS acknowledges support from the Alexander von Humboldt foundation, Clemson University start-up funds, and NSF (CAREER MCB-1749778), and NIH (R01MH081923 and P20GM121342). CS acknowledges support by the European Research Council through the Advanced Grant 2014 hybridFRET (number 671208). TP thanks the International Helmholtz Research School of Biophysics and Soft Matter (IHRS BioSoft).

# Supporting Information:
# Unraveling multi-state molecular dynamics in single-molecule FRET experiments
# Part II: Quantitative analysis of multi-state kinetic networks

Oleg Opanasyuk[1,*], Anders Barth[1,a,*], Thomas-Otavio Peulen[1,b], Suren Felekyan[1], Stanislav Kalinin[1], Hugo Sanabria[2,‡], Claus A.M. Seidel[1,‡]

[1] Institut für Physikalische Chemie, Lehrstuhl für Molekulare Physikalische Chemie, Heinrich Heine Universität, Düsseldorf, Germany

[2] Department of Physics and Astronomy, Clemson University, Clemson, S.C., USA

[a] Present address: Department of Bionanoscience, Kavli Institute of Nanoscience, Delft University of Technology, Delft, The Netherlands

[b] Present address: Department of Bioengineering and Therapeutic Sciences, University of California, San Francisco, California, USA

*Contributed equally

[‡] Corresponding authors: cseidel@hhu.de, hsanabr@clemson.edu

# Table of Contents

# Supplementary Note 1 - Potential ambiguities in multi-state systems

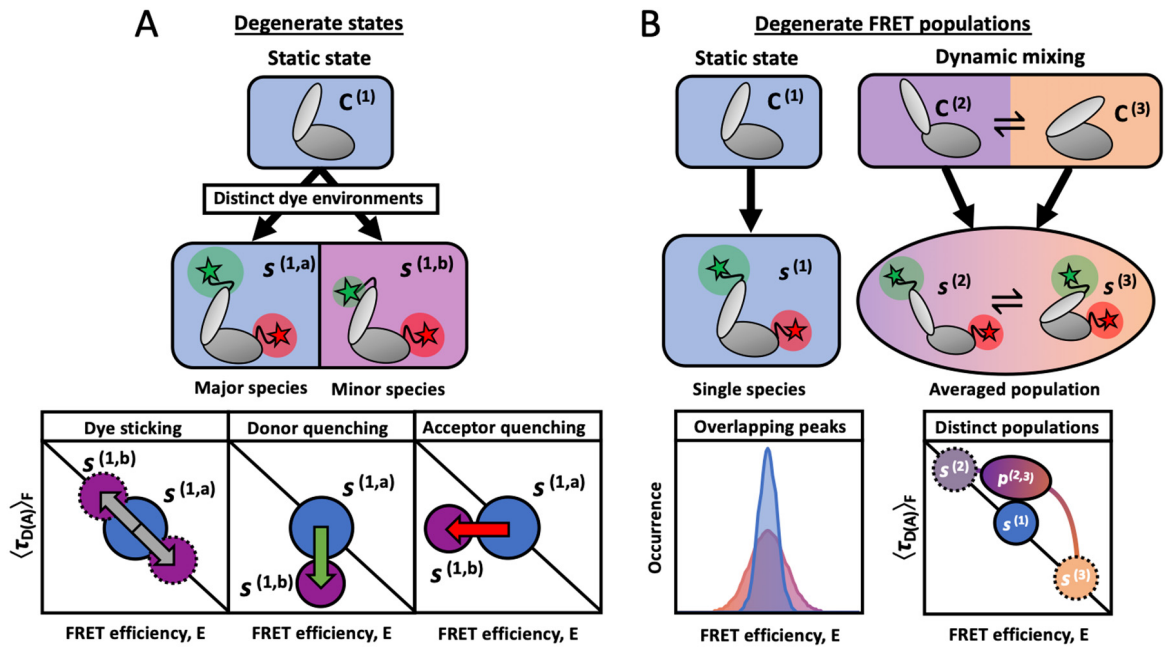

**Figure S1: Degeneracies occurring in smFRET experiments of multi-state systems (extension of Figure 17 in the main text). A) Degeneracy in the assignment of conformational states $C^{(i)}$ to observed fluorescence species $o^{(i)}$. Top:** A single structural state (conformation) of the biomolecule may show multiple observed fluorescence species in the experiment, e.g., due to the interactions of the fluorophores with the biomolecular surface. **Bottom:** Possible artifacts are sticking of the fluorophore to the surface and quenching of the donor or acceptor fluorophore, which show distinct shifts in the two-dimensional plot of the FRET efficiency $E$ against the intensity-averaged donor fluorescence lifetime $\langle \tau_{D(A)} \rangle_F$. Dye sticking affects the orientational factor $\kappa^2$, thus changing the rate of energy transfer. As both $E$ and $\langle \tau_{D(A)} \rangle_F$ are affected, this results in a shift of the population along the static FRET-line (upward or downward). Quenching of the donor fluorophore only affects the lifetime, while the FRET efficiency $E$ remains constant, resulting in a downward shift from the static FRET-line. Contrary, quenching of the acceptor fluorophore affects the FRET efficiency $E$ while the donor fluorescence lifetime $\langle \tau_{D(A)} \rangle_F$ is unchanged, resulting in a leftward shift from the static FRET-line. **B) Degeneracy in the assignment of observed populations ($o^{(i)}$, $p^{(i,j)}$) of identical FRET efficiency to static or dynamic states $C^{(i)}$. Top:** A biomolecule in a static conformation belongs to a single fluorescence species observed as a static population in the experiment. If the biomolecule quickly interconverts between two conformational states (belonging to different fluorescence species), an averaged population is observed in the experiment. **Bottom:** Populations resulting from the static and dynamic conformational states. If only the FRET efficiency $E$ is observed, the populations of the static and dynamic conformational states overlap, i.e., they show a degeneracy of the FRET efficiency. Using the donor fluorescence lifetime $\langle \tau_{D(A)} \rangle_F$, the dynamics are identified, and the static and dynamic populations can be separated. Through the use of the dynamic FRET-line, the species that contributed to the dynamic population are further identified.

## Supplementary Note 2 – Dependence of the dynamic shift on the mixing fractions

In this section, we derive the explicit expression for the dynamic shift as a function of the equilibrium fractions of the dynamic system. In the first part of the paper, we have defined the dynamic shift of a two-state system as the maximum distance of the dynamic FRET-line from the static FRET-line. Specifically, the dynamic shift is measured perpendicular to the static FRET-line in the parameter space defined by the FRET efficiency, $E$, and the normalized donor fluorescence lifetime, $\langle \tau_{D(A)} \rangle_F / \tau_{D(0)}$, where $\langle \tau_{D(A)} \rangle_F$ is the intensity-weighted average donor fluorescence lifetime in the presence of the acceptor and $\tau_{D(0)}$ is the donor-only lifetime.

For the difference between dynamic and static FRET-lines along the FRET efficiency axis, $\Delta_E$, we obtained (compare eq. 19 in the SI of Part I):

$$\Delta_E = \frac{\langle \tau_{D(A)} \rangle_F}{\tau_{D(0)}} - \frac{\left(1 - E^{(1)}\right)\left(1 - E^{(2)}\right)}{\left(2 - E^{(1)} - E^{(2)} - \frac{\langle \tau_{D(A)} \rangle_F}{\tau_{D(0)}}\right)}$$

By replacing $\langle \tau_{D(A)} \rangle_F$ using the expression for the dynamic FRET-line, $E^{(\text{dyn})}$ (compare eq. 17 in the SI of Part I):

$$E^{(\text{dyn})} = 1 - \frac{\left(1 - E^{(1)}\right)\left(1 - E^{(2)}\right)}{\left(2 - E^{(1)} - E^{(2)} - \frac{\langle \tau_{D(A)} \rangle_F}{\tau_{D(0)}}\right)},$$

and using the parametric expression for the average FRET efficiency $E^{(\text{dyn})}$ as a function of the equilibrium fractions $x^{(1)}$ and $x^{(2)} = 1 - x^{(1)}$ of the two states:

$$E^{(\text{dyn})} = x^{(1)} E^{(1)} + \left(1 - x^{(1)}\right) E^{(2)},$$

we obtain:

$$\Delta_E = \left(E^{(2)} - E^{(1)}\right)^2 \frac{x^{(1)}(1 - x^{(1)})}{x^{(1)}(1 - E^{(1)}) + (1 - x^{(1)})(1 - E^{(2)})}$$

Then the "dynamic shift" is then given by:

$$\boxed{\text{ds}(x_1) \stackrel{\text{def}}{=} \frac{\Delta_E}{\sqrt{2}} = \frac{1}{\sqrt{2}} \left(E^{(2)} - E^{(1)}\right)^2 \frac{x^{(1)}(1 - x^{(1)})}{x^{(1)}(1 - E^{(1)}) + (1 - x^{(1)})(1 - E^{(2)})}}.$$

This function reaches its maximum at the value $x_{d,\text{max}}^{(1)}$:

$$\boxed{x_{\text{max}}^{(1)} = \frac{\sqrt{1 - E^{(2)}}}{\sqrt{1 - E^{(1)}} + \sqrt{1 - E^{(2)}}}}$$

The corresponding maximum value of the dynamic shift at $x_{d,\text{max}}^{(1)}$ is then:

$$\boxed{\text{ds}_{\text{max}} = \text{ds}\left(x_{\text{max}}^{(1)}\right) = \frac{1}{\sqrt{2}} \left(\sqrt{1 - E^{(1)}} - \sqrt{1 - E^{(2)}}\right)^2},$$

which is the same as eq. 28 derived in the Part I without consideration of the fraction parameter $x_1$.

# Supplementary Note 3 - The state occupancy distribution for two-state systems

## 3.1 The probability distribution function of the state occupancies for a two-state system

### 3.1.1 Derivation

Let $\boldsymbol{\theta} = \{\theta_1, \theta_2, \ldots \theta_N\}$ are times spent in the states of a system with an exchange during a period $T$ (occupation times). Let the transitions between states be characterized by a transition probability matrix $\boldsymbol{P}(t)$. Bondesson (Bondesson, 1981)[1] gives the expression for the Laplace transform ($T \to s, \boldsymbol{\theta} \to \boldsymbol{s}$) of a probability matrix of occupation times:

$$\boldsymbol{P}(\boldsymbol{s}, s) = \frac{1}{s}(\boldsymbol{I} + \boldsymbol{P}(s)\boldsymbol{S})^{-1}; \quad \boldsymbol{P}(s) = \mathcal{L}^{(T)}[\boldsymbol{P}(T)](s); \quad \boldsymbol{s} = \begin{pmatrix} s_1 \\ \vdots \\ s_1 \end{pmatrix}; \quad \boldsymbol{S} = \begin{pmatrix} s_1 & & 0 \\ & \ddots & \\ 0 & & s_n \end{pmatrix} \quad (1)$$

For a usual time-homogeneous Markov process $\boldsymbol{P}(T) = e^{\boldsymbol{K}T}$, where $\boldsymbol{K}$ is a transition rate matrix. Then, we can write:

$$\begin{aligned}
\boldsymbol{P}(s) &= (s\boldsymbol{I} - \boldsymbol{K})^{-1} \\
\boldsymbol{P}(\boldsymbol{s}, s) &= (s\boldsymbol{I} + \boldsymbol{S} - \boldsymbol{K})^{-1} \\
\boldsymbol{P}(\boldsymbol{s}, T) &= \mathcal{L}^{-(s)}[\boldsymbol{P}(\boldsymbol{s}, s)](\boldsymbol{s}, T) = e^{(\boldsymbol{K}-\boldsymbol{S})T},
\end{aligned} \quad (2)$$

The form $\boldsymbol{P}(\boldsymbol{s}, s)$ is convenient for calculating the joint probability distribution function (pdf) of the occupation times, which can be found by multidimensional inverse Laplace transform (ILT). The form $\boldsymbol{P}(\boldsymbol{s}, T)$ is related to the moment generation function (m.g.f) of this distribution and is convenient for calculating occupation times moments. The moment generating function is given by:

$$\text{m.g.f.} = \boldsymbol{1}\, \boldsymbol{P}(-\boldsymbol{s}, T)\, \boldsymbol{p}(0), \quad (3)$$

where $\boldsymbol{p}(0)$ is the vector of the initial population of the states. For example, the mean occupation time in the first state is:

$$\bar{\theta}_1 = E[\theta_1] = \left(\frac{\partial}{\partial s_1} \boldsymbol{1}\, \boldsymbol{P}(-\boldsymbol{s}, T)\, \boldsymbol{p}(0)\right)\bigg|_{\boldsymbol{s}=0}, \quad (4)$$

where usually $\boldsymbol{p}(0) = \boldsymbol{x}_d$, i.e., equal to the vector of stationary fractions.

If the probability matrix $\boldsymbol{P}(\boldsymbol{\theta}, T)$ of occupation times is known, the joint p.d.f. is found by bracketing the probability matrix by vectors $\boldsymbol{1}$ and $\boldsymbol{x}_d$.

$$p(\boldsymbol{\theta}, T) = \boldsymbol{1}\, \boldsymbol{P}(\boldsymbol{\theta}, T)\boldsymbol{x}_d \quad (5)$$

Let us find the distribution of state occupancies for the 2-state system:

$$\boxed{\begin{array}{c} \theta_1 \quad k_{21} \quad \theta_2 \\ \boxed{1} \rightleftarrows \boxed{2} \\ k_{12} \end{array}}_{\text{During } T} \quad ; \quad K = \begin{pmatrix} -k_{21} & k_{12} \\ k_{21} & -k_{12} \end{pmatrix}; \quad (6)$$

The stationary (equilibrium) fractions:

$$\boldsymbol{x}_d = \frac{1}{k_{12} + k_{21}}\begin{pmatrix} k_{12} \\ k_{21} \end{pmatrix}; \quad (7)$$

The probability matrix:

$$\boldsymbol{P}(\boldsymbol{s}, T) = e^{\begin{pmatrix} -k_{21}-s_1 & k_{12} \\ k_{21} & -k_{12}-s_2 \end{pmatrix}T} \quad (8)$$

The moment generation function:

$$\text{m.g.f.}(\boldsymbol{s}, T) = \boldsymbol{1}\, \boldsymbol{P}(-\boldsymbol{s}, T)\boldsymbol{x}_d = \boldsymbol{1}\, e^{\begin{pmatrix} -k_{21}+s_1 & k_{12} \\ k_{21} & -k_{12}+s_2 \end{pmatrix}T}\boldsymbol{x}_d \quad (9)$$

The Laplace transform of the p.d.f. of the occupation times:

$$P(s,s) = (sI + S - K)^{-1} = \begin{pmatrix} k_{21} + s_1 + s & -k_{12} \\ -k_{21} & k_{12} + s_2 + s \end{pmatrix}^{-1} \tag{10}$$

To obtain the joint p.d.f., $p(\boldsymbol{\theta}, T)$, we need to take a 3D inverse Laplace transform (ILT) of $P(s,s)$:

$$P(\boldsymbol{\theta}, T) = \mathcal{L}^{-(s,s)}[P(s,s)](\boldsymbol{\theta}, T) \tag{11}$$

By the change of variables $u_1 = k_{21} + s_1 + s$; $u_2 = k_{12} + s_2 + s$ the matrix $P(s,s)$ transfers to the matrix:

$$P(u) = \begin{pmatrix} u_1 & -k_{12} \\ -k_{21} & u_2 \end{pmatrix}^{-1}$$
$$= \begin{pmatrix} \frac{1}{u_1} & 0 \\ 0 & \frac{1}{u_2} \end{pmatrix} + \frac{k_{12}k_{21}}{u_1 u_2 - k_{12}k_{21}} \begin{pmatrix} \frac{1}{u_1} & 0 \\ 0 & \frac{1}{u_2} \end{pmatrix} + \frac{1}{u_1 u_2 - k_{12}k_{21}} \begin{pmatrix} 0 & k_{12} \\ k_{21} & 0 \end{pmatrix} \tag{12}$$

The 2D ILT of $P(s,s)$ relative to $s$ expressed trough the ILT of $P(u)$ relative to $u$ as:

$$\mathcal{L}^{-(s)}[P(s,s)](\boldsymbol{\theta}, s) = e^{-(\theta_1 + \theta_2)s} \, e^{-(k_{21}\theta_1 + k_{12}\theta_2)} \mathcal{L}^{-(u)}[P(u)](\boldsymbol{\theta}) \tag{13}$$

Taking ILT relative to $s$ gives matrix $P(\boldsymbol{\theta}, T)$:

$$P(\boldsymbol{\theta}, T) = \mathcal{L}^{-(s,s)}[P(s,s)](\boldsymbol{\theta}, T) = \delta(\theta_1 + \theta_2 - T) e^{-(k_{21}\theta_1 + k_{12}\theta_2)} \mathcal{L}^{-(u)}[P(u)](\boldsymbol{\theta}) \tag{14}$$

Thus, the task of finding joint pdf is reduced to taking the ILT: $\mathcal{L}^{-(u)}[P(u)](\boldsymbol{\theta})$.

This ILT can be calculated using known relationships:

$$\mathcal{L}^{-(u_1, u_2)}\left[\frac{1}{u_i^{\nu}} \frac{1}{u_1 u_2 - a}\right](t_1, t_2) = t_i^{\nu} F_{\nu}(a\, t_1 t_2);$$
$$\mathcal{L}^{-(u_1, u_2)}\left[\frac{1}{u_i}\right](t_1, t_2) = \delta(t_j); \quad i = \{1,2\}, i \neq j \tag{15}$$

Where the functions $F_{\nu}(z) = {}_0\tilde{F}_1(; \nu + 1; z) = z^{-\nu/2} I_{\nu}(2\sqrt{z})$ are regularized hypergeometric functions. Thus, the joint probability of occupation times takes the form:

$$\boxed{P(\boldsymbol{\theta}, T) = \delta(\theta_1 + \theta_2 - T)e^{-(k_{21}\theta_1 + k_{12}\theta_2)} \left[\begin{aligned} &+ \begin{pmatrix} \delta(\theta_2) & 0 \\ 0 & \delta(\theta_1) \end{pmatrix} \\ &+ k_{12}k_{21} F_1(k_{12}k_{21}\, \theta_1\theta_2) \begin{pmatrix} \theta_1 & 0 \\ 0 & \theta_2 \end{pmatrix} + F_0(k_{12}k_{21}\, \theta_1\theta_2) \begin{pmatrix} 0 & k_{12} \\ k_{21} & 0 \end{pmatrix} \end{aligned}\right]} \tag{16}$$

Because of $\delta(\theta_1 + \theta_2 - T)\delta(\theta_i)$ factors in the first term, this formula can be rewritten in the equivalent form in terms of a probability density function:

$$P(\boldsymbol{\theta}, T) = \delta(\theta_1 + \theta_2 - T) \left[\begin{aligned} &+ \begin{pmatrix} e^{-k_{21}T}\delta(\theta_2) & 0 \\ 0 & e^{-k_{12}T}\delta(\theta_1) \end{pmatrix} \\ &+ e^{-(k_{21}\theta_1 + k_{12}\theta_2)}\left(k_{12}k_{21} F_1(k_{12}k_{21}\, \theta_1\theta_2)\begin{pmatrix} \theta_1 & 0 \\ 0 & \theta_2 \end{pmatrix} + F_0(k_{12}k_{21}\, \theta_1\theta_2)\begin{pmatrix} 0 & k_{12} \\ k_{21} & 0 \end{pmatrix}\right) \end{aligned}\right] \tag{17}$$

Note that $\partial_z F_0(z) = F_1(z)$, so the second term can also be rewritten in the differential operator form:

$$P(\boldsymbol{\theta}, T) = \delta(\theta_1 + \theta_2 - T)e^{-(k_{21}\theta_1 + k_{12}\theta_2)} \left[\begin{aligned} &+ \begin{pmatrix} \delta(\theta_2) & 0 \\ 0 & \delta(\theta_1) \end{pmatrix} \\ &+ \tilde{\mathcal{D}}^{(12)} F_0(k_{12}k_{21}\, \theta_1\theta_2) \end{aligned}\right]; \quad \tilde{\mathcal{D}}^{(12)} = \begin{pmatrix} \partial_2 & k_{12} \\ k_{21} & \partial_1 \end{pmatrix}; \quad \partial_i = \frac{\partial}{\partial \theta_i} \tag{18}$$

To obtain the marginal distribution of $\theta_1$, the joint probability should be integrated over $\theta_2$:

$$P(\theta_1, T) = \int_0^{\infty} P(\theta_1, \theta_2, T) d\theta_2 =$$
$$= \left[\begin{aligned} &+ \begin{pmatrix} \delta(\theta_2)e^{-k_{21}T} & 0 \\ 0 & \delta(\theta_1)e^{-k_{12}T} \end{pmatrix} \\ &+ e^{-(k_{21}\theta_1 + k_{12}\theta_2)}\left(k_{12}k_{21} F_1(k_{12}k_{21}\, \theta_1\theta_2)\begin{pmatrix} \theta_1 & 0 \\ 0 & \theta_2 \end{pmatrix} + F_0(k_{12}k_{21}\, \theta_1\theta_2)\begin{pmatrix} 0 & k_{12} \\ k_{21} & 0 \end{pmatrix}\right) \end{aligned}\right] \tag{19}$$

where $\theta_2 = T - \theta_1$. Making the matrix products with vectors $\mathbf{1}$ and $\mathbf{x}_d$ and expressing transition rates through stationary fractions we get the pdf of occupation time $\theta_1$:

$$p(\theta_1, T) = \begin{bmatrix} +x_d^{(1)}\delta(\theta_2)e^{-x_d^{(2)}kT} \\ +x_d^{(2)}\delta(\theta_1)e^{-x_d^{(1)}kT} \\ +x_d^{(1)}\mu_2 k\, e^{-(1-y)kT}\left(y\, kT\, F_1(z) + 2\, F_0(z)\right) \end{bmatrix}; \quad \begin{aligned} y &= \frac{\left(x_d^{(1)}\theta_1 + x_d^{(2)}\theta_2\right)}{T}; \\ z &= x_d^{(1)} x_d^{(2)}\, \theta_1 \theta_2 k^2; \end{aligned} \quad (20)$$

The expression for occupancies (i.e. fractions of time $x^{(i)}$ spent in states $i$) can be obtained by change of variables $\theta_i = x^{(i)} T$:

$$\boxed{\begin{aligned}
p(x^{(1)}, n) &= \delta(1 - x^{(1)}) p_1(x_d^{(1)}, n) + \delta(x^{(1)}) p_2(x_d^{(1)}, n) + \xi_{12}(x^{(1)}; x_d^{(1)}, n); \\
p_1(x_d^{(1)}, n) &= x_d^{(1)} e^{-x_d^{(2)} n}; & y &= x_d^{(1)} x^{(1)} + x_d^{(2)} x^{(2)}; \\
p_2(x_d^{(1)}, n) &= x_d^{(2)} e^{-x_d^{(1)} n}; & z &= n^2 x_d^{(1)} x_d^{(2)}\, x^{(1)} x^{(2)}; \\
\xi_{12}(x^{(1)}; x_d^{(1)}, n) &= n\, x_d^{(1)} x_d^{(2)}\, e^{-(1-y)n}\left(y\, n\, F_1(z) + 2\, F_0(z)\right); & n &= (k_{12} + k_{21})T; \\
F_\nu(z) &= {}_0\tilde{F}_1(; \nu+1; z) = z^{-\nu/2} I_\nu(2\sqrt{z});
\end{aligned}} \quad (21)$$

Where $n$ is the average number of transitions between two states during the observation time $T$.

To our knowledge, Pedler was the first to obtain the same formula for the distribution of the occupation time in one state of a 2-state system[2]. Pedler has used the same 2D ILT technique as presented here but used the moment generating function of the weighted sum of occupation times proposed by Darroch et al.[3]. This formula also coincides with the one derived in Berezhkovskii et al.[4] by two methods different from ours. Kovchegov et al.[5, 6] obtained the same result by another method.

### 3.1.2 Discussion

The distribution of the state occupancies is of a mixed discrete-continuous type, consisting of two discrete $(p_1, p_2)$ and one continuous term $(\xi_{12}(x^{(1)}))$. The discrete terms describe populations of single-molecule events in which the molecule does not leave the initially occupied state. These terms lead to the appearance of two pseudo-static populations on the 2D histograms. The corresponding discrete probabilities $p_1$ and $p_2$ tend to the equilibrium fractions as the burst duration approaches zero, and decay exponentially to zero as T increases:

$$\lim_{kT \to 0} p_1 = x_d^{(1)} \quad ; \quad \lim_{kT \to \infty} p_1 = 0$$
$$\lim_{kT \to 0} p_2 = 1 - x_d^{(1)}; \quad \lim_{kT \to \infty} p_2 = 0 \tag{22}$$

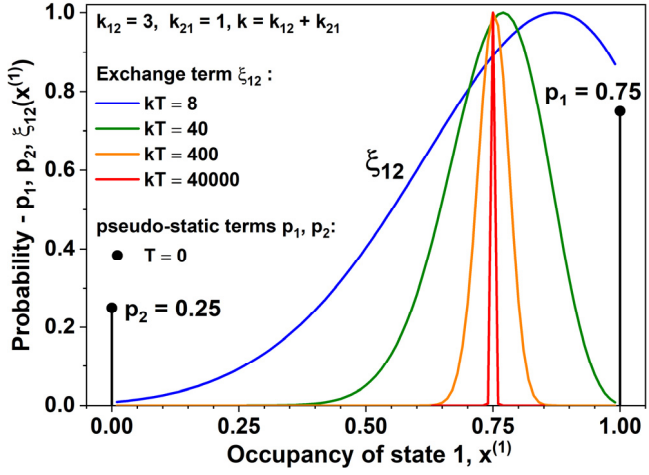

Fig. S2: The probability distribution function of the occupancy of state 1, $x^{(1)}$. Shown are the two pseudo-static terms and the dynamic exchange term $\xi_{12}$ of equation (21), for transition rate constants $k_{12} = 3$ and $k_{21} = 1$ at increasing average number of transitions per observation time $kT$. The dynamic terms are normalized to their maximum value for illustrative purposes. The pseudo-static terms are only shown for $T = 0$ where they represent the equilibrium fractions $x_d^{(1)}$ and $x_d^{(2)}$.

The continuous term of distribution $\xi_{12}$ describes the populations of single-molecule events in which the molecule leaves the initially occupied state at least once. This term leads to the appearance of a dynamic population on the 2D histograms. The total probability of such events becomes zero if the burst duration tends to zero and increases with increasing observation time $T$. At $T \to \infty$ the function tends to the Dirac $\delta$-function located at the equilibrium fraction of state 1 ($x_d^{(1)}$).

$$\lim_{kT \to 0} \int_0^\infty \xi_{12}(x^{(1)}) dx^{(1)} = 0; \quad \lim_{kT \to \infty} \xi_{12}(x^{(1)}) = \delta\left(x - x_d^{(1)}\right) \tag{23}$$

The described behavior of the state occupancy distribution, $p(x^{(1)})$, is illustrated in Fig. S2 for the exchange described by the transition rates constants $k_{12} = 3, k_{21} = 1$, corresponding to an equilibrium fraction $x_d^{(1)} = 0.75$, for different observation times.

Note that the *mode* $x_m^{(1)}$ (i.e., the most probable value) of the dynamic term, which corresponds to the maximum of the dynamic populations observed in the 2D histograms, deviates from the true equilibrium value $x_d^{(1)} = 0.75$ as the burst duration $T$ decreases. This deviation of the mode of the dynamic fraction is shown in Fig. S3 C for arbitrary transition rate constants and burst durations and illustrated in Fig. 3 of the main text. For slow dynamics (relative to the burst duration $T$), the mode shifts toward the limiting fraction (0 and 1) closest to the true equilibrium fraction. This deviation is absent only for $x_d^{(1)} = 0.5$. The deviation vanishes when the duration of the single-molecule events sufficiently exceeds the characteristic exchange time, i.e., when the average number of transitions is large ($kT > 100$). Note that the sum of the transition rates $k$ is equal to the relaxation rate determined by FCS for a two-state system:

$$k_{FCS} = k = k_{12} + k_{21} \tag{24}$$

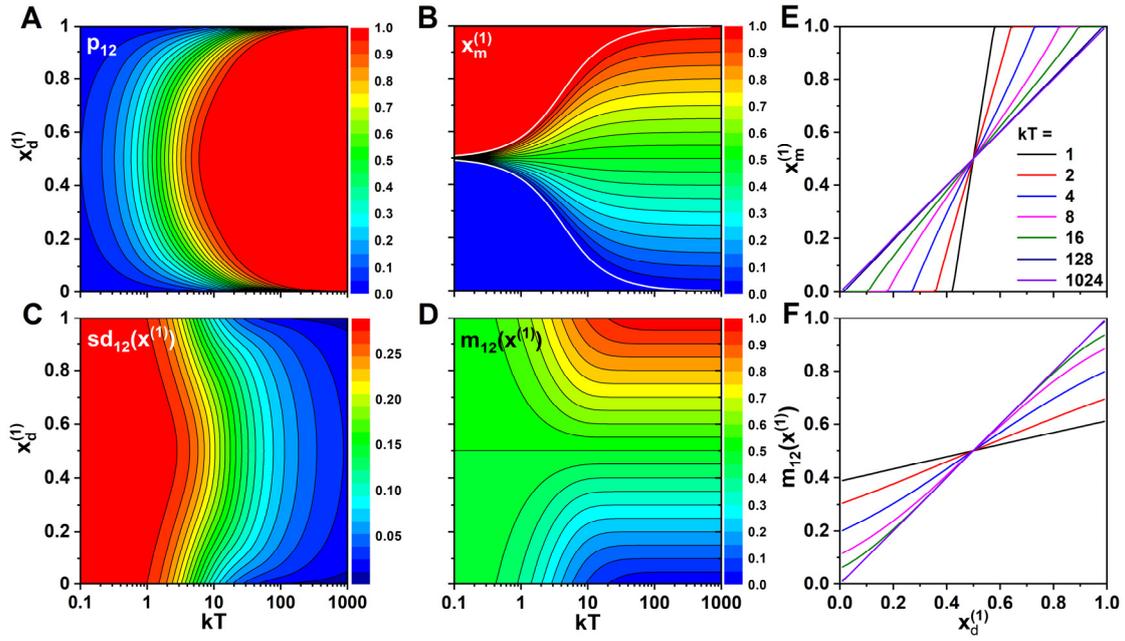

**Fig. S3: Moments of the dynamic distribution term $\xi_{12}(x^{(1)})$ of equation (21).** The different properties of the distribution are scanned over a wide range of equilibrium fractions $x_d^{(1)}$ and transition number $kT$. **A-D:** Shown are contour plots of **(A)** the total contribution of the dynamic term $p_{12}$, **(B)** the mode $x_m^{(1)}$, **(D)** the standard deviation $sd_{12}(x^{(1)})$ and **(E)** mean $m_{12}(x^{(1)})$ of the distribution $\xi_{12}(x^{(1)})$. **E-F:** The mode and the mean show an approximately linear relation to the equilibrium fraction $x_d^{(1)}$.

### 3.2 The moments of the state occupancy distribution

#### 3.2.1 The moments over the total distribution

The mean and variance of the total distribution of the occupation time in state 1 can be derived using the moment generating function presented above (see also reference [2]). By the change of variables $x^{(1)} = \theta_1/T$, where $\theta_1$ is the time spent in state 1, the mean and variance of total distribution are given by:

$$m(x^{(1)}) = x_d^{(1)};$$
$$Var(x^{(1)}) = x_d^{(1)}\left(1 - x_d^{(1)}\right)\frac{2}{n}\left(1 + \frac{e^{-n} - 1}{n}\right); \quad (25)$$

Thus, the mean state occupancy $m(x^{(1)})$ is equal to the equilibrium fraction and its variance is proportional to the product of the equilibrium fractions, with a scaling factor that decays from 1 to 0 as the number of exchanges between states $n = (k_{12} + k_{21})T$ increases:

$$\lim_{n \to 0} Var(x^{(1)}) = x_d^{(1)}(1 - x_d^{(1)}),$$
$$\lim_{n \to \infty} Var(x^{(1)}) = 0 \quad (26)$$

#### 3.2.2 The moments over the dynamic part of the state occupancy distribution

Using the expressions for the total distribution (eq. (21)) and its moments (eq.(25)), we can find the moments of the state occupancy over the dynamical part $\xi_{12}(x^{(1)})$ only.

First, the total fraction of the dynamic term is:

$$p_{12} = \int_0^1 \xi_{12}(x^{(1)})dx^{(1)} = 1 - (p_1 + p_2) = 1 - \left(x_d^{(1)}e^{(1-x_d^{(1)})n} + \left(1 - x_d^{(1)}\right)e^{x_d^{(1)}n}\right) \quad (27)$$

This quantity is presented in Figure C2 A. Then, for the mean and variance of the state occupancy over the distribution $\xi_{12}(x^{(1)})/p_{12}$ we find:

$$m_{12}(x^{(1)}) = \frac{x_d^{(1)} - p_1}{p_{12}},$$
$$Var_{12}(x^{(1)}) = \frac{x^{(1)}(1 - x^{(1)}) - Var(x^{(1)})}{p_{12}} + m_{12}(x^{(1)})\left(1 - m_{12}(x^{(1)})\right) \quad (28)$$

The mean $m_{12}(x^{(1)})$ and standard deviation $sd_{12}(x^{(1)}) = \sqrt{Var_{12}(x^{(1)})}$ are presented in Fig. C2 C-D.

#### 3.2.3 The mode of the dynamic distribution

The mode of the dynamic term of the state occupancy, $x_m^{(1)}$, can be determined numerically by finding the maximum of the term $\xi_{12}(x^{(1)})$. The resulting values for $x_m^{(1)}$ are shown in Fig. C2 B for a wide range of equilibrium fractions $x_d^{(1)}$ and average transition numbers $n = kT$. Note that the relation between the mode $x_m^{(1)}$ and the equilibrium fraction is approximately linear with a slope ($a$) that changes from infinity to 1 as $n$ increases. With increasing $x_d^{(1)}$, the mode $x_m^{(1)}$ remains 0 until some threshold $x_{d,lim}^{(1)}$ (that depends on $kT$), after which it rises almost linearly to the value 1, passing through the point $x_m^{(1)} = x_d^{(1)} = 0.5$, and remains at 1 until $x_d^{(1)}$ reaches the maximum value. This behavior can be well approximated by the following expressions:

$$x_m^{(1)} \approx \begin{cases} 0 & , \quad x_d^{(1)} < x_{d,lim}^{(1)} \\ \dfrac{x_d^{(1)} - x_{d,lim}^{(1)}}{1 - 2\, x_{d,lim}^{(1)}(kT)}, & x_{d,lim}^{(1)} \leq x_d^{(1)} \leq \left(1 - x_{d,lim}^{(1)}\right); \\ 1 & , \quad x_d^{(1)} > \left(1 - x_{d,lim}^{(1)}\right) \end{cases} \qquad (29)$$

where:

$$x_{d,lim}^{(1)}(kT) = \frac{3}{2}\left(1 + \frac{kT}{2}\left(1 + \frac{I_0\left(\frac{kT}{2}\right)}{I_1\left(\frac{kT}{2}\right)}\right)\right)^{-1};$$

The limiting fractions $x_{d,lim}^{(1)}(kT)$ are shown in Fig. C2 B as white curves and Fig. C3 B as a function of the average number of transitions $kT$. While the mean fraction $m_{12}$ (over the exchange term $\xi_{12}$) behaves similarly as the mode $x_m^{(1)}$ (Fig. C2 F), the slope of the dependence changes from 0 to 1 with $kT$, and the linearity is violated for intermediate values of $kT$ (e.g., curves with $kT = 4, 8, 16$ are obviously non-linear).

From the experiment, we can quickly determine the mode of the distribution $x_m^{(1)}$ from the peak of the distribution in the 2D histograms, from which we would like to infer the true equilibrium fraction $x_d^{(1)}$. By rearranging eq. (29), we obtain:

$$x_d^{(1)} \approx x_{d,lim}^{(1)} + \left(1 - 2\, x_{d,lim}^{(1)}\right) x_m^{(1)}, \quad 0 < x_m^{(1)} < 1 \qquad (30)$$

which enables the estimation of the equilibrium fraction from the experimentally determined value of $x_m^{(1)}$, as described in the next section.

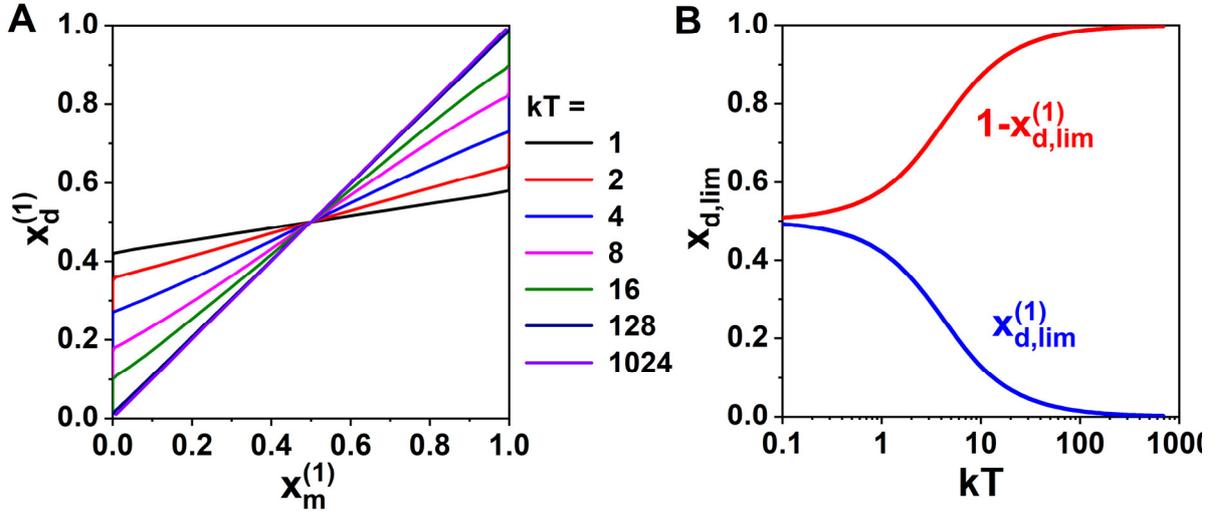

**Fig. S4:** A) The relationship between the equilibrium fraction and the mode of the distribution. B) The limiting equilibrium fraction $x_{d,lim}^{(1)}$ of the linear relation between the mode $x_m^{(1)}$ and the equilibrium fraction $x_d^{(1)}$ given in eq. (29) as a function of the average number of transitions $kT$. The limiting fraction $x_{d,lim}^{(1)}$ corresponds to the y-axis intercept of the curves shown in panel A at $x_m^{(1)} = 0$.

## 3.3 Determining exchange rate constants from the dynamic population

Fig. S3-S4 and the approximation given by eq. (29)-(30) outline the procedure for estimating the true equilibrium fractions $x_d^{(1)}$ from the mode of the dynamic distribution $x_m^{(1)}$. When the equilibrium fraction is known, the transition rates $k_{12}$ and $k_{21}$ can then be estimated if the relaxation rate from FCS, $k_{FCS} = k_{12} + k_{21}$, and average burst duration $T$ are known.

As an example, consider a mean burst duration $T = 2$ ms, a relaxation time determined from FCS of $\tau_{FCS} = 1/k_{FCS} = 0.5$ ms, and an observed mode of the occupancy of the first state $x_m^{(1)}$ of 0.3. Then $k_{FCS} = 2$ ms$^{-1}$ and the average number of transitions is $k_{FCS}T = 4$. The true equilibrium fraction of state 1, $x_d^{(1)}$, may then be determined either from a graphical analysis of the presented contour plot in Fig. C4, or from the approximate linear relation given in eq (30).

a) **Using the contour plots (see Fig. S5):**
   By finding the intersection with the contour line corresponding to $x_m^{(1)} = 0.3$ (white line), we determine that $x_d^{(1)} \approx 0.4$.

b) **Using the linear approximation (eq. (29)-(30)):**
   We find that $x_{d,lim}(kT = 4) \approx 0.26$, which yields $x_d^{(1)} = x_{d,lim}^{(1)} + \left(1 - 2 x_{d,lim}^{(1)}\right) x_m^{(1)} \approx 0.4$.

From the equilibrium fraction $x_d^{(1)}$, we obtain the transition rate constants by $k_{12} = x_d^{(1)} k_{FCS}$ and $k_{21} = (1 - x_d^{(1)}) k_{FCS}$ as $k_{12} = 0.8$ $ms^{-1}$ and $k_{21} = 1.2$ $ms^{-1}$. If instead, we had erroneously used the mode of the state occupancy $x_m^{(1)}$ as an estimate of the true equilibrium fraction, the resulting error would have exceeded 25%.

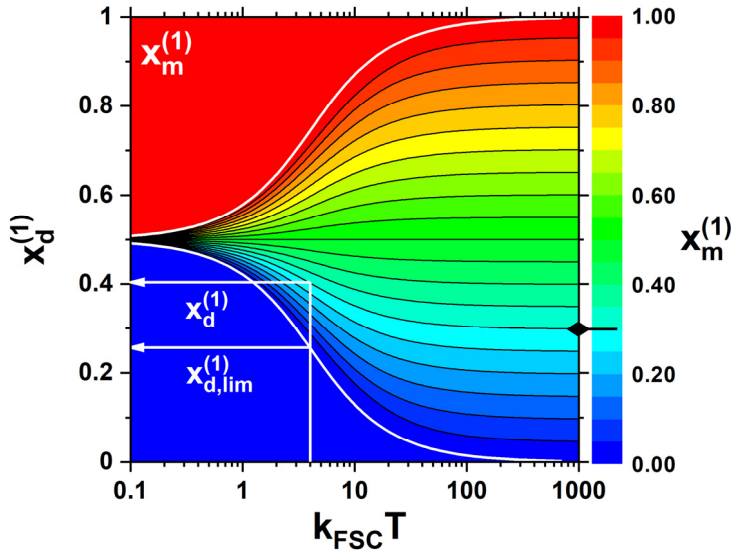

**Fig. S5: Determining the true equilibrium fraction from the mode.** The equilibrium fraction $x_d^{(1)}$ is obtained by finding the intersection of the contour line of the mode $x_m^{(1)}$ with the corresponding value of $k_{FCS}T$. The relaxation rate determined by FCS is given by $k_{FCS} = k_{12} + k_{21}$.

### 3.4 Qualitative description of the dynamic distribution

Three regimes can be defined based on the average number of transitions during a single molecule event, $n = kT$, ($k = k_{12} + k_{21} \approx k_{FCS}$): very fast dynamics where $kT > 100$; very slow dynamics where $kT < 1$; and intermediate dynamics for $1 < kT < 100$. The following properties characterize these regimes:

***Very fast dynamics: $kT > 100$***
- The distribution of the state occupancy $x^{(1)}$ is close to a $\delta$-function.
- The relative width of the distribution is <15%.
- Almost all single-molecule events are located in the dynamic population ($p_{12} \approx 1$). Thus, the observed static populations can be attributed to additional "truly static" species, i.e., not pseudo-static populations of a dynamic system. If it is possible to separate these static populations from the dynamic population, one can determine the total fraction of dynamic molecules, i.e., the parameter $p_d$ of our model. The knowledge of this parameter allows avoiding the ambiguity described in Supplementary Note 6.
- The observed mode and mean of the state occupancy calculated over all dynamical population are close to the true equilibrium fractions ($x_m = m_{12}(x^{(1)}) = x_d^{(1)}$).
- The kinetic exchange and diffusion contributions in the FCS curves are separated by 2 orders of magnitude. Thus, the kinetic relaxation rate can be determined accurately from FCS, and the exchange rates $k_{12}$ and $k_{21}$ can be determined by the product of the relaxation rate with the observed mode fractions ($x^{(1)}, 1 - x^{(1)}$), even without applying the correction described in section 3.3.

***Very slow dynamics: $kT < 1$***
- The distribution of the state occupancy is highly skewed towards the edges ($x^{(1)} = \{0,1\}$).
- The distribution stretches over the whole range of state occupancies.
- A considerable fraction of single-molecule events is located in the pseudo-static populations, i.e. $p_{12} < 0.5$ and $p_1 + p_2 > 0.5$. As the dynamic distribution $\xi_{12}$ is broad and the pseudo-static populations are distributed narrowly, the dynamic population is challenging to detect due to its low amplitude.
- The mode of the state occupancy $x_m^{(1)}$ exists only if the transition rates are close to equal, $k_{12} \approx k_{21}$. If the rates are different, the observed mode fractions fall onto the nearest edge ($x^{(1)} = \{0,1\}$), i.e., to the location of the static populations. The mean dynamic fractions ($m_{12}(x^{(1)})$) deviate from the center of the distribution ($x^{(1)} = 0.5$).
- As the kinetic relaxation time and the diffusion time are in the same order of magnitude, the corresponding kinetic and diffusional contributions to the FCS curves overlap. Thus, the reliable determination of kinetic relaxation rates from FCS becomes challenging.

***Intermediate dynamics: $1 < kT < 100$***
- If the transition rate constants are similar, $k_{12} \approx k_{21}$, the distribution is unimodal with a pronounced maximum. On the other hand, if $k_{ij} \gg k_{ji}$ the distribution takes a monotonic shape and the mode of the state occupancy collapses to 0 or 1.
- If the mode $x_m^{(1)}$ and the kinetic relaxation rate $k = k_{12} + k_{21}$ (from FCS) can be determined, the true equilibrium fractions can be estimated by the method described in the section 3.3.
- If the dynamic and static populations are separable, the total fraction of dynamic molecules (i.e., the parameter $p_d$) can be estimated from the calculated value of $p_{12}$.
- In the intermediate regime, the true values of $k_{12}$ and $k_{21}$ can be estimated using the conversion of the mode into the true equilibrium fraction as described above if the difference between the transition rates is such that equilibrium fraction $x_d^{(1)}$ is within range $\left(x_{d,lim}^{(1)}, \left(1 - x_{d,lim}^{(1)}\right)\right)$.

# Supplementary Note 4 - Description of FCS model

## 4.1 Derivation of the correlation function

The time correlation of two time-dependent variables $a(t)$ and $b(t)$ is defined as:

$$G_{ab}(t_c) = \int a(t+t_c)b(t)\,dt = \overline{a(t+t_c)b(t)}, \qquad (31)$$

where an upper bar denotes the time average.

Let suppose that observed species can be in several states (indexed by $i$), and each state is characterized by its pair of signal values $a_i$ and $b_i$. We can write the observed signals $a(t)$ and $b(t)$ as:

$$a(t) = \sum_i n_i(t)a_i; \quad b(t) = \sum_i n_i(t)b_i, \qquad (32)$$

where $n_i(t)$ is the number of species in state $i$ at time $t$.

We can write a correlation function of the total signals from $N$ species using the definition eq. (31):

$$G_{ab}(t_c) = \overline{a(t+t_c)b(t)} = \overline{\sum_i n_i(t+t_c)a_i \sum_j n_j(t)b_j} = \sum_{i,j} a_i b_j \overline{n_i(t+t_c)n_j(t)} \qquad (33)$$

Expression eq. (33) in the more compact matrix form is:

$$G_{ab}(t_c) = \boldsymbol{a}^T \mathbf{G}^s(t_c)\boldsymbol{b}, \qquad (34)$$

$$\mathbf{G}^s(t_c) = \overline{\boldsymbol{n}(t+t_c)\boldsymbol{n}^T(t)} = \overline{\boldsymbol{n}(t+t_c)\otimes\boldsymbol{n}(t)}$$

where $\boldsymbol{a}, \boldsymbol{b}, \boldsymbol{n}$ are the vectors of the corresponding quantities in the different states, $\mathbf{G}^s(t_c)$ is the matrix whose $ij$-th element is the correlation function of the number of species in states $i$ and $j$ and the symbol "$\otimes$" designates the dyadic product (also called the outer product) of two vectors: $\boldsymbol{x}\otimes\boldsymbol{y} = \boldsymbol{xy}^T$.

If the species can transit from one state to another, the kinetics of this exchange is characterized by the transition rate matrix $\mathbf{K}$. Then, the time evolution of the numbers $\boldsymbol{n}(t)$ follows the equation:

$$\frac{\partial \boldsymbol{n}(t)}{\partial t} = \mathbf{K}\,\boldsymbol{n}(t) \qquad (35)$$

The general solution of this equation is:

$$\boldsymbol{n}(t+t_c) = e^{\mathbf{K}t_c}\boldsymbol{n}(t) \qquad (36)$$

After sufficient time the system reaches the stationary state in which the number of species in the different states $\boldsymbol{n}(t)$ fluctuates around the stationary value $\overline{\boldsymbol{n}}$.

$$\delta\boldsymbol{n}(t) = \boldsymbol{n}(t) - \overline{\boldsymbol{n}} \qquad (37)$$

We can thus define the *fluctuation* correlation functions of signals $a$ and $b$ by:

$$\mathbf{G}^f(t_c) = \overline{\delta\boldsymbol{n}(t+t_c)\otimes\delta\boldsymbol{n}(t)} \qquad (38)$$
$$G^f_{ab}(t_c) = \boldsymbol{a}^T \mathbf{G}^f(t_c)\boldsymbol{b}$$

Using the result from eq. (36), the signal and fluctuation correlation functions can be rewritten as:

$$\mathbf{G}^f(t_c) = e^{\mathbf{K}t_c}\overline{\delta\boldsymbol{n}(t)\otimes\delta\boldsymbol{n}(t)} \qquad (39)$$
$$\mathbf{G}^s(t_c) = e^{\mathbf{K}t_c}\,\overline{\boldsymbol{n}(t)\otimes\boldsymbol{n}(t)} = \mathbf{G}^f(t_c) + \overline{\boldsymbol{n}}\otimes\overline{\boldsymbol{n}}$$

As we consider a system in the stationary state, the averaging over time in eq. (39) can be replaced by averaging over the ensemble and the quantities $\overline{\boldsymbol{n}}, \overline{\boldsymbol{n}\otimes\boldsymbol{n}}$ and $\overline{\delta\boldsymbol{n}\otimes\delta\boldsymbol{n}}$ acquire the meaning of first two moments and variance of the vector random variable $\boldsymbol{n}$:

$$\overline{\boldsymbol{n}(t)} \to \overline{\boldsymbol{n}} \qquad (40)$$
$$\overline{\delta\boldsymbol{n}(t)\otimes\delta\boldsymbol{n}(t)}^{(\text{time})} \to \overline{\delta\boldsymbol{n}\otimes\delta\boldsymbol{n}}^{(\text{ensemble})} = \overline{\boldsymbol{n}\otimes\boldsymbol{n}} - \overline{\boldsymbol{n}}\otimes\overline{\boldsymbol{n}} = \text{Var}[\boldsymbol{n}]$$

If the total number of observed species is stationary (i.e., if there is no diffusion of species into or out of the confocal volume), the distribution of the vector quantity $\boldsymbol{n}$ is multinomial:

$$\boldsymbol{n} \sim \text{multinomial}(\boldsymbol{\mu}, N), \qquad (41)$$

where $\boldsymbol{\mu}$ is the vector of the stationary fractions. The mean and variance of multinomial distribution are:

$$\overline{n} = N\mu \tag{42}$$
$$\mathbf{Var}[n] = N(\mathbf{M} - \mu\otimes\mu); \quad \mathbf{M} = \text{diagonal matrix}(\mu)$$

Then, the second moment $\overline{n\otimes n}$ can be found from the relation of the first two moments of the distribution (see eq.(40)):

$$\overline{n\otimes n} = \mathbf{Var}[n] + \overline{n}\otimes\overline{n} = N(\mathbf{M} - \mu\otimes\mu) + N^2\mu\otimes\mu \tag{43}$$

In the stationary state, the net flux from and to each state is zero.

$$\mathbf{K}\mu = 0 \tag{44}$$

Equivalently, starting from the equilibrium fractions $\mu$, the system will remain in this state and thus not evolve in time, i.e.:

$$e^{\mathbf{K}t_c}\mu = \mu \tag{45}$$

Using this property and the expressions of the moments, the fluctuation correlation matrices can be rewritten in the form:

$$\begin{aligned}\mathbf{G}^f(t_c) &= e^{\mathbf{K}t_c}\mathbf{Var}[n] = Ne^{\mathbf{K}t_c}(\mathbf{M}-\mu\otimes\mu) = N(e^{\mathbf{K}t_c}\mathbf{M} - \mu\otimes\mu)\\ \mathbf{G}^s(t_c) &= \mathbf{G}^f(t_c) + N^2\mu\otimes\mu\end{aligned} \tag{46}$$

In practice, the correlation functions are usually normalized by the product of the total values of the quantities $a$ and $b$ in the stationary state, i.e. by:

$$a_{eq}b_{eq} = \left(\sum \overline{n}_i a_i\right)\left(\sum \overline{n}_j b_j\right) = \overline{n}^T a\, \overline{n}^T b = a^T \overline{n\otimes n}\, b = N^2 a^T \mu\otimes\mu\, b = N^2 \overline{a}\, \overline{b} \tag{47}$$

Thus, the kinetic signal correlation function in the absence of diffusion takes the form:

$$\begin{aligned}G_{ab}(t_c) &= \frac{1}{N}\left(G_{k,ab}(t_c) - 1\right) + 1;\\ G_{k,ab}(t_c) &= \frac{a^T \mathbf{G}(t_c) b}{\overline{a}\,\overline{b}};\\ \mathbf{G}(t_c) &\stackrel{\text{def}}{=} e^{\mathbf{K}t_c}\mathbf{M} = e^{\mathbf{K}t_c}\mathbf{\Sigma} + \mu\otimes\mu = \frac{1}{N}\mathbf{G}^f(t_c) + \mu\otimes\mu,\end{aligned} \tag{48}$$

where we have introduced the matrix $\mathbf{\Sigma} = \mathbf{M} - \mu\otimes\mu$ describing the variance of the fractions of the contributing species. $\mathbf{\Sigma}$ relates to the covariance of the correlated quantities $a$ and $b$ in the stationary state by $\text{Cov}(a,b) = a^T \mathbf{\Sigma}\, b$.

In the presence of a diffusion process that is slower than the kinetic exchange process, the correlation functions acquire an additional term defined by the fluctuation of the total number of species in the confocal volume. In other words, even if the kinetics reach equilibrium, the correlation function does not decay to zero. Mathematically this lead to the replacement $\left(G_{k,ab}(t_c) - 1\right) \to G_{diff}(t_c)G_{k,ab}(t_c)$ in eq. (48) and the expression for the full correlation function takes the form:

$$\boxed{\begin{aligned}G_{ab}(t_c) &= \frac{1}{N}G_{diff}(t_c)G_{k,ab}(t_c) + 1;\\ G_{diff}(t_c) &= \left(1 + \frac{t_c}{t_{\text{diff}}}\right)^{-1}\left(1 + \left(\frac{w_0}{z_0}\right)^2 \frac{t_c}{t_{\text{diff}}}\right)^{-1/2};\\ G_{k,ab}(t_c) &= \frac{a^T \mathbf{G}(t_c) b}{\overline{a}\,\overline{b}};\\ \mathbf{G}(t_c) &\stackrel{\text{def}}{=} e^{\mathbf{K}t_c}\mathbf{M}\end{aligned}} \tag{49}$$

where $t_{\text{diff}}$ is the diffusion time. The parameters $w_0$ and $z_0$ are the width of the focal and the axial plane of the detection volume, respectively, where the intensity decays to $1/e^2$ of the maximum value.

## 4.2 The correlation function of two isolated subsets of species

Suppose that the total set of states of molecules consists of two subsets with no exchange between them, designated by the indexes $d$ and $s$. (In the following, they will be considered to be dynamic or static, but for now they just represent two isolated sets.) Transitions in such a system are described by the total transition rate matrix that has block-diagonal form. The matrix exponential of a block-diagonal matrix is also block-diagonal:

$$\mathbf{K} = \begin{pmatrix} \mathbf{K}_d & 0 \\ 0 & \mathbf{K}_s \end{pmatrix}$$
$$e^{\mathbf{K}t_c} = \begin{pmatrix} e^{\mathbf{K}_d t_c} & 0 \\ 0 & e^{\mathbf{K}_s t_c} \end{pmatrix} \tag{50}$$

Let $\boldsymbol{x}_d$ and $\boldsymbol{x}_s$ be the vectors of the stationary fractions in subsystems $d$ and $s$, correspondingly. Then, the vector of the stationary fractions of the whole system $\boldsymbol{\mu}$ and the corresponding diagonal matrix $\mathbf{M}$ are:

$$\boldsymbol{\mu} = \begin{pmatrix} p_d \boldsymbol{x}_d \\ (1-p_d)\boldsymbol{x}_s \end{pmatrix}, \quad \mathbf{M} = \begin{pmatrix} p_d \mathbf{X}_d & 0 \\ 0 & (1-p_d)\mathbf{X}_s \end{pmatrix}, \tag{51}$$

where $p_d$ is the total fraction of molecules in subsystem $d$. As there is no exchange between the subsystems, the number of molecules (in the ensemble) in subsystems $d$ and $s$ remains constant. Therefore, the fraction $p_d$ is also stationary and a characteristic of the whole system, independent of the transition rate matrices $\mathbf{K}_d$ and $\mathbf{K}_s$.

If the correlated quantities $a$ and $b$ have different values in the subsystems $d$ and $s$, the vectors of these values for the whole system can be written as:

$$\boldsymbol{a} = \begin{pmatrix} \boldsymbol{S}_{a,d} \\ \boldsymbol{S}_{a,s} \end{pmatrix}, \boldsymbol{b} = \begin{pmatrix} \boldsymbol{S}_{b,d} \\ \boldsymbol{S}_{b,s} \end{pmatrix} \tag{52}$$

The correlation function of the whole system is then obtained from eq. (49) by substitution of eqs. (50)-(52). If the correlated quantities $a$ and $b$ additionally have the same values in subsystems $d$ and $s$ the dimensionality of the involved quantities can be reduced by the substitutions $\boldsymbol{a} \to \boldsymbol{a}'$, $\boldsymbol{b} \to \boldsymbol{b}'$ and $\boldsymbol{\mu} \to \boldsymbol{\mu}'$:

$$\begin{aligned} \boldsymbol{a} \to \boldsymbol{a}' &= \boldsymbol{S}_a = \boldsymbol{S}_{a,d} = \boldsymbol{S}_{a,s} \\ \boldsymbol{b} \to \boldsymbol{b}' &= \boldsymbol{S}_b = \boldsymbol{S}_{b,d} = \boldsymbol{S}_{b,s} \\ \boldsymbol{\mu} \to \boldsymbol{\mu}' &= \boldsymbol{x} = p_d \boldsymbol{x}_d + (1-p_d)\boldsymbol{x}_s \end{aligned} \tag{53}$$

Then, the corresponding expression for the correlation matrix in the reduced form is:

$$\begin{aligned} \mathbf{G}(t_c) &= p_d \mathbf{G}_d(t_c) + (1-p_d)\mathbf{G}_s(t_c) \\ \mathbf{G}_d(t_c) &\stackrel{\text{def}}{=} e^{\mathbf{K}_d t_c} \mathbf{X}_d = e^{\mathbf{K}_d t_c} \boldsymbol{\Sigma}_d + \boldsymbol{x}_d \otimes \boldsymbol{x}_d \\ \mathbf{G}_s(t_c) &\stackrel{\text{def}}{=} e^{\mathbf{K}_s t_c} \mathbf{X}_s = e^{\mathbf{K}_s t_c} \boldsymbol{\Sigma}_s + \boldsymbol{x}_s \otimes \boldsymbol{x}_s \end{aligned} \tag{54}$$

where $\mathbf{G}_d(t_c)$ is the correlation matrix of subsystem $d$ and $\mathbf{G}_s(t_c)$ is the correlation matrix of subsystem $s$. Thus, the correlation matrix of the whole system (i.e., the fluctuation correlation matrix for a single molecule) is the linear superposition of the correlation functions of the subsystems.

The expression for kinetic part of the correlation function in eq. (49) takes the form:

$$G_{k,ab}(t_c) = \frac{\boldsymbol{S}_a^T \mathbf{G}(t_c) \boldsymbol{S}_b}{\overline{\boldsymbol{S}_a} \, \overline{\boldsymbol{S}_b}} \tag{55}$$

Here, the matrices $\mathbf{X}$, $\mathbf{X}_d$ and $\mathbf{X}_s$ are the diagonal matrices of the corresponding fractions $\boldsymbol{x}$, $\boldsymbol{x}_d$ and $\boldsymbol{x}_s$ and produce the *averages of the products* of the correlated quantities in the final correlation functions through multiplication with the signal vectors:

$$\overline{S_a S_b}^d = \boldsymbol{S}_a^T \mathbf{X}_d \boldsymbol{S}_b \tag{56}$$

On the other hand, the matrices $\boldsymbol{x} \otimes \boldsymbol{x}$, $\boldsymbol{x}_d \otimes \boldsymbol{x}_d$ and $\boldsymbol{x}_s \otimes \boldsymbol{x}_s$ produce the *products of the averages* of the correlated quantities in the final correlation functions:

$$\overline{S_a}^d \, \overline{S_b}^d = \boldsymbol{S}_a^T (\boldsymbol{x}_d \otimes \boldsymbol{x}_d) \boldsymbol{S}_b \tag{57}$$

The normalization factor of the correlation functions can then be written as (compare eq. (47)):

$$\overline{S_a}\,\overline{S_b} = \boldsymbol{S}_a^T(\boldsymbol{x}\otimes\boldsymbol{x})\boldsymbol{S}_b \tag{58}$$

The matrices $\boldsymbol{\Sigma}$, $\boldsymbol{\Sigma}_d$ and $\boldsymbol{\Sigma}_s$ are the variance matrices of the corresponding fractions and yield the covariances of the correlated quantities in the final correlation functions:

$$\text{cov}(S_a, S_b) = \boldsymbol{S}_a^T\,\boldsymbol{\Sigma}\,\boldsymbol{S}_b \tag{59}$$

As an example, $\boldsymbol{S}_a^T\,\boldsymbol{\Sigma}_d\,\boldsymbol{S}_b = \text{cov}_d(S_a, S_b)$ is the covariance of the quantities $S_a$ and $S_b$ in the stationary state of the subsystem $d$.

### 4.3 The correlation matrix in the presence of static states

If the species in the subsystem $s$ are static, i.e., there are no transitions between them, the corresponding transition matrix is a zero matrix ($\mathbf{K}_s \to \mathbf{0}$), and its matrix exponential is the identity matrix ($e^{\mathbf{K}_s t_c} \to \mathbf{I}$). Thus, we obtain the following expression for the dynamic system in the presence of static states:

$$\boxed{\mathbf{G}(t_c) = p_d\, e^{\mathbf{K} t_c}\,\mathbf{X}_d + (1 - p_d)\mathbf{X}_s} \tag{60}$$

where $\boldsymbol{x}_d$ is the vector of the *equilibrium* fractions of the dynamic states, $\boldsymbol{x}_s$ is the fractions of the static states, and $p_d$ is the fraction of molecules participating in the dynamic exchange.

If static species are absent (i.e. $p_d \to 1$), the correlation matrix converges to the correlation matrix $\mathbf{G}_d(t_c)$ of the dynamic species (see eq. (53)). Otherwise, if dynamic species are absent (i.e. $p_d \to 0$), the correlation matrix converges to the diagonal matrix of the static fractions $\mathbf{X}_s$. As the correlation lag time approaches zero ($t_c \to 0$), the correlation matrix converges to the total fraction matrix $\mathbf{X}$, that is, the initial amplitudes of the correlation function are proportional to the product of the correlated quantities $S_a$ and $S_b$ averaged over all states (both static and dynamic). The asymptotic value of the correlation functions at very long correlation lag times ($t_c \to \infty$) can be obtained using the eigenvalue decomposition of the matrix exponential $e^{\mathbf{K} t_c}$ (see the matrix $\mathbf{A}^{(0)}$ defined in the next section).

$$\begin{aligned}
\mathbf{G}(t_c)|_{p_d \to 1} &= e^{\mathbf{K} t_c}\,\mathbf{X}_d \stackrel{\text{def}}{=} \mathbf{G}_d(t_c)\,; \\
\mathbf{G}(t_c)|_{p_d \to 0} &= \mathbf{X}_s\,; \\
\mathbf{G}(t_c)|_{t_c \to 0} &= \mathbf{X} = \boldsymbol{x}\otimes\boldsymbol{x} + \boldsymbol{\Sigma}\,; \\
\mathbf{G}(t_c)|_{t_c \to \infty} &= \mathbf{X} - p_d \boldsymbol{\Sigma}_d = \boldsymbol{x}\otimes\boldsymbol{x} + \boldsymbol{\Sigma} - p_d \boldsymbol{\Sigma}_d\,;
\end{aligned} \tag{61}$$

### 4.4 Calculation of the matrix exponential using eigen-value decomposition (EVD)

The matrix $e^{\mathbf{K} t_c}$ is the matrix exponential of the product of the transition rate matrix $\mathbf{K}$ and the correlation time $t_c$. The matrix exponential can be expanded using the eigenvalue decomposition of the matrix $\mathbf{K}$:

$$\begin{aligned}
\mathbf{K} &= \sum_{l=0}^{n-1} \boldsymbol{\Gamma}^{(l)}\lambda^{(l)}\,;\ \mathbf{K}\,\boldsymbol{\Gamma}^{(l)} = \boldsymbol{\Gamma}^{(l)}\mathbf{K} = \boldsymbol{\Gamma}^{(l)}\lambda^{(l)};\\
e^{\mathbf{K} t_c} &= \sum_{l=0}^{n-1} \boldsymbol{\Gamma}^{(l)} e^{\lambda^{(l)} t_c}
\end{aligned} \tag{62}$$

where $\boldsymbol{\Gamma}^{(l)}$ are the eigen-matrices and $\lambda^{(l)}$ the eigenvalues of $\mathbf{K}$, i.e., the solutions of the characteristic polynomial $\det(\mathbf{K} - \lambda \mathbf{I}) = 0$, where $\det(\dots)$ denotes the determinate of a matrix. For a transition rate matrix, one eigenvalue is always zero, so we can set $\lambda^{(0)} = 0$.

The matrices $\boldsymbol{\Gamma}^{(l)}$ can be expressed in terms of the tensor (dyadic) products ($\otimes$) of the right and left eigenvectors $\boldsymbol{u}^{(l)}$ and $\boldsymbol{v}^{(l)}$:

$$\begin{aligned} \mathbf{K}\boldsymbol{u}^{(l)} &= \lambda^{(l)}\boldsymbol{u}^{(l)}; \quad \boldsymbol{v}^{(l)}\mathbf{K} = \lambda^{(l)}\boldsymbol{v}^{(l)}; \\ \boldsymbol{\Gamma}^{(l)} &= \boldsymbol{u}^{(l)}\otimes\boldsymbol{v}^{(l)} \quad \text{or} \quad \Gamma_{ij}^{(l)} = u_i^{(l)} v_j^{(l)} \end{aligned} \tag{63}$$

Note that eigen-matrices are orthogonal and idempotent, i.e.:

$$\boldsymbol{\Gamma}^{(l)}\boldsymbol{\Gamma}^{(m)} = \delta_{lm}\boldsymbol{\Gamma}^{(l)} \tag{64}$$

The vector of the equilibrium (or stationary) dynamic state fractions $\boldsymbol{x}_d$ is proportional to the left eigenvector $\boldsymbol{u}^{(0)}$ corresponding to the zero eigenvalue ($\lambda^{(0)} = 0$) and is any column (or diagonal) of the matrix $\boldsymbol{\Gamma}^{(0)}$:

$$\begin{aligned} \mathbf{K}\boldsymbol{x}_d &= 0; \\ \boldsymbol{x}_d &= \frac{\boldsymbol{u}^{(0)}}{\sum_i u_i^{(0)}} = \forall\, col(\boldsymbol{\Gamma}^{(0)}) \quad \text{or} \quad x_d^{(i)} = \frac{u_i^{(0)}}{\sum_i u_i^{(0)}} = \Gamma_{ij}^{(0)}, j = \forall \end{aligned} \tag{65}$$

where $\forall$ denotes "any".

Therefore, we can write the eigen-matrix $\boldsymbol{\Gamma}^{(0)}$ in the form:

$$\boldsymbol{\Gamma}^{(0)} = \boldsymbol{x}_d \otimes \mathbf{1} \quad \text{or} \quad \Gamma_{ij}^{(0)} = x_d^{(i)} \tag{66}$$

Equation (63) is convenient for computer calculations of the correlation functions, as eigenvalues and eigenvectors are effectively obtained by a numerical eigenvalue decomposition (EVD).

The alternative relation for the matrices $\boldsymbol{\Gamma}^{(l)}$ is:

$$\boldsymbol{\Gamma}^{(l)} = \prod_{l \neq m}\left(\frac{\lambda^{(m)}\mathbf{I} - \mathbf{K}}{\lambda^{(m)} - \lambda^{(l)}}\right) \tag{67}$$

This alternative expression for the eigen-matrices is more convenient for deriving analytical expressions for particular kinetic schemes (see section 4.8), as it defines the explicit relations between the elements of the eigen-matrices and the elements of the transition rate matrix.

### 4.5 The form of the correlation functions using EVD

Substituting eq. (62) into eq. (60) and then into eq. (48), we obtain the following expression for the kinetic part of the correlation function in the presence of static states:

$$\begin{aligned} G_{k,ab}(t_c) &= \frac{\boldsymbol{S}_a^{\mathrm{T}}\left[\mathbf{G}^{(0)} + \sum_{l=1}^{n-1}\mathbf{G}^{(l)}e^{\lambda^{(l)}t_c}\right]\boldsymbol{S}_b}{\overline{S_a}\,\overline{S_b}} + 1; \\ \mathbf{G}^{(0)} &= \boldsymbol{\Sigma} - p_d\boldsymbol{\Sigma}_d = \boldsymbol{\Sigma} - \sum_{l=1}^{n-1}\mathbf{G}^{(l)}; \\ \mathbf{G}^{(l)} &= p_d\boldsymbol{\Gamma}^{(l)}\mathbf{X}_d = p_d\boldsymbol{\Gamma}^{(l)}\boldsymbol{\Sigma}_d \end{aligned} \tag{68}$$

Note that the second form of the $\mathbf{G}^{(l)}$ matrices arises from the orthogonality and idempotency of the eigen-matrices (eq.(64)).

## 4.6 The signal contrast factors

We require that transitions between states fulfill the detailed balance condition. The detailed balance condition states that the inward and outward flux of each state are equal, i.e., there is no net flux in the network. That is, for a given state $i$, the sums of all inward and outward pathways must be equal:

$$\underbrace{\sum_j k_{ij} x_d^{(j)}}_{\text{flux } i \leftarrow j} = \underbrace{\sum_j k_{ji} x_d^{(i)}}_{\text{flux } j \leftarrow i} \tag{69}$$

The detailed balance condition fulfills this condition:

$$k_{ij} x_d^{(j)} = k_{ji} x_d^{(i)} \tag{70}$$

In matrix notation, this condition can be expressed as:

$$\mathbf{K} \mathbf{X}_d = \mathbf{X}_d \mathbf{K}^T \tag{71}$$

Under detailed balance, the matrices $\mathbf{G}^{(l)}$ are symmetrical and each of their columns sums to zero. To prove the symmetry, that is $\mathbf{G}^{(l)} = \mathbf{G}^{(l)T}$, we obtain using eq. (67) and (68):

$$\begin{aligned}
\mathbf{G}^{(l)} = p_d \mathbf{\Gamma}^{(l)} \mathbf{X}_d &= p_d \prod \left( \frac{\lambda^{(m)} \overbrace{\mathbf{I} \mathbf{X}_d}^{Diag.} - \overbrace{\mathbf{K} \mathbf{X}_d}^{DB}}{\lambda^{(m)} - \lambda^{(l)}} \right) \\
&= p_d \prod \left( \frac{\lambda^{(m)} \overbrace{\mathbf{X}_d \mathbf{I}}^{Diag.} - \overbrace{\mathbf{X}_d \mathbf{K}^T}^{DB}}{\lambda^{(m)} - \lambda^{(l)}} \right) \\
&= p_d \mathbf{X}_d \left( \prod \left( \frac{\lambda^{(m)} \mathbf{I} - \mathbf{K}}{\lambda^{(m)} - \lambda^{(l)}} \right) \right)^T \\
&= p_d \mathbf{X}_d \mathbf{\Gamma}^{(l)T} = \left( p_d \mathbf{\Gamma}^{(l)} \mathbf{X}_d \right)^T = \mathbf{G}^{(l)T}
\end{aligned} \tag{72}$$

where $DB$ denotes that the detailed balance condition has been used and $Diag.$ designates a diagonal matrix. To prove that the columns of the matrices $\mathbf{G}^{(l)}$ sum to zero, that is $\mathbf{1}^T \mathbf{G}^{(l)} = \mathbf{0}$, we use the matrix symmetry proved above:

$$\begin{aligned}
\mathbf{1}^T \mathbf{G}^{(l)} = \mathbf{1}^T \mathbf{G}^{(l)T} &= \left( \mathbf{G}^{(l)} \mathbf{1} \right)^T = p_d \mathbf{\Gamma}^{(l)} \mathbf{\Sigma}_d \mathbf{1} \\
&= p_d \mathbf{\Gamma}^{(l)} \left( \underbrace{\mathbf{X}_d \mathbf{1}}_{x_d} - \underbrace{x_d \overbrace{x_d^T \mathbf{1}}^{1}}_{x_d} \right) = \mathbf{0}
\end{aligned} \tag{73}$$

The variance matrices $\mathbf{\Sigma}$ introduced above have the same properties. Matrices with these properties can be decomposed in terms of the *contrast matrices* $\mathbf{\Delta}^{(ij)}$:

$$\begin{aligned}
\mathbf{\Sigma} &= \sum_{i<j} x^{(i)} x^{(j)} \mathbf{\Delta}^{(ij)} ; \\
\mathbf{G}^{(l)} &= \sum_{i<j} G_{ij}^{(l)} \mathbf{\Delta}^{(ij)} ;
\end{aligned} \tag{74}$$

The elements of the contrast matrices are given by:

$$[\mathbf{\Delta}^{(ij)}]_{i'j'} = \delta^{ij}_{i'j'} = \begin{cases} 1 & \text{if } i' \neq j' \text{ and } i'j' = ij \\ -1 & \text{if } i' \neq j' \text{ and } i'j' = ji \\ 0 & \text{in all other cases} \end{cases}. \tag{75}$$

Here, $\delta^{ij}_{i'j'}$ is the generalized Kronecker delta function. The product of the matrix $\mathbf{\Delta}^{(ij)}$ with the correlated quantities is equal to the contrast of the states $i$ and $j$, i.e., the product of the differences of $S_a$ and $S_b$ in the states $i$ and $j$:

$$\mathbf{S}_a^T \mathbf{\Delta}^{(ij)} \mathbf{S}_b = \left(S_a^{(i)} - S_a^{(j)}\right)\left(S_b^{(i)} - S_b^{(j)}\right) \tag{76}$$

Therefore, we obtain for the covariance of the correlated signals:

$$\text{cov}(S_a, S_b) = \mathbf{S}_a^T \mathbf{\Sigma} \mathbf{S}_b = \sum_{i<j} x^{(i)} x^{(j)} \left(S_a^{(i)} - S_a^{(j)}\right)\left(S_b^{(i)} - S_b^{(j)}\right) \tag{77}$$

In each matrix $\mathbf{G}^{(l)}$ and $\mathbf{\Sigma}$, there are only $\frac{1}{2}(n-1)n$ independent elements. Let us illustrate this decomposition on the example of a three-state system, which yields the following decomposition of the covariance matrices:

$$\begin{aligned}
\mathbf{\Sigma} = \mathbf{X} - \mathbf{x} \otimes \mathbf{x} &= \begin{pmatrix} x^{(1)} & 0 & 0 \\ 0 & x^{(2)} & 0 \\ 0 & 0 & x^{(3)} \end{pmatrix} - \begin{pmatrix} x^{(1)}x^{(1)} & x^{(1)}x^{(2)} & x^{(1)}x^{(3)} \\ x^{(2)}x^{(1)} & x^{(2)}x^{(2)} & x^{(2)}x^{(3)} \\ x^{(3)}x^{(1)} & x^{(3)}x^{(2)} & x^{(3)}x^{(3)} \end{pmatrix} \\
&= x^{(1)}x^{(2)} \underbrace{\begin{pmatrix} -1 & 1 & 0 \\ 1 & -1 & 0 \\ 0 & 0 & 0 \end{pmatrix}}_{\mathbf{\Delta}^{(12)}} + x^{(1)}x^{(3)} \underbrace{\begin{pmatrix} -1 & 0 & 1 \\ 0 & 0 & 0 \\ 1 & 0 & -1 \end{pmatrix}}_{\mathbf{\Delta}^{(13)}} + x^{(2)}x^{(3)} \underbrace{\begin{pmatrix} 0 & 0 & 0 \\ 0 & -1 & 1 \\ 0 & 1 & -1 \end{pmatrix}}_{\mathbf{\Delta}^{(23)}}
\end{aligned} \tag{78}$$

Analogously, for the matrices $\mathbf{G}$ we obtain:

$$\begin{aligned}
\mathbf{G} &= \begin{pmatrix} -G_{12} - G_{13} & G_{12} & G_{13} \\ G_{12} & -G_{12} - G_{23} & G_{23} \\ G_{13} & G_{23} & -G_{13} - G_{23} \end{pmatrix} \\
&= G_{12} \underbrace{\begin{pmatrix} -1 & 1 & 0 \\ 1 & -1 & 0 \\ 0 & 0 & 0 \end{pmatrix}}_{\mathbf{\Delta}^{(12)}} + G_{13} \underbrace{\begin{pmatrix} -1 & 0 & 1 \\ 0 & 0 & 0 \\ 1 & 0 & -1 \end{pmatrix}}_{\mathbf{\Delta}^{(13)}} + G_{23} \underbrace{\begin{pmatrix} 0 & 0 & 0 \\ 0 & -1 & 1 \\ 0 & 1 & -1 \end{pmatrix}}_{\mathbf{\Delta}^{(23)}}
\end{aligned} \tag{79}$$

Thus, if detailed balance is fulfilled, we can rewrite the correlation functions to separate the contribution of the signals $S_a$ and $S_b$ from the kinetics. By factoring out the *normalized contrast factors* $\partial_{ab}^{(ij)}$ using eq. (74), we obtain:

$$\boxed{\begin{aligned}
G_{k,ab}(t_c) &= 1 + \sum_{i<j} \partial_{ab}^{(ij)} \left( x^{(i)} x^{(j)} - \sum_{l=1}^{n-1} G_{ij}^{(l)} \left(1 - e^{\lambda^{(l)} t_c}\right) \right) \\
G_{ij}^{(l)} &= p_d [\mathbf{\Gamma}^{(l)} \mathbf{X}_d]_{ij}
\end{aligned}} \tag{80}$$

where the factors $\partial_{ab}^{(ij)}$ are defined by:

$$\begin{aligned}
\partial_{ab}^{(ij)} &= \frac{\left(S_a^{(i)} - S_a^{(j)}\right)}{\overline{S_a}} \frac{\left(S_b^{(i)} - S_b^{(j)}\right)}{\overline{S_b}} \\
\overline{S_u} &= \sum x^{(i)} S_u^{(i)} \quad \text{for } u \in \{a, b\}
\end{aligned} \tag{81}$$

In this form, all features of the correlated quantities (i.e., the type of the FCS experiment – for example, color-FCS or filtered-FCS) are accounted for by the normalized contrast factors $\partial_{ab}^{(ij)}$. These contrast factors depend only on the total fractions of the species and their photophysical properties, such as their brightness, quantum yield, or FRET efficiency. On the other hand, the terms describing the initial amplitude, $x^{(i)}x^{(j)}$, and the time-evolution, $G_{ij}^{(l)}\left(1 - e^{\lambda^{(l)}t_c}\right)$, do not depend on the photophysical features of the species (as there are no indices $a$ and $b$), but only on the total fractions of the species $x^{(i)}$, the fraction of dynamic molecules $p_d$ and the transition rate matrix $\mathbf{K}$.

Note that the normalized contrast factors $\partial_{ab}^{(ij)}$ are not independent. The cross-correlation contrast can always be expressed in terms of the corresponding auto-correlation contrasts: $\partial_{a\neq b}^{(ij)} = \partial_{b\neq a}^{(ij)} = \pm\sqrt{\partial_{aa}^{(ij)}\partial_{bb}^{(ij)}}$. If the correlated variables $S_a$ and $S_b$ depend linearly on some common variable, but the slope of the two dependencies are opposite, then the amplitudes of the kinetic exchange, $\Delta G_{k,ab}$, will be positive for autocorrelation functions and negative for cross-correlation functions (anti-correlation), as in the latter case the differences $(S_a^{(i)} - S_a^{(j)})$ and $(S_a^{(i)} - S_a^{(j)})$ in eq.(81) will have opposite signs. This is the case in color-FCS where $S_a \sim E$ and $S_b \sim (1-E)$ (see section 4.7.1 below). On the other hand, if there are no dynamics in the system, both autocorrelation and cross-correlation kinetic amplitudes will vanish, and no anti-correlation will be detected. This feature can be used for the detection of dynamics in the system.

In practice, for example for the fitting of experimental curves, the kinetic part of correlation functions is often described in the formal multiexponential form. Using the derivation above, we can express the exponential amplitudes in terms of physically meaningful parameters:

$$\boxed{\begin{aligned}
G_{k,ab}(t_c) &= 1 + A_{ab}^{(0)} + \sum_{l=1}^{n-1} A_{ab}^{(l)} e^{\lambda^{(l)}t_c} ; \\
A_{ab}^{(0)} &= \sum_{i<j} \partial_{ab}^{(ij)} \left( x^{(i)}x^{(j)} - p_d x_d^{(i)} x_d^{(j)} \right); \\
A_{ab}^{(l)} &= \sum_{i<j} \partial_{ab}^{(ij)} G_{ij}^{(l)}, l = \{1 \ldots n-1\}; \\
\partial_{ab}^{(ij)} &= \frac{\left(S_a^{(i)} - S_a^{(j)}\right)}{\overline{S_a}} \frac{\left(S_b^{(i)} - S_b^{(j)}\right)}{\overline{S_b}} \\
G_{ij}^{(l)} &= p_d[\mathbf{\Gamma}^{(l)}\mathbf{X}_d]_{ij}
\end{aligned}} \qquad (82)$$

The offset amplitudes $A_{ab}^{(0)}$ can be expressed in terms of covariances of correlated signals:

$$A_{ab}^{(0)} = \frac{\text{cov}(S_a, S_b)}{\overline{S_a}\,\overline{S_b}} - p_d \frac{\text{cov}_d(S_a, S_b)}{\overline{S_a}\,\overline{S_b}} = \frac{\text{cov}(S_a, S_b)}{\overline{S_a}\,\overline{S_b}} - \sum_{l=1}^{n-1} A_{ab}^{(l)}; \qquad (83)$$

The expression above is convenient to consider the asymptotic behavior of the kinetic part of correlation functions.

$$G_{k,ab}(t_c = 0) = 1 + \frac{\text{cov}(S_a, S_b)}{\overline{S_a}\,\overline{S_b}};$$

$$G_{k,ab}(t_c = \infty) = 1 + \frac{\text{cov}(S_a, S_b)}{\overline{S_a}\,\overline{S_b}} - p_d \frac{\text{cov}_d(S_a, S_b)}{\overline{S_a}\,\overline{S_b}}$$

$$= 1 + (1 - p_d)\frac{\text{cov}_s(S_a, S_b)}{\overline{S_a}\,\overline{S_b}} + \frac{\text{cov}_{ds}(\overline{S_a}, \overline{S_b})}{\overline{S_a}\,\overline{S_b}};$$

$$\Delta G_{k,ab} = G_{k,ab}(0) - G_{k,ab}(\infty) = p_d \frac{\text{cov}_d(S_a, S_b)}{\overline{S_a}\,\overline{S_b}};$$

(84)

The initial amplitude of the kinetic correlation function is the covariance of the correlated quantities over all states (i.e., static <u>and</u> dynamic) plus one. The amplitude of the time-dependent part of the correlation function is the covariance of the correlated quantities over the dynamic states, scaled by the total fraction of dynamic states. The residual values of the correlation functions are defined by the sum of the covariance of the signals in the static states and the covariance of the average signals in the dynamic and static states: $\text{cov}_{ds}(\overline{S_a}, \overline{S_b}) = p_d(1 - p_d)\left(\overline{S_a}^d - \overline{S_a}^s\right)\left(\overline{S_b}^d - \overline{S_b}^s\right)$. If the fractions of the dynamic and static states are identical, the covariance of the averages $\text{cov}_{ds}(\overline{S_a}, \overline{S_b})$ goes to 0.

## 4.7 Correlation functions for specific correlated quantities

### 4.7.1 Color Fluorescence Correlation Spectroscopy (cFCS)

To calculate the generalized model function for cFCS describing a kinetic network in the presence of static states, we need to define the correlated signal vectors $S_a$ and $S_b$. In cFCS, these signals are the detected "green" (donor) and "red" (acceptor) signal intensities $S_G$ and $S_R$. These vectors can be written as

$$\begin{aligned} S_G &= Q_0(\mathbf{1} - E) \\ S_R &= Q_0(\gamma E + \alpha(\mathbf{1} - E)) \end{aligned} \quad (85)$$

where, $E$ is a vector whose elements are the FRET efficiencies of the fluorescence species, $Q_0$ is the molecular brightness of the donor in the absence of FRET, $\alpha$ is the crosstalk from the donor fluorophore into the red detection channel, $\gamma$ is a combined correction parameter relating the donor and acceptor fluorescence quantum yields and the detection efficiencies of the green and red channels[8, 9]. Here, we do not consider the correction factors $\alpha$ and $\gamma$, in which case the signal vectors are defined solely by the FRET efficiency, i.e. $S_G = Q_0(\mathbf{1} - E)$ and $S_R = Q_0 E$. These signal vectors define the normalized contrast factors $\partial_{ab}^{(ij)}$ in equation (80) as:

$$\partial_{ab}^{(12)} = \begin{cases} \left(E^{(1)} - E^{(2)}\right)^2 / (1 - \bar{E})^2 & , \ ab = GG \\ \left(E^{(1)} - E^{(2)}\right)^2 / \bar{E}^2 & , \ ab = RR \\ -\left(E^{(1)} - E^{(2)}\right)^2 / \left(\bar{E}(1 - \bar{E})\right), & ab = RG, GR \end{cases} \quad (86)$$

### 4.7.2 Filtered Fluorescence Correlation Spectroscopy (fFCS)

Analogously, we derive an analytical expression for the species correlation functions obtained from a filtered-FCS analysis. Due to the orthogonality of the statistical filters, the filtered signals represent the pure contributions of the different species. In this case, the signal vectors $S_a$ and $S_b$ are zero except for the element corresponding to the particular species:

$$S_a, S_b = \{q^{(i)}\}, \quad q_j^{(i)} = Q^{(i)} \delta_{ij}, \quad \{i,j\} = \{1,\ldots,n\}, \quad (87)$$

where $\delta_{ij}$ is Kronecker's delta symbol: $\delta_{ij} = \{0, i \neq j; 1, i = j\}$. The vectors $q^{(i)}$ and $q^{(j)}$, applied to any matrix $\mathbf{G}$ from the left and the right, yield the $ij$-th element of this matrix, scaled by the product of the brightnesses in the states $i$ and $j$:

$$S_a^T \mathbf{G} S_b = q^{(i)T} \mathbf{G} q^{(j)} = Q^{(i)} Q^{(j)} G_{ij} \quad (88)$$

In the same way, the normalization factor of the correlated functions converges to the product of the fractions in the states $i$ and $j$:

$$\overline{S_a}\,\overline{S_b} = S_a^T (x \otimes x) S_b = (S_a^T x)(S_b x) = (q^{(i)T} x)(q^{(j)T} x) = Q^{(i)} Q^{(j)} x^{(i)} x^{(j)} \quad (89)$$

Therefore, the kinetic correlation functions (eq. (48) with eq.(60)) for fFCS take the form:

$$G_{k,ab}(t_c) = \frac{S_a^T \mathbf{G}(t_c) S_b}{\overline{S_a}\,\overline{S_b}} \to G_{k,ij}(t_c) = \frac{q^{(i)T} \mathbf{G}(t_c) q^{(j)}}{q^{(i)T} x \otimes x \, q^{(j)}} = \frac{Q^{(i)} Q^{(j)} G_{ij}(t_c)}{Q^{(i)} Q^{(j)} x^{(i)} x^{(j)}} = \frac{G_{ij}(t_c)}{x^{(i)} x^{(j)}} \quad (90)$$

Using EVD, we can write the multiexponential form as (see eq.(82)):

$$G_{k,ij}(t_c) = 1 + A_{ij}^{(0)} + \sum_{l=1}^{n-1} A_{ij}^{(l)} e^{\lambda^{(l)} t_c};$$

$$A_{ij}^{(0)} = \left(\frac{\delta_{ij}}{x^{(i)}} - 1\right) - \sum_{l=1}^{n-1} A_{ij}^{(l)}; \quad (91)$$

$$A_{ij}^{(l)} = p_d \frac{\Gamma_{ij}^{(l)} x_d^{(j)}}{x^{(i)} x^{(j)}}$$

and the asymptotical amplitudes (see eq.(84)) as:

$$G_{k,ij}(0) = \frac{\delta_{ij}}{x^{(i)}}$$

$$\Delta G_{k,ij} = p_d \frac{\left(\delta_{ij} x_d^{(j)} - x_d^{(i)} x_d^{(j)}\right)}{x^{(i)} x^{(j)}} \quad (92)$$

Therefore, the species correlation functions are independent of the brightness or FRET efficiencies of the states and depend only on their transition rates and fractions. This result is valid for the case that the photophysical effects are decoupled from the transitions between states.

Species-FCS functions can be obtained using filters defined by the fluorescence lifetime, anisotropy, or spectrum of the specific states. Contrary to traditional color-FCS, with filtered-FCS, it is possible to reconstruct the full correlation matrix $\mathbf{G}(t_c)$ from the species cross-correlation functions, because the optimal selection of the filters maximizes the contrast between states. In practice, a complete set of filters for each species is needed for the computation of the experimental species correlation function. The generation and the practical computation of species filters are beyond the scope of this paper and described in detail elsewhere[10, 11]. Note that for fFCS, the correlation function $G_{ab}(t_c)$ is a pair correlation of the signal of two species-specific vectors ($\mathbf{S}_a$ and $\mathbf{S}_b$). These vectors correspond to the recorded photon stream weighted by the species-specific filters, and their time average corresponds to the fractional intensity of the respective species.

## 4.8 Examples

As an example, we derive expressions for the correlation functions for two- and three-state cases in the following.

### 4.8.1 Two-state case

The transition rate matrix is:

$$\mathbf{K} = \begin{pmatrix} -k_{21} & k_{12} \\ k_{21} & -k_{12} \end{pmatrix} \quad (93)$$

The eigen-expansion of the transition rate matrix and the eigen-values (see eq.(62)) are:

$$\begin{aligned} \mathbf{K} &= \mathbf{\Gamma}^{(0)}\lambda^{(0)} + \mathbf{\Gamma}^{(1)}\lambda^{(1)} ; \\ e^{\mathbf{K}t_c} &= \mathbf{\Gamma}^{(0)}e^{\lambda^{(0)}t_c} + \mathbf{\Gamma}^{(1)}e^{\lambda^{(1)}t_c}; \\ \lambda^{(0)} &= 0 \; ; \lambda^{(1)} = -k_{12} - k_{21} \end{aligned} \quad (94)$$

The stationary fractions are the columns or the diagonal of $\mathbf{\Gamma}^{(0)}$, equal to $\mathbf{I} - \mathbf{K}/\lambda^{(1)}$ (see eq.(67)):

$$\begin{aligned} \mathbf{\Gamma}^{(0)} &= \mathbf{I} - \frac{\mathbf{K}}{\lambda^{(1)}} = \frac{1}{k_{12}+k_{21}}\begin{pmatrix} k_{12} & k_{12} \\ k_{21} & k_{21} \end{pmatrix} = \frac{1}{k_{12}+k_{21}}\begin{pmatrix} k_{12} \\ k_{21} \end{pmatrix}\otimes(1\;1) \\ \mathbf{\Gamma}^{(0)} &= x_d\otimes\mathbf{1} \;\Rightarrow\; x_d = \frac{1}{k_{12}+k_{21}}\begin{pmatrix} k_{12} \\ k_{21} \end{pmatrix} \end{aligned} \quad (95)$$

The eigen-matrix $\mathbf{\Gamma}^{(1)}$ is obtained using eq. (67), or from the EVD above:

$$\mathbf{\Gamma}^{(1)} = \mathbf{I} - \mathbf{\Gamma}^{(0)} = \frac{\mathbf{K}}{\lambda^{(1)}} = -\frac{\mathbf{K}}{k_{12}+k_{21}} \quad (96)$$

The matrix defining the amplitudes of the exponents of the correlation functions is:

$$\mathbf{G}^{(1)} \stackrel{\text{def}}{=} p_d\mathbf{\Gamma}^{(1)}\mathbf{X}_d = p_d x_d^{(1)} x_d^{(2)} \mathbf{\Delta}^{(12)}; \quad \mathbf{\Delta}^{(12)} = \begin{pmatrix} 1 & -1 \\ -1 & 1 \end{pmatrix} \quad (97)$$

The matrix $\mathbf{\Delta}^{(12)}$ above produces the product of the differences of the correlated quantities ($\partial_{ab}^{(12)}$ in eq.(98)).

Finally, we can account for the presence of static states and write the kinetic correlation function of the quantities $S_a$ and $S_b$ in the two-state system (see eq.(82)):

$$\begin{aligned} G_{k,ab}(t_c) &= 1 + A_{ab}^{(0)} + A_{ab}^{(1)} e^{-k\,t_c}; \\ k &= k_{12} + k_{21}; \\ \partial_{ab}^{(12)} &= \frac{S_a^{(1)} - S_a^{(2)}}{\overline{S_a}} \frac{S_b^{(1)} - S_b^{(2)}}{\overline{S_b}}; \\ A_{ab}^{(0)} &= \quad x^{(1)}\bigl(1-x^{(1)}\bigr)\partial_{ab}^{(12)} - A_{ab}^{(1)} \\ A_{ab}^{(1)} &= p_d\, x_d^{(1)}(1-x_d^{(1)})\,\partial_{ab}^{(12)}; \end{aligned} \quad (98)$$

In the explicit form, the correlation function is given by:

$$\begin{aligned} G_{k,ab}(t_c) &= 1 + \partial_{ab}^{(12)}\left(x^{(1)}\bigl(1-x^{(1)}\bigr) + p_d\, x_d^{(1)}(1-x_d^{(1)})\left(e^{-k\,t_c}-1\right)\right); \\ x^{(1)} &= p_d x_d^{(1)} + (1-p_d) x_s^{(1)}; \\ k &= k_{12} + k_{21}; \\ x_d^{(1)} &= \frac{k_{12}}{k_{12}+k_{21}} \end{aligned} \quad (99)$$

Substituting the correlated signals $S_a^{(i)}$ and $S_b^{(i)}$ in the cFCS form (eq.(85)), we obtain the following expression for the contrast factors $\partial_{ab}^{(12)}$ (see eq.(80)):

$$\partial_{ab}^{(12)} = \begin{cases} \left(E^{(1)} - E^{(2)}\right)^2 / (1 - \bar{E})^2 & , \ ab = GG \\ \left(E^{(1)} - E^{(2)}\right)^2 / \bar{E}^2 & , \ ab = RR \\ -\left(E^{(1)} - E^{(2)}\right)^2 / \left(\bar{E}(1 - \bar{E})\right), & ab = RG, GR \end{cases} \quad (100)$$

Note that the contrast factors also depend on the total fractions of the different species $x^{(i)}$ through the average FRET efficiency $\bar{E} = \sum x^{(i)} E^{(i)}$. From eq. (100) it follows that the cross-correlation amplitude is proportional to the square root of the product of the autocorrelation amplitudes:

Likewise, for the correlation functions between two species 1 and 2 by fFCS (eq.(87)), we obtain:

$$\partial_{ab}^{(12)} = \begin{cases} 1/x^{(1)2} & , \ ab = 11 \\ 1/\left(1 - x^{(1)}\right)^2 & , \ ab = 22 \\ -1/\left(x^{(1)}(1 - x^{(1)})\right), & ab = 12, 21 \end{cases} \quad (101)$$

In the case of fFCS, the kinetic part of the species cross-correlation function has the simple form:

$$G_{k,12}(t_c) = G_{k,21}(t_c) = p_d \frac{x_d^{(1)}(1 - x_d^{(1)})}{x^{(1)}(1 - x^{(1)})} \left(1 - e^{-k\,t_c}\right) \quad (102)$$

In both FCS experimental variants, the cross-correlation contrast factors can be expressed in terms of the two auto-correlation factors.

$$\partial_{ab}^{(12)} = \partial_{ba}^{(12)} = -\sqrt{\partial_{aa}^{(12)} \partial_{bb}^{(12)}} \quad (103)$$

This result has important implications for the data analysis. While in total, four amplitudes can be fit, they represent only two independent parameters that can be estimated from the data.

### 4.8.2 Three-state case

The transition rate matrix is given by:

$$\mathbf{K} = \begin{pmatrix} -k_{21} - k_{31} & k_{12} & k_{13} \\ k_{21} & -k_{12} - k_{32} & k_{23} \\ k_{31} & k_{32} & -k_{13} - k_{23} \end{pmatrix}$$

The eigen-expansion of the three-state transition rate matrix is:

$$\mathbf{K} = \mathbf{\Gamma}^{(0)} \lambda^{(0)} + \mathbf{\Gamma}^{(1)} \lambda^{(1)} + \mathbf{\Gamma}^{(2)} \lambda^{(2)} ;$$
$$e^{\mathbf{K} t_c} = \mathbf{\Gamma}^{(0)} e^{\lambda^{(0)} t_c} + \mathbf{\Gamma}^{(1)} e^{\lambda^{(1)} t_c} + \mathbf{\Gamma}^{(2)} e^{\lambda^{(2)} t_c};$$

$$\lambda^{(0)} = 0 \quad ;$$
$$\lambda^{(1)} = \tfrac{1}{2}\left(\operatorname{Tr}(\mathbf{K}) - \sqrt{\operatorname{Tr}(\mathbf{K})^2 - 4z}\right); \quad \operatorname{Tr}(\mathbf{K}) = -(k_{21} + k_{31} + k_{12} + k_{32} + k_{13} + k_{23})$$
$$\lambda^{(2)} = \tfrac{1}{2}\left(\operatorname{Tr}(\mathbf{K}) + \sqrt{\operatorname{Tr}(\mathbf{K})^2 - 4z}\right); \quad z = z_1 + z_2 + z_3$$
$$\qquad\qquad z_1 = k_{32}k_{13} + k_{12}k_{13} + k_{12}k_{23}$$
$$\lambda^{(1)} + \lambda^{(2)} = \operatorname{Tr}(\mathbf{K}) \qquad z_2 = k_{31}k_{23} + k_{21}k_{13} + k_{21}k_{23}$$
$$\lambda^{(1)} \lambda^{(2)} = z \qquad z_3 = k_{21}k_{32} + k_{31}k_{12} + k_{31}k_{32}$$

where $\operatorname{Tr}(\mathbf{K})$ is the trace of the transition rate matrix, i.e. the sum of the diagonal elements.

The eigen-matrices are calculated using eq. (67):

$$\mathbf{\Gamma}^{(0)} = \frac{\lambda^{(1)}\mathbf{I} - \mathbf{K}}{\lambda^{(1)}} \frac{\lambda^{(2)}\mathbf{I} - \mathbf{K}}{\lambda^{(2)}} \Rightarrow \mathbf{K}^2 = \left(\lambda^{(1)} + \lambda^{(2)}\right)\mathbf{K} - \lambda^{(1)}\lambda^{(2)}(\mathbf{I} - \mathbf{\Gamma}^{(0)});$$

$$\mathbf{\Gamma}^{(1)} = \frac{\mathbf{K}}{\lambda^{(1)}} \frac{\lambda^{(2)}\mathbf{I} - \mathbf{K}}{\lambda^{(2)} - \lambda^{(1)}} = \frac{(\mathbf{K} - \lambda^{(2)}\mathbf{I})(\mathbf{I} - \mathbf{\Gamma}^{(0)})}{\lambda^{(1)} - \lambda^{(2)}}$$

$$\mathbf{\Gamma}^{(2)} = \frac{\mathbf{K}}{\lambda^{(2)}} \frac{\lambda^{(1)}\mathbf{I} - \mathbf{K}}{\lambda^{(1)} - \lambda^{(2)}} = \frac{(\mathbf{K} - \lambda^{(1)}\mathbf{I})(\mathbf{I} - \mathbf{\Gamma}^{(0)})}{\lambda^{(2)} - \lambda^{(1)}}$$

The equilibrium fractions are:

$$\mathbf{x}_d = \frac{1}{z}\begin{pmatrix} z_1 \\ z_2 \\ z_3 \end{pmatrix}$$

Then, the matrices defining the amplitudes of the exponents of the correlation functions are:

$$\mathbf{G}^{(1)} = \mathbf{\Gamma}^{(1)} \mathbf{X}_d = \frac{(\mathbf{K} - \lambda^{(2)}\mathbf{I})\mathbf{\Sigma}_d}{\lambda^{(1)} - \lambda^{(2)}} = \frac{\mathbf{K}\mathbf{X}_d - \lambda^{(2)}\mathbf{\Sigma}_d}{\lambda^{(1)} - \lambda^{(2)}};$$

$$\mathbf{G}^{(2)} = \mathbf{\Gamma}^{(2)} \mathbf{X}_d = \frac{(\mathbf{K} - \lambda^{(1)}\mathbf{I})\mathbf{\Sigma}_d}{\lambda^{(2)} - \lambda^{(1)}} = \frac{\mathbf{K}\mathbf{X}_d - \lambda^{(1)}\mathbf{\Sigma}_d}{\lambda^{(2)} - \lambda^{(1)}};$$

$$\mathbf{\Sigma}_d \stackrel{\text{def}}{=} \mathbf{X}_d - \mathbf{x}_d \otimes \mathbf{x}_d;$$

Both the covariance matrix $\mathbf{\Sigma}_d$ and the matrix $\mathbf{K}\mathbf{X}_d$ can be expressed using the contrast factors:

$$\mathbf{\Sigma}_d = \sum_{i<j} x_d^{(i)} x_d^{(j)} \mathbf{\Delta}^{(ij)} ;$$

$$\mathbf{K}\mathbf{X}_d = -\sum_{i<j} k_{ij} x_d^{(j)} \mathbf{\Delta}^{(ij)} ;$$

$$\mathbf{\Delta}^{(13)} = \begin{pmatrix} 1 & 0 & -1 \\ 0 & 0 & 0 \\ -1 & 0 & 1 \end{pmatrix};$$

The matrices $\mathbf{\Delta}^{(ij)}$ above (one example matrix $\mathbf{\Delta}^{(13)}$ is shown) produce the product of the differences of the correlated quantities in states $i$ and $j$:

$$\mathbf{S}_a^T \mathbf{\Delta}_{ij} \mathbf{S}_b = \left(S_a^{(i)} - S_a^{(j)}\right)\left(S_b^{(i)} - S_b^{(j)}\right)$$

Thus, the expression for the $\mathbf{G}^{(l)}$ matrices can also be expressed using the contrast factors:

$$\mathbf{G}^{(1)} = \frac{1}{\lambda^{(2)} - \lambda^{(1)}} \sum_{i<j} \left(k_{ij} + \lambda^{(2)} x_d^{(i)}\right) x_d^{(j)} \mathbf{\Delta}^{(ij)}$$

$$\mathbf{G}^{(2)} = \frac{1}{\lambda^{(1)} - \lambda^{(2)}} \sum_{i<j} \left(k_{ij} + \lambda^{(1)} x_d^{(i)}\right) x_d^{(j)} \mathbf{\Delta}^{(ij)}$$

The complete kinetic correlation function then takes the form:

$$G_{k,ab}(t_c) = 1 + A_{ab}^{(0)} + A_{ab}^{(1)} e^{-k^{(1)} t_c} + A_{ab}^{(2)} e^{-k^{(2)} t_c};$$

$$k^{(1)} = -\lambda^{(1)}; \quad k^{(2)} = -\lambda^{(2)}; \quad \partial_{ab}^{(ij)} = \frac{S_a^{(i)} - S_a^{(j)}}{S_a} \frac{S_b^{(i)} - S_b^{(j)}}{S_b};$$

$$A_{ab}^{(1)} = p_d \frac{1}{k^{(1)} - k^{(2)}} \sum_{i<j} \partial_{ab}^{(ij)} \left(k_{ij} x_d^{(j)} - k^{(2)} x_d^{(i)} x_d^{(j)}\right);$$

$$A_{ab}^{(2)} = p_d \frac{1}{k^{(2)} - k^{(1)}} \sum_{i<j} \partial_{ab}^{(ij)} \left(k_{ij} x_d^{(j)} - k^{(1)} x_d^{(i)} x_d^{(j)}\right);$$

$$A_{ab}^{(0)} = -A_{ab}^{(1)} - A_{ab}^{(2)} + \sum_{i<j} \partial_{ab}^{(ij)} x^{(i)} x^{(j)} = \sum_{i<j} \partial_{ab}^{(ij)} \left(x^{(i)} x^{(j)} - p_d x_d^{(i)} x_d^{(j)}\right);$$

(104)

Expressions for a higher number of states can be derived similarly.

## Supplementary Note 5 - Ambiguity arising in the cFCS analysis

Let us consider the simplest exchange scheme – a two-state dynamic system in the absence of static states. The kinetic part of the FCS curves for such a system are given by eq. (99) by setting the parameter $p_d$ to 1:

$$G_{k,ab}(t_c) = 1 + A_{ab}\, e^{-k\, t_c}$$
$$k = k_{12} + k_{21}$$
$$A_{ab} = \partial_{ab}^{(12)} x_d^{(1)} \left(1 - x_d^{(1)}\right)$$
$$\partial_{ab}^{(12)} = \frac{S_a^{(1)} - S_a^{(2)}}{\overline{S_a}} \frac{S_b^{(1)} - S_b^{(2)}}{\overline{S_b}}$$
$$x_d^{(1)} = \frac{k_{12}}{k}$$

Phenomenologically, each FCS curve is fully described by three parameters: the relaxation rate $k$ and the auto-correlation amplitudes $A_{aa}$ and $A_{bb}$, which define the cross-correlation amplitudes as $A_{ab} = A_{ba} = \sqrt{A_{aa} A_{bb}}$. On the other hand, the full characterization of the system requires knowledge of six parameters: two transition rates $(k_{12}, k_{21})$ and the four components of the correlated signals $(S_a^{(1)}, S_a^{(2)}, S_b^{(1)}, S_b^{(2)})$. Therefore, the system remains underdetermined as there are more unknowns then independent experimental parameters.

**For fFCS experiment,** the auto-correlation amplitudes do not depend on the properties of the states and are not independent:

$$A_{11} = \frac{1}{A_{22}} = \frac{1 - x_d^{(1)}}{x_d^{(1)}} = \frac{k_{21}}{k_{12}}$$

Thus, the two physical parameters of the system $(k_{12}, k_{21})$ are fully determined by the two fitted parameters $(A_{11}^{fit}, k^{fit})$:

$$\begin{cases} A_{11}^{fit} = \dfrac{k_{21}}{k_{12}} \\ k^{fit} = k_{12} + k_{21} \end{cases} \Rightarrow \begin{cases} k_{12} = k^{fit} \dfrac{1}{A_{11}^{fit} + 1} \\ k_{12} = k^{fit} \left(1 - \dfrac{1}{A_{11}^{fit} + 1}\right) \end{cases}$$

**For cFCS experiment,** the auto-correlation amplitudes are independent and defined by the FRET efficiencies of the two states. Thus, the number of parameters that can be fit is three $(A_{GG}, A_{RR}, k)$ while the number of physical parameters is four $(k_{12}, k_{21}, E^{(1)}, E^{(2)})$. The system remains underdetermined, and there is an infinite number of solutions (this corresponds to Ambiguity I discussed in the main text and the next section). The explicit dependence of the estimated physical parameters can be obtained by varying one of the physical parameters and solving the system of equations:

$$\begin{cases} A_{GG}^{fit} = \dfrac{\left(E^{(1)} - E^{(2)}\right)^2}{(1 - \bar{E})^2} \dfrac{k_{12} k_{21}}{(k_{12} + k_{21})^2} \\ A_{RR}^{fit} = \dfrac{\left(E^{(1)} - E^{(2)}\right)^2}{\bar{E}^2} \dfrac{k_{12} k_{21}}{(k_{12} + k_{21})^2} \\ k^{fit} = k_{12} + k_{21} \end{cases}$$

The dependence of the fitted physical parameters for a particular system is illustrated in Figure S6. Thus, the analysis of FCS experiments alone does not allow to recover all physical parameters of the system, even in the simplest case. The information of the state properties (i.e., their FRET efficiencies) should be introduced into the analysis. In the case of fFCS, this is done at the step of filtering. For cFSC, the joined fitting of FCS and TCSPC can be used. However, as the number of states is rising and additional complications are introduced (for example, the presence of static states), the number of physical parameters rises faster than the number of observed phenomenological parameters, and the system becomes underdetermined again.

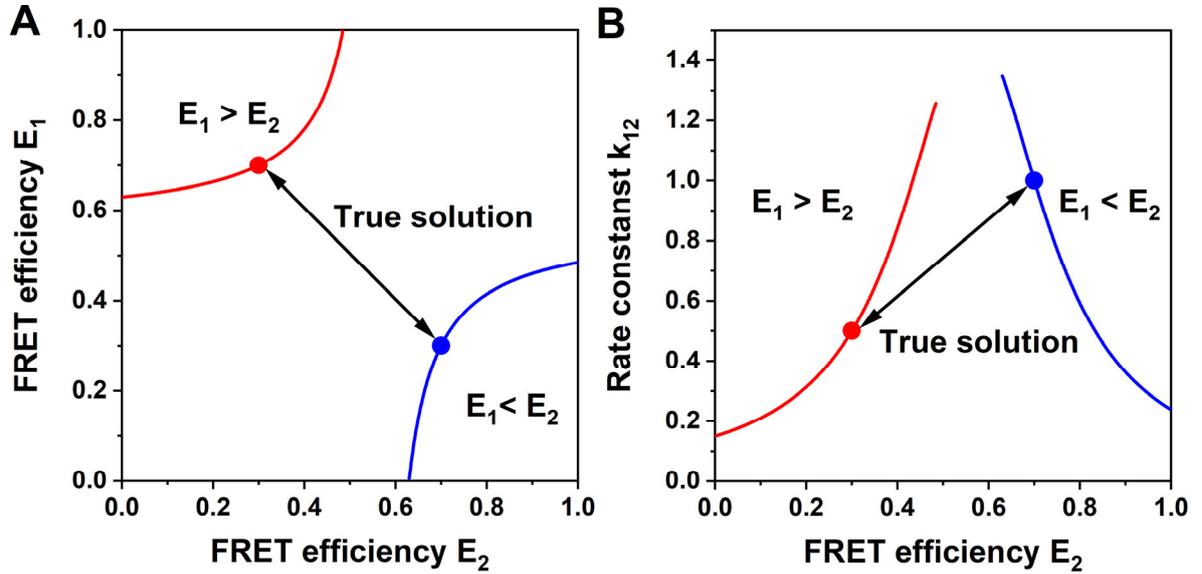

**Fig. S6: Ambiguities in the color-FCS analysis for two-state systems in the absence of static states.** The fitted FRET efficiency in the first state ($E^{(1)\text{fit}}$) and the transition rate ($k_{12}^{\text{fit}}$) are shown as functions of the free parameter $E^{(2)\text{fit}}$. The parameters of the system are $k_{12}^{\text{true}} = 1 \text{ ms}^{-1}$, $k_{21}^{\text{true}} = 0.5 \text{ ms}^{-1}$, $E^{(1)\text{true}} = 0.3$, $E^{(2)\text{true}} = 0.7$. The two branches of the space of possible solutions, indicated in red ($E^{(1)} > E^{(2)}$) and blue ($E^{(1)} < E^{(2)}$), are equivalent and can be transformed into another by exchange of the indexes $1 \leftrightarrows 2$ for all parameters of the system.

# Supplementary Note 6 - Ambiguities arising in the cFCS+TCSPC analysis

In the analysis of the simulations, we found ambiguous solutions for fitting two-state systems in the presence of static states. In the following, we will provide a theoretical description of the observed ambiguity. Specifically, we found an infinite number of solutions that resulted in equal values for the reduced chi-square, $\chi_r^2$. The solutions are undefined because the system is underdetermined, i.e., there are more parameters of the model than independent experimental observables (Ambiguity I). Additionally, the solution space is split into several branches because there is an inherent ambiguity for the system regarding the assignment of the FRET efficiencies to the individual states (Ambiguity II).

## 6.1.1 Ambiguity I – Underdetermined system

We consider a two-state dynamic system in the presence of static states of the same FRET efficiencies and one additional purely static state. We can treat such a system as a three-state system with transition rates $k_{13}, k_{31}, k_{23}, k_{32}$ set to zero. The equations for the kinetic part of the cFCS curves can be derived from eq. (105):

$$G_{k,ab}(t_c) = 1 + A_{ab}^{(\text{off})} + A_{ab}^{(1)}(e^{-k_{FCS} t_c} - 1);$$

$$A_{ab}^{(1)} = \partial_{ab}^{(12)} p_d x_d^{(1)} x_d^{(2)};$$
$$A_{ab}^{(\text{off})} = \partial_{ab}^{(12)} x^{(1)} x^{(2)};$$
$$\partial_{ab}^{(12)} = \frac{S_a^{(1)} - S_a^{(2)}}{\overline{S_a}} \frac{S_b^{(1)} - S_b^{(2)}}{\overline{S_b}};$$
$$k_{FCS} = k_{12} + k_{21};$$
$$a, b \in \{G, R\}$$
(105)

For now, we assume that we know which FRET states participate in dynamic exchange (states 1+2) and which are purely static (state 3). This implies that we know the correct assignment of the different FRET efficiencies to the different states, which we classify as low-FRET (*L*), medium-FRET (*M*) and high-FRET (*H*):

$$\begin{array}{lllll}
\text{Dynamic state fractions:} & x_d^{(1)} & x_d^{(2)} & & \\
\text{Total state fractions:} & x^{(1)} & x^{(2)} & x^{(3)} & \\
\text{FRET efficiencies:} & E^{(1)} < & E^{(2)} < & E^{(3)} & \\
\text{Label:} & L & M & H &
\end{array} \quad (106)$$

The model function for such a system is described by three FRET efficiencies $E^{(i)}$, three total fractions $x^{(i)}$, the total fraction of dynamic molecules $p_d$ and two transition rate constants $k_{12}$ and $k_{21}$. Since the fractions are normalized to one, they represent only two independent parameters of the model. We thus obtain a total of 8 independent model parameters.

The analysis of the fluorescence decays obtained from TCSPC yields a set of three FRET efficiencies ($E^{(i)}$) together with the corresponding total fractions ($x^{(i)}$). Since the fractions sum to 1, we thus determine 5 independent parameters from TCSPC. From the FCS analysis, we obtain one relaxation time ($k_{FCS} = k_{12} + k_{21}$) and three pairs of amplitudes of the correlation function of the dynamic and static contributions, $A_{ab}^{(1)}$ and $A_{ab}^{(0)}$.

However, as described above in section 4.8.1, the cross-correlation amplitude is related to the auto-correlation amplitudes and contains no unique information (see eq. (108) below). Additionally, the offset FCS amplitude, $A_{ab}^{(0)}$, is determined entirely from the quantities determined from the TCSPC analysis, that is, the FRET efficiencies and the total fractions. Thus, the only new information obtained from the FCS amplitudes is the amplitudes of the kinetic contribution $A_{ab}^{(1)}$ of the autocorrelation functions, i.e. $A_{GG}^{(1)}$ and $A_{RR}^{(1)}$, given by eq.(105).

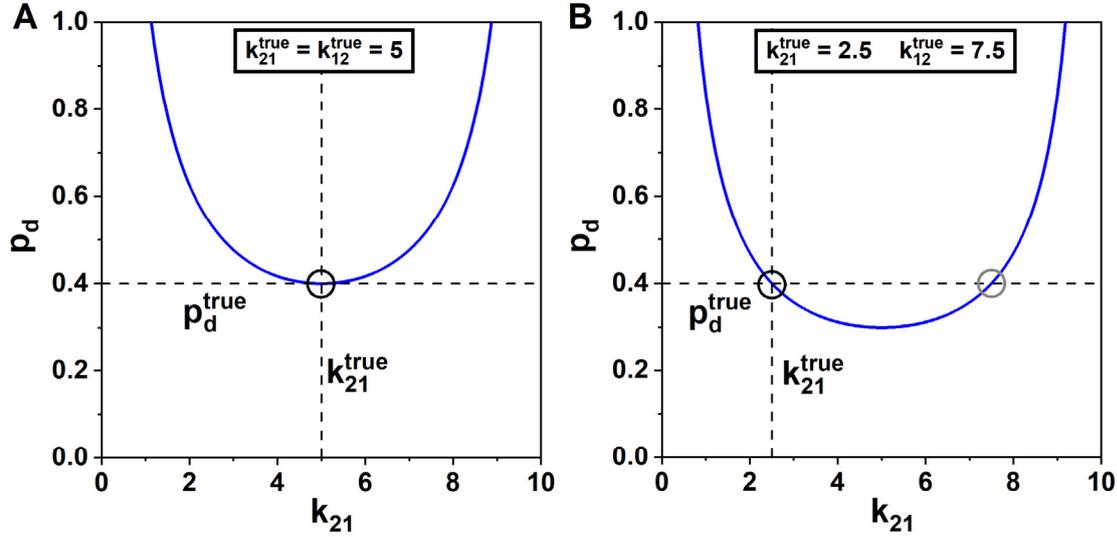

**Figure S7: Ambiguity between the total fraction of dynamic molecules $p_d$ and the transition rate constant $k_{21}$.** The relation as given by eq. (111). The values of the parameters are $p_d^{true} = 0.4$, $k_{12}^{true} = 5$ and $k_{21}^{true} = 5$.

The normalized contrast factors $\partial_{ab}^{(12)}$ depend only on the parameters obtained from the TCSPC analysis (i.e., the FRET efficiencies $E^{(i)}$ and the total state fractions $x^{(i)}$) and are hence known (see section 4.8.1 and eq. (100)):

$$\partial_{ab}^{(12)} = \begin{cases} \left(E^{(1)} - E^{(2)}\right)^2/(1-\bar{E})^2 & , ab = GG \\ \left(E^{(1)} - E^{(2)}\right)^2/\bar{E}^2 & , ab = RR \\ -\left(E^{(1)} - E^{(2)}\right)^2/\left(\bar{E}(1-\bar{E})\right), & ab = RG, GR \end{cases} \quad (107)$$

The cross-correlation amplitudes are related to the autocorrelation amplitudes by:

$$A_{GR}^{(1)} = A_{RG}^{(1)} = -\sqrt{A_{GG}^{(1)} A_{RR}^{(1)}} \quad (108)$$

Therefore, the only independent information that is obtained from the correlation amplitudes is the term $p_d\, x_d^{(1)}(1 - x_d^{(1)})$, which is identical for the two amplitudes $A_{GG}^{(1)}$ and $A_{RR}^{(1)}$. Thus, we obtain only a single independent parameter from the FCS amplitudes, leaving us with a total of 7 independent observed parameters and thus an *underdetermined* system with respect to the 8 model parameters.

Specifically, as seen in equation (105), the ambiguity arises between the quantities $p_d$ and $x_d^{(1)}$ which relates to the transition rate by $x_d^{(1)} = k_{12}/k_{FCS}$. Thus, different combinations of the total fraction of dynamic molecules, $p_d$, and the transition rate $k_{12}$ can result in identical amplitudes of the FCS curves, which results in ambiguous solutions subject to the boundary condition given in equation (106).

In the following, we will derive an expression to describe all possible solutions for a particular value of the FRET efficiencies $E^{(i)}$ and total fractions $x^{(i)}$, i.e., for a given contrast factor $\partial_{ab}^{(12)}$. The fractions $x_d^{(i)}$ are functions of the transition rates $k_{12}, k_{21}$:

$$x_d^{(1)} = \frac{k_{12}}{k_{21} + k_{12}}; \quad x_d^{(2)} = 1 - x_d^{(1)} = \frac{k_{21}}{k_{21} + k_{12}} \quad (109)$$

The same rates define the relaxation rates of correlation functions, $k = k_{12} + k_{21}$. From the fitting of the FCS curves, we estimate the exponential rate $k = k_{FCS}^{fit}$. We also determine the amplitude $A_{ab}^{(1),fit}$. The fitted amplitude must correspond to the true value $A_{ab}^{(1),true}$.

$$A_{ab}^{(1),fit} = A_{ab}^{(1),true}$$
$$p_d^{fit} \frac{k_{12}^{fit} k_{21}^{fit}}{k_{12}^{fit} + k_{21}^{fit}} \partial_{ab}^{(12),fit} = p_d^{true} \frac{k_{12}^{true} k_{21}^{true}}{k_{12}^{true} + k_{21}^{true}} \partial_{ab}^{(12),true} \tag{110}$$

Assuming that we know which states are dynamic (in this case states 1 and 2), the contrast factor is known from the TCSPC information and correct ($\partial_{ab}^{(12),fit} = \partial_{ab}^{(12),true}$). Moreover, we know the correct value for the sum of the rates from the relaxation time of the FCS curves. Thus, we obtain the following dependency of the total fraction of dynamic molecules $p_d$ on the transition rate constant $k_{21}$:

$$\boxed{p_d(k_{21}) = p_d^{true} \frac{k_{12}^{true} k_{21}^{true}}{k_{21}(k_{12}^{true} + k_{21}^{true} - k_{21})}} \tag{111}$$

Hereby, the value of the transition rate $k_{21}$ varies in the range from 0 to $k$. An example of the possible solutions for $p_d$ is given in Figure S7 for equal transition rate constants of $k_{12}^{true} = k_{21}^{true} = 5$ (A) and $k_{12}^{true} = 2.5 / k_{21}^{true} = 7.5$ (B).

In practice, it may be possible to determine $p_d$ from other sources such as the two-dimensional histograms (see Appendix E and main text) or dynamic-PDA analysis. For the case of equal backward and forward rates (Fig. S7 A), the ambiguity is then resolved. However, in the general case of $k_{12}^{true} \neq k_{21}^{true}$, one obtains two principal solutions for the rates which differ by the assignment of the rates to the states (Fig S7 B). This ambiguity arises because $p_d$ only depends on the product of the transition rate constants, which is unchanged under inversion of the state assignment.

On the other hand, one could determine the transition rate constants (or the dynamic fractions $x_d^{(1)}$) from the experimental data, e.g. from the mode of the dynamic distribution (see Appendix E). In this case, the ambiguity is always resolved.

### 6.1.2 Ambiguity II – State assignment

In the previous section, we have assumed that the FRET efficiencies to the different static and dynamic states are known *a priori*. In the experiment, this condition is not fulfilled if only the information from TCSPC and FCS is considered. However, as discussed in the main text, FRET-lines are a powerful tool to address this problem. Unfortunately, ambiguity I remains even in the case that the assignment between dynamic and static states is known because the assignment of the rates to the dynamic states remains ambiguous.

The missing information about the state assignment leads to another ambiguity in the analysis. If $n_d$ is the number of dynamic states and $n$ is the total number of states with different FRET efficiencies, then the number of permutations to assign $n_d$ out of $n$ states as dynamic and $n - n_d$ as static is given by:

$$P(n_d, n) = \frac{n!}{(n - n_d)!'} \tag{112}$$

Therefore, in principle there are $P(n_d, n)$ possibilities to assign FRET-states to the dynamic states. As the amplitude of the dynamic term of the correlation functions, $A_{ab}^{(1)}$, depend on the contrast factor of the correlated quantities (see eq. (107) above), different state assignments may have the same covariance and correlation amplitude and may thus not be differentiable.

**Table E1: Dynamic fractions $x_d$ and dynamic correlation amplitudes $A_{ab}^{(1)}$ for different state assignments of the dynamic system.**

| # | $E^{(L)}$ | $E^{(M)}$ | $E^{(H)}$ | $A_{ab}^{(1)}$ |
|---|---|---|---|---|
| 1 | $x_d^{(1)}$ | $x_d^{(2)}$ | 0 | $p_d x_d^{(1)} x_d^{(2)} \partial_{ab}^{(LM)}$ |
| 1' | $x_d^{(2)}$ | $x_d^{(1)}$ | 0 | |
| 2 | $x_d^{(1)}$ | 0 | $x_d^{(2)}$ | $p_d x_d^{(1)} x_d^{(2)} \partial_{ab}^{(LH)}$ |
| 2' | $x_d^{(2)}$ | 0 | $x_d^{(1)}$ | |
| 3 | 0 | $x_d^{(1)}$ | $x_d^{(2)}$ | $p_d x_d^{(1)} x_d^{(2)} \partial_{ab}^{(MH)}$ |
| 3' | 0 | $x_d^{(2)}$ | $x_d^{(1)}$ | |

For the given example of two dynamic states of three FRET-states, there are $3! = 6$ possibilities to assign the states defined by the FRET efficiencies $\{E^{(L)}, E^{(M)}, E^{(H)}\}$ as given in Table E1. The permutations 1↔1', 2↔2' and 3↔3' differ only by the assignment of the FRET efficiencies to the dynamic states, while the FRET efficiency of the purely static is unchanged. Due to the uncertainty introduced by ambiguity 1, they have the same correlation amplitudes $A_{ab}^{(1)}$ and are thus indistinguishable.

On the other hand, the state assignments 1, 2 or 3, i.e. the choice of the dynamic states to be *L+M*, *L+H* or *M+H*, results in different amplitudes $A_{ab}^{(1)}$. In principle, there could thus be three possible solutions for the system based on the state assignment. Not all of these solutions are physical due to the boundary conditions:

$$0 \leq p_d \leq 1$$
$$0 \leq k_{21} \leq k \tag{113}$$

The possible solutions additionally have to fulfill the condition:

$$A_{ab}^{(1),\text{fit}} = p_d x_d^{(1)} x_d^{(2)} \partial_{ab}^{(ij)}; \quad \{ij\} = \{LM, LH, MH\} \tag{114}$$

Thus, there is an ambiguity of the recovered parameters of the dynamic states caused by the fact that the contrasts factors $\partial_{ab}^{(ij)}$ defined by the different connectivity of states can be compensated by the factor $p_d x_d^{(1)} x_d^{(2)}$ defined by the different transition rates such way that resulting amplitude of correlation curves is the same. This is described by the following condition (compare eq. (110)):

$$A_{ab}^{(1),\text{fit}} = A_{ab}^{(1),\text{true}}$$
$$p_d^{\text{fit}} \frac{k_{12}^{\text{fit}} k_{21}^{\text{fit}}}{k_{12}^{\text{fit}} + k_{21}^{\text{fit}}} \partial_{ab}^{(ij)} = p_d^{\text{true}} \frac{k_{12}^{\text{true}} k_{21}^{\text{true}}}{k_{12}^{\text{true}} + k_{21}^{\text{true}}} \partial_{ab}^{(\text{true})} \tag{115}$$

where $\partial_{ab}^{(\text{true})}$ is the contrast factor of the correct state assignment. Thus, we obtain three branches of solutions for the three different assignments:

$$p_d(k_{21}|ij) = p_d(k_{21}) \frac{\partial_{ab}^{(\text{true})}}{\partial_{ab}^{(ij)}} \tag{116}$$

Where $p_d(k_{21})$ describes the general dependency given by eq. (116). The relation can be further simplified by considering that the contrast factors have the same normalization (see eq. (107)) to:

$$\boxed{p_d(k_{21}|ij) = p_d(k_{21}) \frac{\Delta E_{\text{true}}^2}{\Delta E_{ij}^2}; \quad \{ij\} = \{LM, LH, MH\}} \tag{117}$$

For the systems described in the previous section where the true exchange is between the low-FRET and medium-FRET states (eq. (106) and Fig. F2), an additional solution is found by assigning the dynamic exchange to the low-FRET and high-FRET states (Fig. S8 A and D). The third solution for exchange between the medium-FRET and high-FRET states is non-physical and results in values of $p_d > 1$. Similar solutions are obtained for true exchange between low-FRET/high-FRET and medium-FRET/high-FRET states (Fig. S8 B-C and E-F), whereby all three state assignments result in physical solutions the MH case.

In the case of fFCS, the ratio of the contrast factors $\partial_{ab}^{(ij)}$ is equal to 1 (see eq. (101)). Thus, the ambiguity of type II disappears, however the ambiguity I remains. In conclusion, the fitting of FCS together with TCSPC does not allow to recover the kinetic parameters even for a simple 2-state exchange when static species are present.

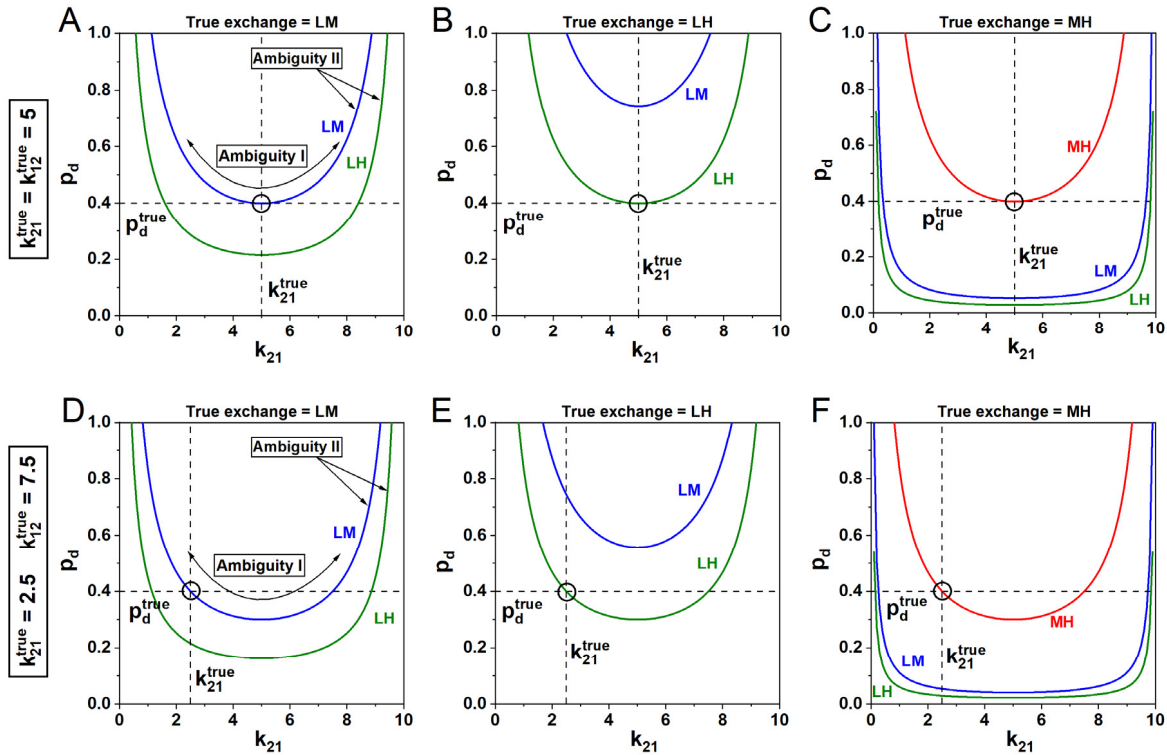

**Fig. S8: Ambiguity arising due to the assignment of dynamic states for true exchange between low-FRET and medium-FRET (LM), low-FRET and high-FRET (LH) and medium-FRET and high-FRET states (MH).** All possible solutions belonging to the different state assignments (blue: LM, green: LH, red: MH) are shown for the combinations of transition rate constants given in Fig. F2 (A-C: $k_{12}^{true} = k_{21}^{true} = 5$, D-F: $k_{12}^{true} = 2.5$, $k_{21}^{true} = 7.5$). For true exchange between LM and LH, only two physical solutions are found, while for exchange between MH valid solutions are obtained for all three state assignments. The remaining parameters are $p_d^{true} = 0.4$, $E^{(L)} = 0.1$, $E^{(M)} = 0.65$ and $E^{(H)} = 0.85$.

## Supplementary Materials and Methods

### 6.2 Simulations of freely diffusing single molecules

Simulations of single-molecule measurements were implemented using a Brownian dynamics approach as described previously[12]. The spatial intensity distribution of the observation volume was modeled by a 3D Gaussian distribution. To simulate the kinetic exchange, the time that a molecule spends in state $i$ is assumed to be exponentially distributed:

$$P(t_i) = k_i^{-1} \exp(-k_i t_i) \tag{118}$$

where the rate $k_i$ represents the sum of all rates depopulating state $i$:

$$k_i = \sum_{j \neq i} k_{ij} \tag{119}$$

Simulated photon counts are saved in SPC-132 data format (Becker & Hickel GmbH, Berlin, Germany) and treated as experimental data.

## 6.3 Simulation parameters

**Table S1:** Fluorescence parameters of the different FRET species used in the simulations for the high-FRET (HF), medium-FRET (MF/MF') and low-FRET (LF) species. Given are the FRET efficiency E, the intensity-weighted average donor fluorescence lifetime $\langle \tau_{D(A)} \rangle_F$ and the brightnesses of the donor fluorophore and the FRET-sensitized acceptor fluorophore $Q_G$ and $Q_R$. The MF species has different parameters in simulations 1-3 compared to simulation 4. In simulation 1, an additional MF' species is included. In all simulations the donor-only lifetime was $\tau_0$= 4 ns, the donor-only brightness was $Q_{D0}$= 50 KHz, the Förster radius was $R_0$= 52 Å, the quantum yield of the donor and acceptor were $\Phi_D$=0.8 and $\Phi_A$=0.32, the spectral crosstalk was $\alpha$=0.017, the diffusion time was t =5 ms and the detection efficiencies ratio between the green and red detection channels was $g_G/g_R$=0.8.

| Parameter | Species | | | | |
|---|---|---|---|---|---|
| | HF | MF (1-3, 5-7) | MF (4) | LF | MF' (1) |
| E | 0.85 | 0.65 | 0.48 | 0.1 | 0.375 |
| $\langle \tau_{D(A)} \rangle_F$, ns | 0.6 | 1.40 | 2.08 | 3.6 | 2.5 |
| $Q_G$, kHz | 7.5 | 17.50 | 26.00 | 45.0 | 31.25 |
| $Q_R$, kHz | 21.38 | 16.55 | 12.44 | 3.27 | 9.91 |

**Table S2:** Kinetic parameters and species fractions of the different FRET species used in the simulations. Given are the average number of molecules in the confocal volume $N_{tot}$, the species fractions of the different species $x$, the total fraction of dynamic molecules $p_d$, the states involved in dynamic exchange and the respective interconversion rates. The total fraction of a given species is related to the total fraction of dynamic molecules, $p_d$, and the static and dynamic fractions of the species, $x_s$ and $x_d$, by $x = p_d x_d + (1 - p_d) x_s$. The MF species has different parameters in simulations 1-3 compared to simulation 4 (see Table S1). In simulation 1, an additional MF' species is included. In all simulations the donor-only lifetime was $\tau_0$= 4 ns, the donor-only brightness was $Q_{D0}$= 50 KHz, the Förster radius was $R_0$= 52 Å, the quantum yield of the donor and acceptor were $\Phi_D$=0.8 and $\Phi_A$=0.32, the spectral crosstalk was $\alpha$=0.017, the diffusion time was $t_d$=5 ms and the detection efficiencies ratio between the green and red detection channels was $g_G/g_R$=0.8.

| Simulation | $N_{tot}$ | Total species fractions $x$ | | | | Dynamic fraction, $p_d$ | Dynamic states | Rates, ms$^{-1}$ |
|---|---|---|---|---|---|---|---|---|
| | | HF | MF | MF' | LF | | | |
| 1 | 0.016 | 0.25 | 0.25 | 0.25 | 0.25 | 0 | - | - |
| 2 | 0.015 | 0.2 | 0.2 | 0 | 0.2 | 0.4 | MF⇌LF | MF→LF: 5.0 / LF→MF: 5.0 |
| 3 | 0.015 | 0.2 | 0.2 | 0 | 0.2 | 0.4 | HF⇌LF | HF→LF: 6.3 / LF→HF: 3.7 |
| 4 | 0.015 | 0.2 | 0.2 | 0 | 0.2 | 0.4 | HF⇌LF | HF→LF: 5.0 / LF→HF: 5.0 |
| 5 | 0.015 | 0.2 | 0.2 | 0 | 0.2 | 0.4 | LF⇌HF⇌MF | All rates: 5.0 |
| 6 | 0.015 | 0.2 | 0.2 | 0 | 0.2 | 0.4 | LF⇌MF⇌HF | All rates: 5.0 |
| 7 | 0.015 | 0.2 | 0.2 | 0 | 0.2 | 0.4 | MF⇌LF⇌HF | All rates: 5.0 |

## 6.4 Analysis of simulation 1

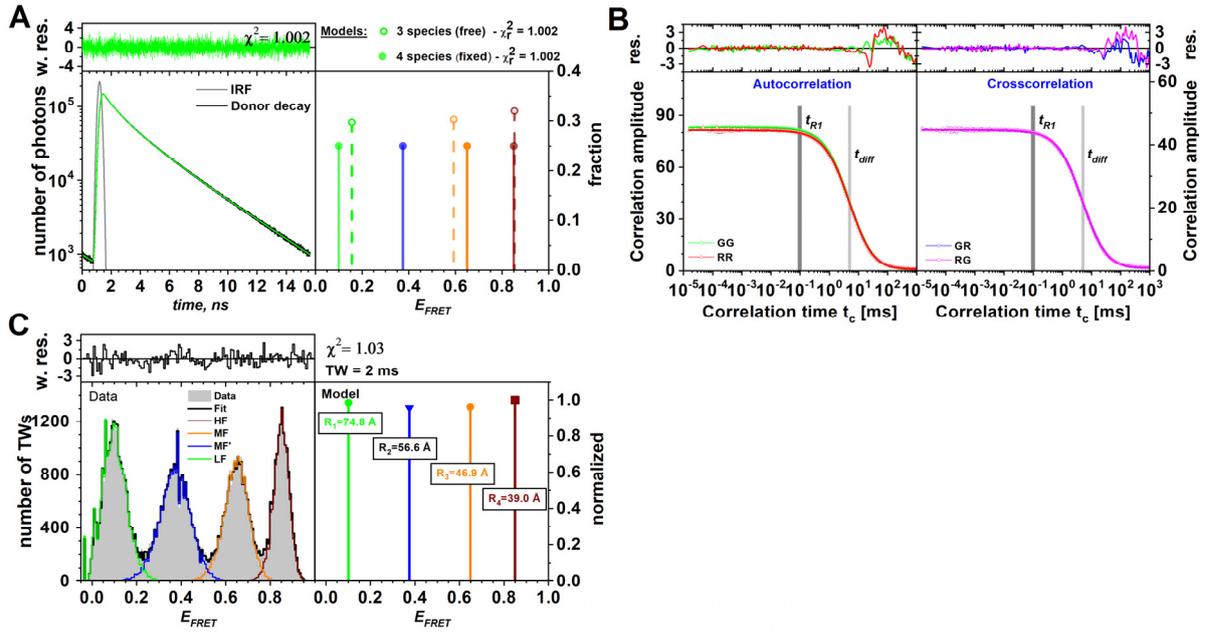

**Fig. S9: Quantitative analysis of simulation 1 containing four static species of different FRET efficiencies. A)** Ensemble fluorescence decays of the donor fluorophore (left) and recovered FRET efficiency components (right). As a fit with three FRET species results in identical $\chi_r^2$ compared to the ground truth model, it is impossible to identify the number of states correctly using only the TCSPC information. **B)** Color-FCS autocorrelation and cross-correlation functions reveal no kinetic contribution as is visible from the plateau of the correlation functions up to the timescale of diffusion ($t_{\text{diff}}$ = 5 ms). **C)** Photon distribution analysis (PDA) reveals the four FRET species and correctly recovers the distances. Weighted residuals (w. res.) of the fits are given above. The simulation parameters are given in Table S1 and S2.

## 6.5 Species autocorrelation functions of simulated datasets 4-7

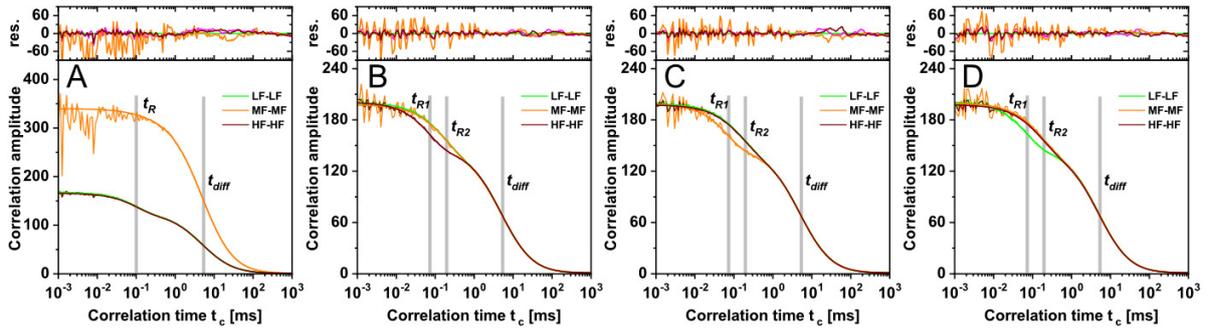

**Fig. S10: Species autocorrelation functions of the three species LF, MF and HF for the simulated datasets 4-7 shown in Fig. 20 of the main text.** For the binary exchange (A), an additional bunching term is obtained only for the autocorrelation functions of the exchanging species (LF,HF). For the linear three-state kinetic networks (B-D), the fastest decay of the bunching term exchange is detected for the center state of the network (B: LF⇌HF⇌MF, C: HF⇌MF⇌LF, D: MF⇌LF⇌HF). The auto- and cross-correlation curves are fitted to a kinetic model involving one (E) or two (F-H) relaxation times $t_R$ and a diffusion model with a global diffusion time $t_{\text{diff}}$. Weighted residuals of the fits are given above. The simulation parameters are given in Table S1.